\documentclass[global]{svjour}

\usepackage[a4paper, left=25mm, width =160mm, top =25mm, height=245mm]{geometry}

\usepackage{amsmath}
\usepackage{natbib} 
\usepackage{graphics}
\usepackage[font={small,it}]{caption}
\usepackage{subcaption}
\usepackage{dcolumn}
\usepackage{bm}
\usepackage{hyperref}
\usepackage{float}
\usepackage{xcolor}
\journalname{Computational Mechanics}

\begin{document}
	
\title{Prediction of Aerodynamic Flow Fields Using Convolutional Neural Networks}
\author{ Saakaar Bhatnagar\inst{1} \and Yaser Afshar \inst{1}  \and Shaowu Pan\inst{1} \and Karthik Duraisamy \inst{1} \and Shailendra Kaushik\inst{2}
}                     
%
%
\institute{\inst{1} Dept of Aerospace Engineering, \\ University of Michigan, \\ Ann Arbor, MI 48109, United States \\ \inst{2}
	General Motors Global R\&D, \\ Warren, MI 48092, United States \\
	\email{yafshar@umich.edu,kdur@umich.edu}}
%
\date{Received: date / Revised version: date}
\maketitle
\begin{abstract}
An approximation model based on convolutional neural networks (CNNs) is proposed for flow field predictions. The CNN is used to predict the velocity and pressure field in unseen flow conditions and geometries given the pixelated shape of the object. In particular, we consider Reynolds Averaged Navier-Stokes (RANS) flow solutions over airfoil shapes. The CNN can automatically detect essential features with minimal human supervision and shown to effectively estimate the velocity and pressure field orders of magnitude faster than the RANS solver,  making it possible to study the impact of the airfoil shape and operating conditions on the aerodynamic forces and the flow field in near-real time.  The use of specific convolution operations, parameter sharing, and robustness to noise are shown to enhance the predictive capabilities of CNN.  We explore the network architecture and its effectiveness in predicting the flow field for different airfoil shapes, angles of attack, and Reynolds numbers.
\end{abstract}

\section{Introduction}
\label{intro}
With advances in computing power and computational algorithms, 
simulation-based design and optimization has matured to a 
level that it plays a significant role in an industrial setting.
In many practical engineering applications, however, the analysis of the flow field tends to be the most computationally intensive and time-consuming part of the process. 
These drawbacks make the design process tedious, time consuming, and costly, requiring a significant amount of user intervention in design explorations, thus proving to be a barrier between designers from the engineering process. 
	
Data-driven methods have the potential to augment~\citep{Duraisamy:2019} or replace~\citep{Guo:2016} these expensive high-fidelity analyses with less expensive approximations. Learning representations from the data, especially in the presence of spatial and temporal dependencies, have traditionally been limited to hand-crafting of features by domain experts.  Over the past few years, deep learning approaches~\citep{Bengio:2009, Schmidhuber:2015} have shown significant successes in learning from data, and have been successfully used in the development of novel computational approaches~\citep{Raissi:2018, Raissi:2018b, Raissi:2019}.

Deep learning presents a fast alternative solution as an efficient function approximation technique in high-dimensional spaces. 
Deep learning architectures such as deep neural networks (DNNs), routinely used in data mining,  are well-suited for application on big, high-dimensional data sets, to extract multi-scale features.
	
\begin{figure}[H]
	\centering
	\resizebox{\textwidth}{!}{%
		\includegraphics{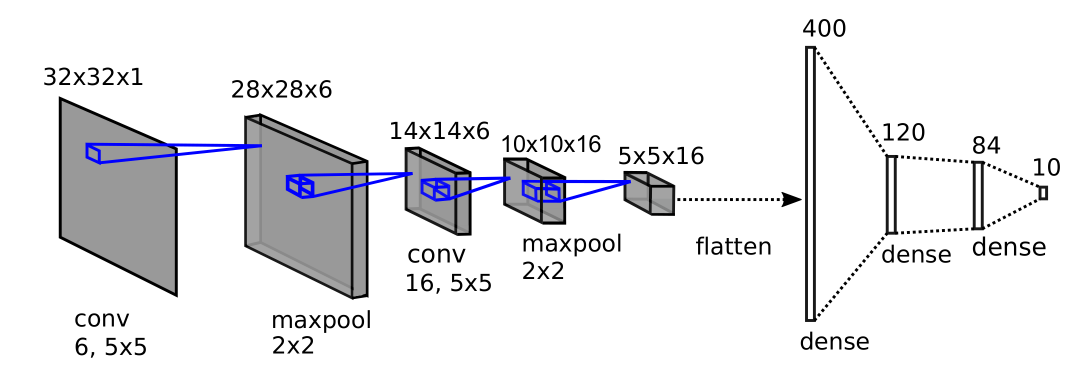}
	}
	\caption{An example of a typical CNN architecture is a LeNet-5 architecture~\citep{Lecun:1998} in identifying handwritten digits for zip code recognition in the postal service.
    The architecture consists of two sets of convolutional and pooling layers. Convolutional layers use a subset of the previous layers for each filter and followed by a flattening convolutional layer, and two fully-connected layers and finally a classifier~\citep{Lecun:1998}.}
	\label{fig:fig1}
\end{figure}

Deep convolutional neural networks (CNN) belong to a class of DNNs, most commonly applied to the analysis of visual imagery.
Previous works~\citep{Lecun:1998, Taylor:2010, Zuo:2015} have illustrated the promise of CNNs to learn high-level features even when the data has strong spatial and temporal correlations. 
Increasing attention being received by CNNs in  fluid mechanics partly originates from their potential benefit of flexibility in the shape representation and scalability for 3D and transient problems.
Figure~\ref{fig:fig1} illustrates the simplified layout of a typical CNN, LeNet-5~\citep{Lecun:1998} applied to the handwritten digit recognition task.
	
The main advantage of a CNN is that it exploits the low dimensional high-level abstraction by convolution. 
The key idea of CNN is to learn the representation and then to use a fully connected standard layer to fit the relationship between the high-level representation and output.
	
\subsection{State of the art in application of CNNs in fluid dynamics}
\label{State_of_the_art}
The use of deep neural networks in computational fluid dynamics recently has been explored in some rudimentary contexts.

\cite{Guo:2016} reported the analysis and prediction of non-uniform steady laminar flow fields around bluff body objects by employing a convolutional neural network (CNN). The authors reported a computational cost lower than that required for numerical simulations by GPU-accelerated CFD solver. Though this work was pioneering in the sense that it demonstrated generalization capabilities, and that CNNs can enable a rapid estimation of the flow field, emphasis was on qualitative estimates of the velocity field, rather than on precise aerodynamic characteristics. 
	
\cite{Miyanawala:2017} used a CNN to predict aerodynamic force coefficients of bluff bodies at a low Reynolds number for different bluff body shapes. They presented a data-driven method using  CNN and the stochastic gradient-descent for the model reduction of the Navier-Stokes equations in unsteady flow problems.  
	
\cite{Lee:2017, Lee:2018} used a generative adversarial network (GAN) to predict unsteady laminar vortex shedding over a circular cylinder.  They presented the capability of successfully learning and predicting both spatial and temporal characteristics of the laminar vortex shedding phenomenon.
	
\cite{hennigh:2017b} presented an approach to use a DNN to compress both the computation time and memory usage of the Lattice Boltzmann flow simulations. The author employed convolutional autoencoders and residual connections in an entirely differentiable scheme to shorten the state size of simulation and learn the dynamics of this compressed form. 
	
\cite{Tompson:2016} proposed a data-driven approach for calculating numerical solutions to the inviscid Euler equations for fluid flow.  In this approach, an approximate inference of the sparse linear system used to enforce the Navier-Stokes incompressibility condition,  the  “pressure projection”  step. This approach cannot guarantee an exact solution pressure projection step, but they showed that it empirically produces very stable divergence-free velocity fields whose runtime and accuracy is better than the Jacobi method while being orders of magnitude faster. 
	
\cite{Zhang:2017} employed a CNN as feature extractor for a low dimensional surrogate modeling.  They presented the potential of learning and predicting lift coefficients using the geometric information of airfoil and operating parameters like Reynolds number, Mach number, and angle of attack.  However, the output is not the flow field around the airfoil but the pressure coefficients at several locations. It is unclear whether this model would have good performance in predicting the drag and pressure coefficient when producing the flow field at the same time.

The primary contribution of the present work is a framework that can be used to predict the flow field around  different geometries under variable flow conditions.
Towards this goal and following~\cite{Guo:2016}, we propose a framework with a general and flexible approximation model for near real-time prediction of non-uniform steady RANS flow in a domain based on convolutional neural networks. In this framework, the flow field can be extracted from simulation data by learning the relationship between an input feature extracted from geometry and the ground truth from a RANS simulation. Then without standard convergence requirements of the RANS solver, and its number of iterations and runtime, which are irrelevant to the prediction process, we can directly predict the flow behavior in a fraction of the time. In contrast to  previous studies, the present work is focused on a more rigorous characterization of aerodynamic characteristics. The present study also improves on computational aspects. For instance,  \cite{Guo:2016} use an separated decoder, whereas the present work employs  shared-encoding and decoding layers, which are  computationally  efficient compared to the separated alternatives.

\section{ Methodology}
\label{Methodology}

\subsection{CFD Simulation}
\label{CFD} 

In this work, flow computations and analyses are performed using the OVERTURNS CFD code~\citep{Duraisamy:2005,Lakshminarayan:2010}.
This code solves the compressible RANS equations using a preconditioned dual-time scheme~\citep{Pandya:2003}. Iterative solutions are pursued using the implicit approximate factorization method~\citep{Pulliam:1981}. Low Mach preconditioning~\citep{Turkel:1999} is used to improve both convergence properties and the accuracy of the spatial discretization. 
A third order Monotonic Upwind Scheme for Conservation Laws (MUSCL)~\citep{VanLeer:1979} with Koren's limiter~\citep{Koren:1993} and Roe's flux difference splitting~\citep{Roe:1986} is used to compute the inviscid terms. Second order accurate central differencing is used for the viscous terms. 
The RANS closure is the SA~\citep{Spalart:1992} turbulence model and $\gamma - \overline{Re_{\theta t}}$ model~\citep{Medida:2011} is used to capture the effect of the flow transition. 
No-slip boundary conditions imposed on the airfoil surface. The  governing equations are provided in the Appendix. 
	
\begin{figure}[H]
	\centering
	\begin{subfigure}{0.45\textwidth}
		\centering
		\resizebox{\textwidth}{!}{%
			\includegraphics{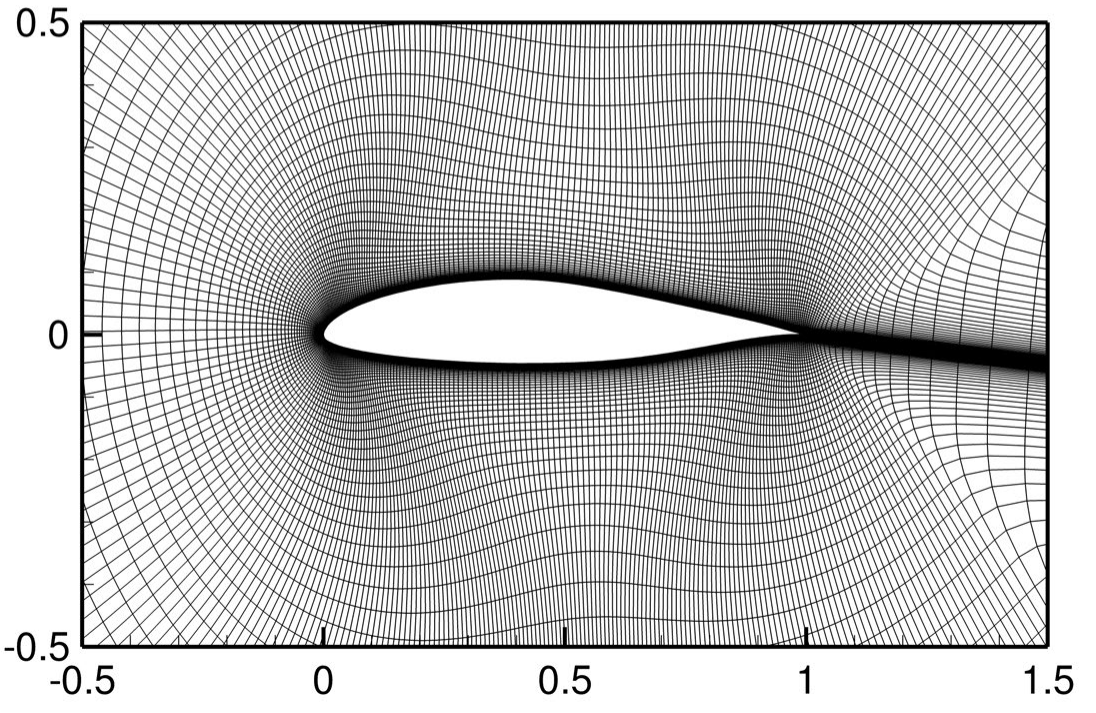}
		}
		\caption{S805 airfoil.}
	\end{subfigure}
	\begin{subfigure}{0.45\textwidth}
		\centering
		\resizebox{\textwidth}{!}{%
			\includegraphics{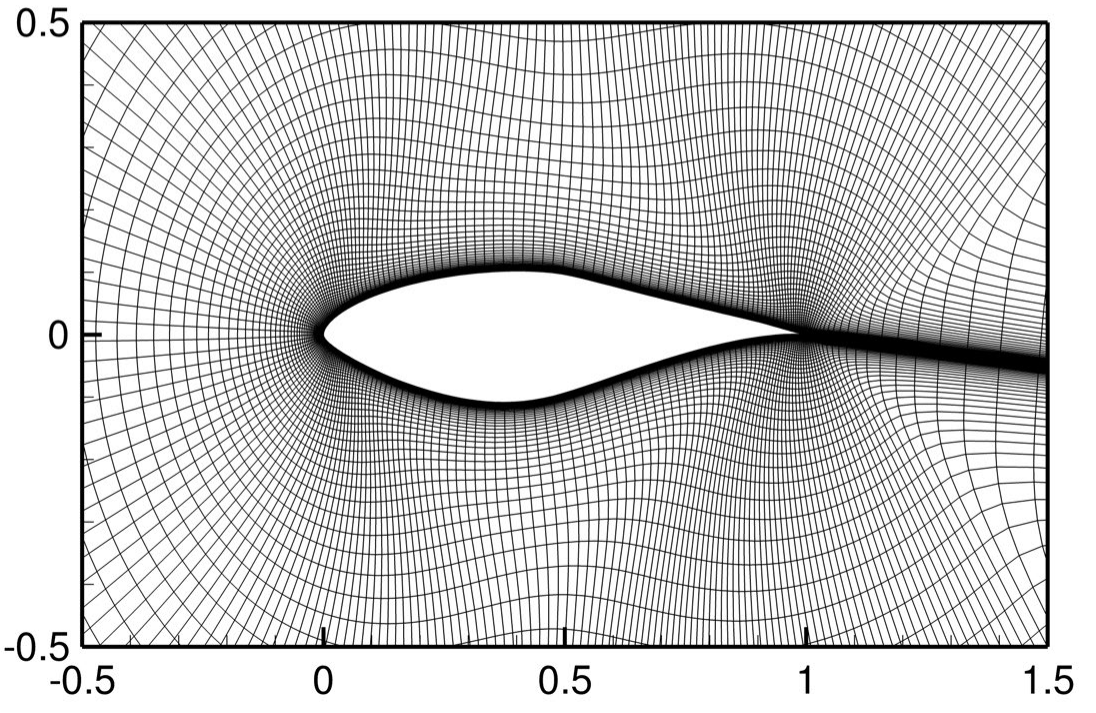}
		}
		\caption{S809 airfoil.}
	\end{subfigure}
	\begin{subfigure}{0.45\textwidth}
		\centering
		\resizebox{\textwidth}{!}{%
			\includegraphics{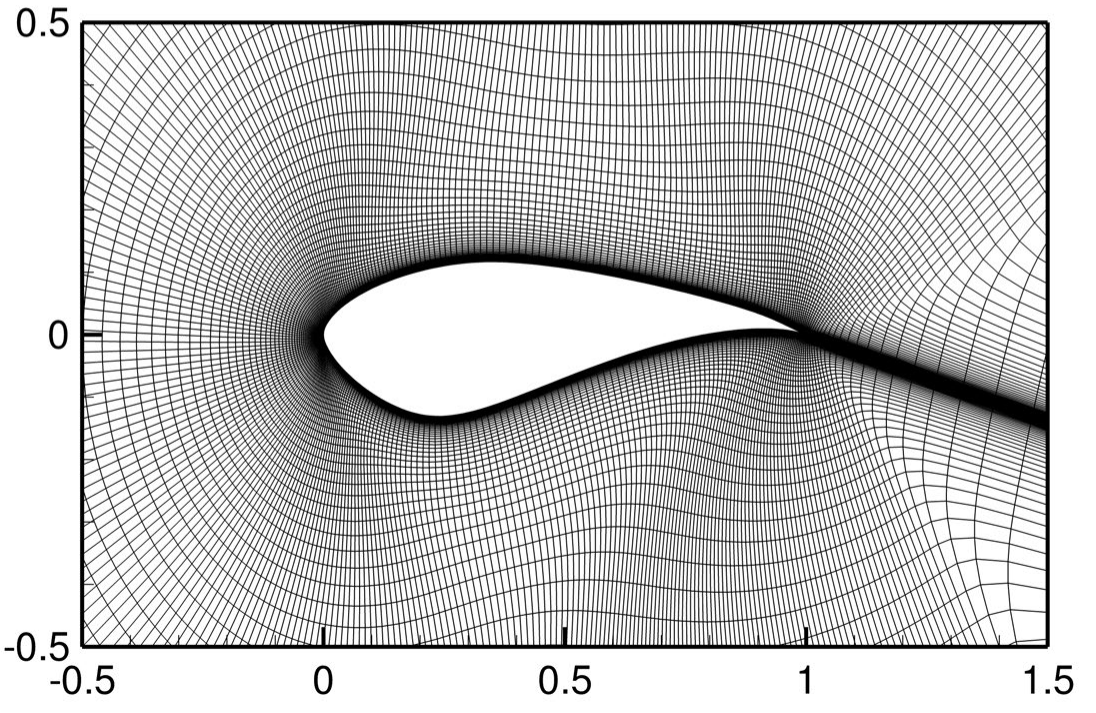}
		}
		\caption{S814 airfoil.}
	\end{subfigure}
	\caption{Zoomed-in view of the structured C-mesh adjacent to the airfoil surface.}
	\label{fig:fig2}
\end{figure}

Simulations are performed over the S805~\citep{Somers:1997a}, S809~\citep{Somers:1997b}, and S814~\citep{Somers:2004} airfoils.
S809 and S814 are among a family of airfoils which contain a region of pressure recovery along the upper surface which induces a smooth transition from laminar to turbulent flow (so-called ``transition-ramp'').  These airfoils are utilized in wind turbines~\citep{Aranake:2012}.
Computations are performed using structured C-meshes with dimensions $394 \times 124$ in the wrap-around and normal directions respectively. Figure~\ref{fig:fig2} shows the airfoils and their near-body meshes.

Simulations are performed at Reynolds numbers $0.5,~1,~2,~\text{and}~3 \times 10^6$, respectively, and a low Mach number of $0.2$ is selected to be representative of wind turbine conditions. At each Reynolds number, the simulation is performed for different airfoils with a sweep of angles of attack from $\alpha=0^{\circ}$  to $\alpha=20^{\circ}$. The OVERTURNS CFD code has been validated for relevant wind turbine applications in~\citep{Aranake:2012}.
	
	
\subsection{Convolutional Neural Networks}
\label{CNN} 
	
In this study, we consider the convolutional neural network to extract relevant features from fluid dynamics data and to predict the entire flow field in near real-time.
The objective is a properly trained CNN which can construct the flow field around an airfoil in a non-uniform turbulence field, using only the shape of the airfoil and fluid flow characteristics of the free stream in the form of the angle of attack and Reynolds number.
In this section, we describe the structure and components of the  proposed CNN.
	
\subsection{Network Structure}
\label{Structure} 
\begin{figure}[H]
	\centering
	\begin{subfigure}{\textwidth}
		\centering
		\resizebox{\textwidth}{!}{%
			\includegraphics{fig3_sep}
		}
		\caption{Shared-encoder but separated decoder.}
		\label{fig:fig3a}
	\end{subfigure}
	\begin{subfigure}{\textwidth}
		\centering
		\resizebox{\textwidth}{!}{%
			\includegraphics{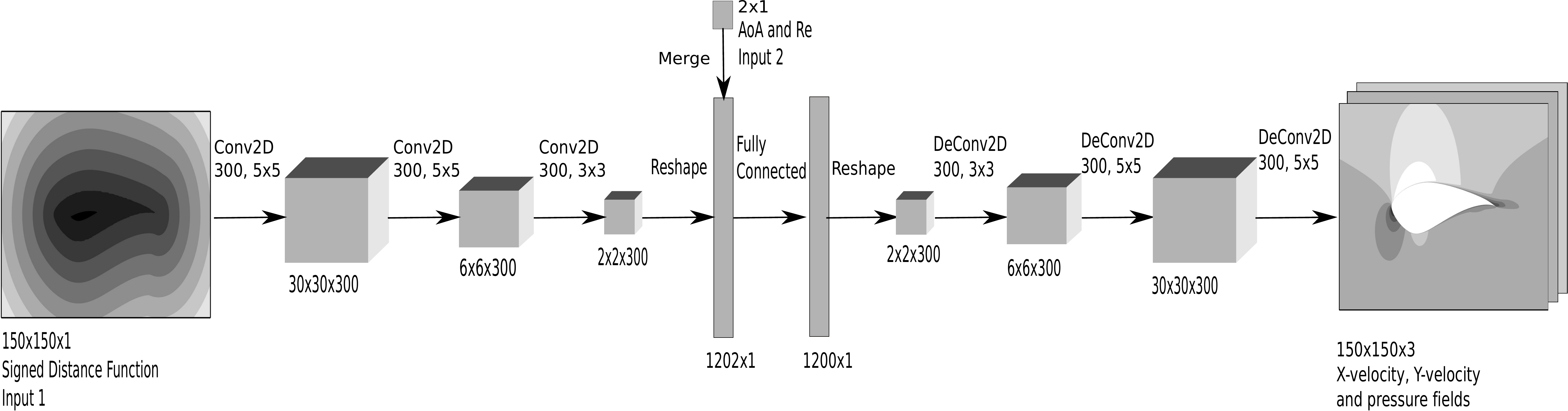}
		}
		\caption{Shared-encoder and decoder.}
		\label{fig:fig3b}
	\end{subfigure}
	\caption{CNN based network architecture for airfoil geometry in the prediction of aerodynamic flow fields. Arrows indicate the forward operation directions. Below each operation,  is the kernel size and the number of filters. The strides for convolutional and deconvolutional layers are with the size of one in each direction. Following each layer is a Swish activation function~\citep{Ramachandran:2017} except in the output layer.}
	\label{fig:fig3}
\end{figure}

To develop suitable CNN architectures for variable flow conditions and airfoil shapes, we build our model based on an encoder-decoder CNN, similar to the model proposed by~\cite{Guo:2016}. Encoder-decoder CNNs are most widely used for machine translation from a source language to a target language~\citep{Chollampatt:2018}. The encoder-decoder CNN has three main components:  a stack of convolution layers, followed by a dense layer and subsequently another stack of convolution layers. 
Figure~\ref{fig:fig3} illustrates the proposed CNN architecture designed in this work.
	
\cite{Guo:2016} used a shared-encoder but separated decoder. We conjecture that the separated decoder may be a limiting performance factor. To address this issue, we designed shared-encoding and decoding layers in our configuration, which save computations compared to the separated alternatives. Explicitly, the weights of the layers of the decoder are shared where they are responsible for extracting high-level representations of pressure and different velocity components. This design provides the same accuracy of the separated decoders but, it is almost utilized fifty percent fewer parameters compared to the separated alternatives. Also, in the work of~\cite{Guo:2016}, the authors used only one low Reynolds number for all the experiments, but here, the architecture is trained with four high Reynolds numbers, three airfoils with different shapes and 21 different angles of attacks. In this architecture, we use three convolution layers both in the shared-encoding and decoding parts. 
	
The inputs to the network are the airfoil shape and the free stream conditions of the fluid flow. 
We use the convolution layers to extract the geometry representation from the inputs. 
The decoding layers use this representation in convolution layers and generate the mapping from the extracted geometry representation to the pressure field and different components of the velocity.
The network uses the Reynolds number, the angle of attack, and the shape of the airfoil in the form of $150 \times 150$ 2D array created for each data entry. The geometry representation has to be extracted from the RANS mesh and fed to the network with images. Using images in CNNs allows encoding specific properties into the architecture, and reducing the number of parameters in the network.
	
\subsection{ Geometry Representation}
\label{Geometry}
A wide range of approaches are employed to capture shape details and to classify points into a learnable format. Among popular examples are methods like implicit functions in image reconstruction~\citep{Hoppe:1992, Carr:2001, Kazhdan:2006, Fuhrmann:2014}, or shape representation and classification~\citep{Zhang:2004a, Ling:2007, Xu:2015, Fernando:2015}.
In applications such as rendering and segmentation and in extracting structural information of different shapes, signed distance functions (SDF) are  widely used. SDF provides a universal representation of different geometry shapes and represents  a grid sampling of the minimum distance to the surface of an object. It also works efficiently with neural networks for shape learning. In this study, to capture shape details in different object representations, and following~\citep{Guo:2016,Prantl:2017}, we use the SDF sampled on a Cartesian grid. \cite{Guo:2016} reported the effectiveness of SDF in representing the geometry shapes for CNNs. The authors empirically showed that the values of SDF on the Cartesian grid provide not only local geometry details but also contain additional information on the global geometry structure.
	
\subsection{Signed Distance Function}
\label{SDF}
\begin{figure}[H]
	\centering
	\resizebox{0.75\textwidth}{!}{%
		\includegraphics{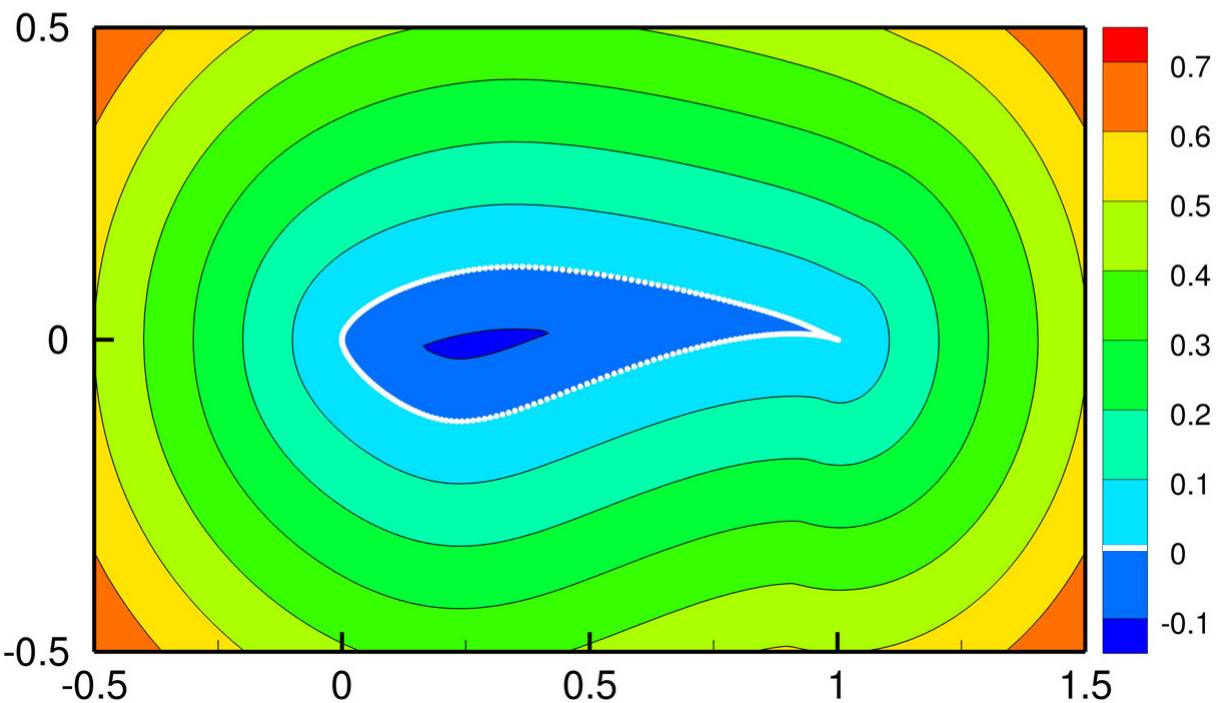}
	}
	\caption{A signed distance function contour plotted for the S814 airfoil in a $150 \times 150$ Cartesian grid. The magnitude of the SDF values on the Cartesian grid equals the shortest distance to the airfoil. The airfoil boundary points are in white.}
	\label{fig:fig4}
\end{figure}

A mathematical definition of the signed distance function of a set of points $\textbf{X}$ determines the minimum distance of each given point $\textbf{x} \in \textbf{X}$ from the boundary of an object $\partial\Omega$. 
\begin{align}
\textrm{SDF}(\textbf{x}) = \left\{\begin{matrix}
d(\textbf{x}, \partial \Omega) & ~\textbf{x}\notin\Omega \\
0 & ~~~\textbf{x}\in\partial\Omega\\
-d(\textbf{x}, \partial \Omega) & ~\textbf{x}\in\Omega 
\end{matrix}\right.,
\label{eq:eq1}
\end{align}
where, $\Omega$ denotes the object, and $d(\textbf{x}, \partial \Omega) = \min_{\textbf{x}_I \in \partial \Omega}{\left( |\textbf{x} - \textbf{x}_I |\right )}$ measures the shortest distance of each given point $\textbf{x}$ from the object boundary points.
The distance sign determines whether the given point is inside or outside of the object.
Figure~\ref{fig:fig4} illustrates the signed distance function contour plot for a S814 airfoil.
	
\begin{figure}[H]
	\centering
	\resizebox{0.75\textwidth}{!}{%
		\includegraphics{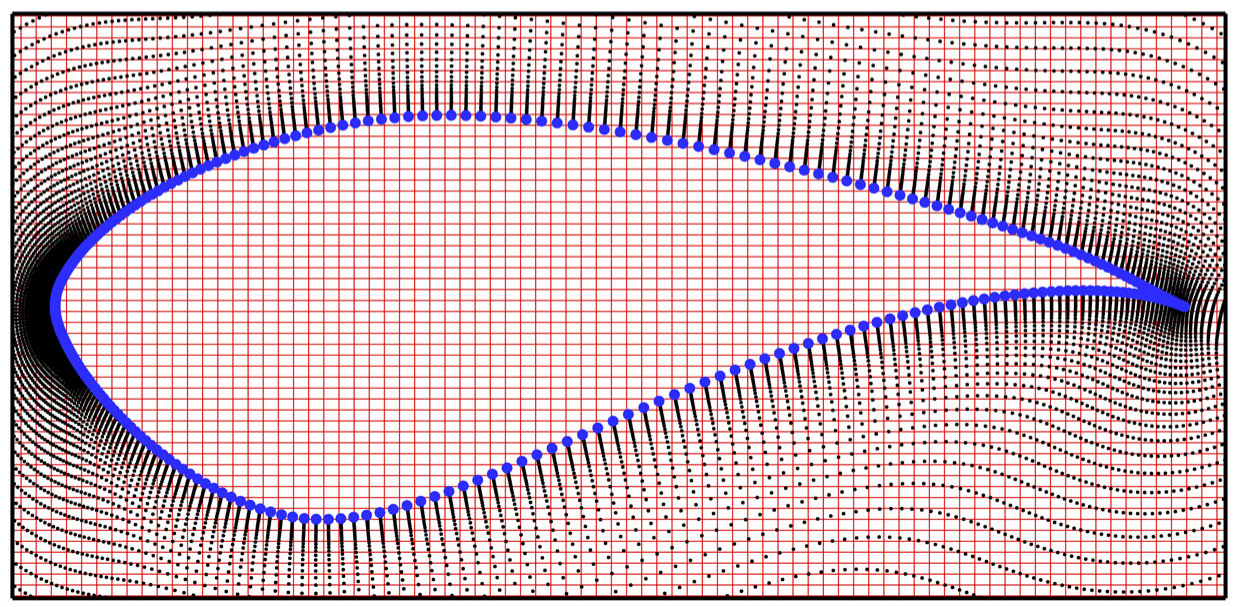}
	}
	\caption{Zoomed-in view of the discrete structured C-mesh representation of an airfoil (S814) (in black) on a Cartesian grid (in red). The airfoil boundary points are in blue.}
	\label{fig:fig5}
\end{figure}

Here, the SDF has positive values at points which are outside of the airfoil, and it decreases as the point approaches the boundary of the airfoil where the SDF is zero, and it takes negative values inside the airfoil.
Fast marching method~\citep{Sethian:1996} and fast sweeping method~\citep{Zhao:2005} are among the popular algorithms for calculating the signed distance function.
To generate a signed distance function, we use the CFD input structured C-mesh information and define the points around the object (airfoil). 
Figure~\ref{fig:fig5} shows the C-mesh representation of an airfoil (S814) and its boundary points on a Cartesian grid.
	
We find the distance of Cartesian grid points from the object boundary points, using the fast marching method~\citep{Sethian:1996}.
To find out whether a given point is inside, outside, or just on the surface of the object, we search the boundary points and compute the scalar product between the normal vector at the nearest boundary points and the vector from the given point to the nearest one and judge the function sign from the scalar product value. For other non-convex objects, one can also use different approaches of crossing number or winding number method which are common in ray casting~\citep{Foley:1995}.
	
After pre-processing the CFD mesh files, we use the SDF as an input to feed the encoder-decoder architecture with multiple layers of convolutions.
Convolution layers in the encoding-decoding part extract all the geometry features from the SDF.
	
\subsection{Convolutional Encoder-Decoder Approach}
\label{Encoder}
To learn all the geometry features from an input SDF, we compose the encoder and decoder with convolution layers and convolutional filters. Every convolutional layer is composed of 300 convolutional filters. Therefore, a convolution produces a set of 300  activation maps. Every convolution in our design is wrapped by a non-linear Swish activation function \citep{Ramachandran:2017}.
Swish is defined as $x.\sigma(\beta x)$ where $\sigma(z)=(1+exp(-z))^{-1}$ is the sigmoid function and $\beta$ is either a constant or a trainable parameter.
The resulting activation maps are the encoding of the input in a low dimensional space of parameters to learn.
The decoding operation is a convolution as well, where the encoding architecture fixes the hyper-parameters of the decoding convolution. Compared to the encoding convolution layer, here a convolution layer has reversed forward and backward passes. This inverse operation is sometimes referred to ``deconvolution''.
The decoding operation unravels the high-level features encoded and transformed by the encoding layers and generates the mapping to the pressure field and different components of the velocity. When we use the CNN, neurons in the same feature map plane have identical weights so that the network can study concurrently, and it learns implicitly from the training data. The training phase of the CNN comprises the input function, the feed-forward process, and the back-propagation process.
	
\subsection{Data Preparation}
\label{Data}
\begin{figure}[H]
	\centering
	\resizebox{0.75\textwidth}{!}{%
		\includegraphics{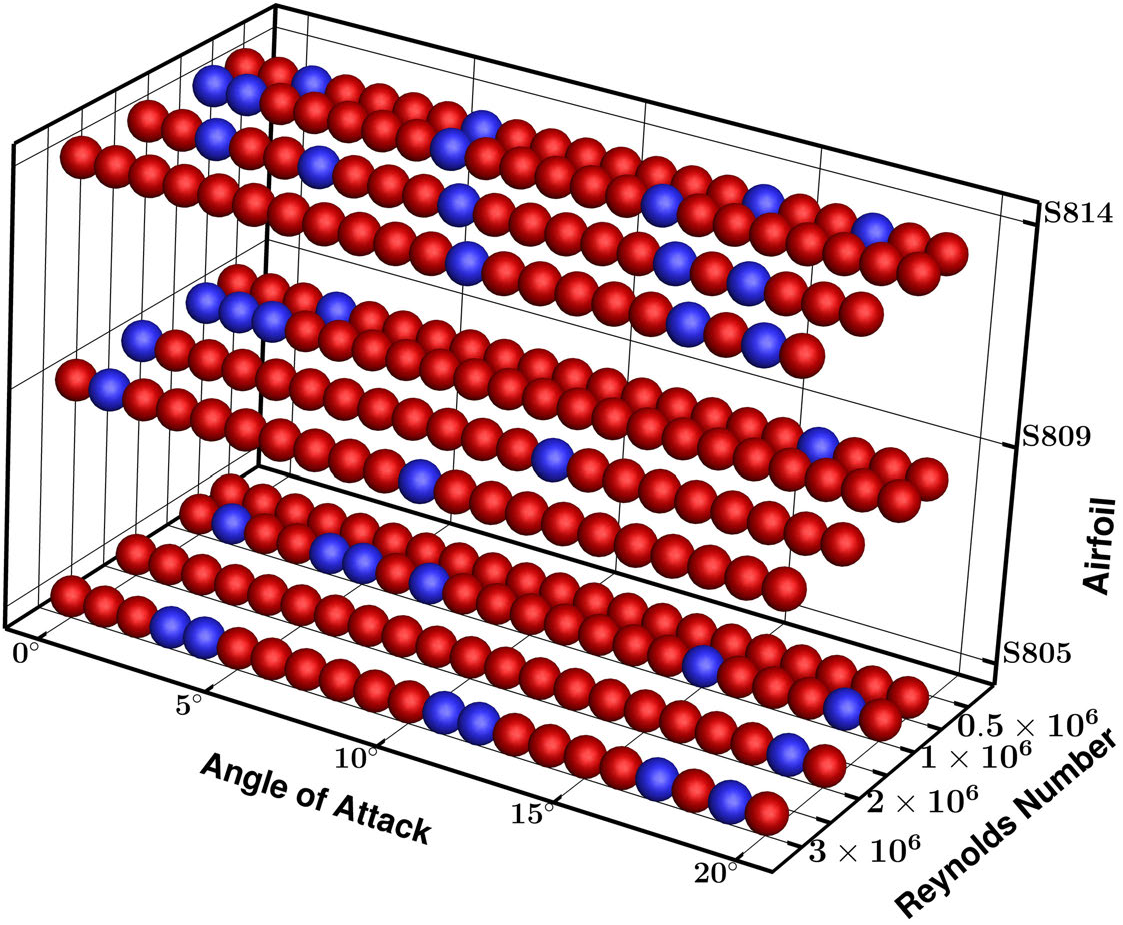}
	}
	\caption{Three-dimensional scatter plots of feature space for training and testing of the network. Blue points are chosen uniformly at random as test set on the feature space, and Red points are the training set.}
	\label{fig:fig6}
\end{figure}

In total, a set of 252 RANS simulations were performed. This data includes our CFD predictions for three different S805, S809, and S814 airfoils.
The training data-set consists of $85$ percent of the full set, and the remaining data sets are used for testing, as shown in Fig.~\ref{fig:fig6}.

The test points are chosen uniformly at random on the feature space, providing an unbiased evaluation of a model fit on the training data-set while tuning the model's hyper-parameters.
	
\begin{figure}[H]
	\centering
	\resizebox{0.75\textwidth}{!}{%
		\includegraphics{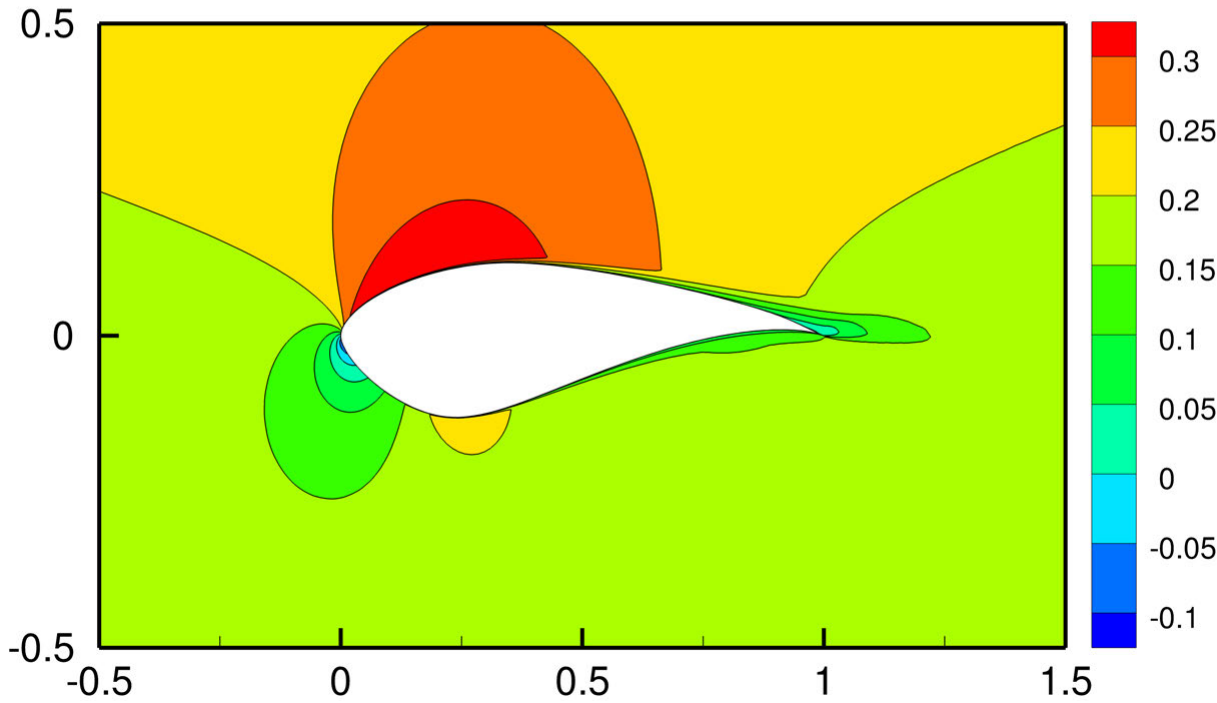}
	}
	\caption{x-component of the velocity field ($U$) around S814 airfoil. RANS simulations are performed at the angle of attack of $\alpha = 9^\circ$ and with the Reynolds number (Re) of $3\times10^6$.}
	\label{fig:fig7}
\end{figure}

Figure~\ref{fig:fig7} shows the x-component of the velocity field ($U$) around the S814 airfoil on the structured C-mesh. The simulation is performed at an angle of attack of $\alpha = 9^\circ$ and with the Reynolds number of $3\times10^6$.

\begin{figure}[H]
	\centering
	\resizebox{0.75\textwidth}{!}{%
		\includegraphics{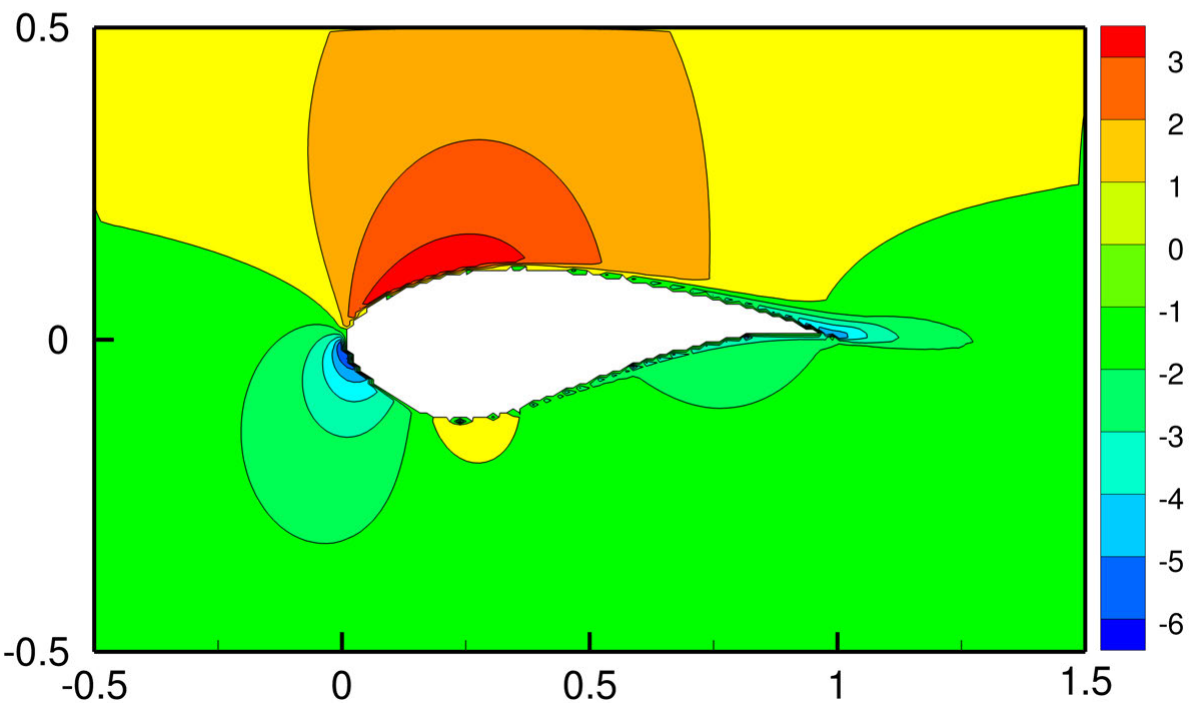}
	}
	\caption{x-component of the velocity field ($U$) around S814 airfoil interpolated from the structured C-mesh initial data in Fig.~\ref{fig:fig7} onto a $150 \times 150$ Cartesian grid and normalized using the standard score normalization. Cartesian grid points inside the airfoil are set to zero.}
	\label{fig:fig8}
\end{figure}

The CFD data has to be interpolated onto a $150\times 150$ Cartesian grid which contains the SDF.  A triangulation-based scattered data interpolation method~\citep{Amidror:2002} is used. 
After the interpolation of the data to the Cartesian grid, the interior points masked and the velocity is set to zero. 
The comparison of the reconstructed data in Fig.~\ref{fig:fig8} and the CFD data in Fig.~\ref{fig:fig7} shows evidence of interpolation errors.

The interpolated data is normalized using the standard score normalization by subtracting the mean from the data and dividing the difference by the standard deviation of the data. Scaling the data causes each feature to contribute approximately proportionately to the training, and also results in a faster convergence of the network~\citep{Aksoy:2000}.
	
\subsection{Network Training and Hyper-parameter Study}
\label{Training}
The network learns different weights during the training phase to predict the flow fields. 
In each iteration, a batch of data undergoes the feed-forward process followed by a back-propagation (see Sec.~\ref{Encoder}).
For a given set of input and ground truth data, the model minimizes a total loss function which is a combination of two specific loss functions and an L2 regularization as follows:
\begin{align}
\label{eq:MSE}
&\text{MSE}_\text{shared} = \frac{1}{m (n_x-2) (n_y-2)}~\sum_{l=1}^{m} \sum_{j=2}^{n_y-1} \sum_{i=2}^{n_x-1}\\
&\nonumber \quad \big[(U^l_{{ij}_\text{truth}}-U^l_{{ij}_\text{pred}})^2~+~(V^l_{{ij}_\text{truth}}-V^l_{{ij}_\text{pred}})^2~+~\\
&\nonumber \quad ~(P^l_{{ij}_\text{truth}}-P^l_{{ij}_\text{pred}})^2 \big], \\
\label{eq:GSshared}
&\text{GS}_\text{shared} = \frac{1}{6m(n_x-2)(n_y-2)}~\sum_{l=1}^{m} \sum_{j=2}^{n_y-1} \sum_{i=2}^{n_x-1} \\
&\nonumber \quad [(\frac{\partial P^l}{\partial x}_{{ij}_\text{truth}} - \frac{\partial P^l}{\partial x}_{{ij}_\text{pred}})^2 ~+~(\frac{\partial P^l}{\partial y}_{{ij}_\text{truth}} - \frac{\partial P^l}{\partial y}_{{ij}_\text{pred}})^2 ~+~ \\
&\nonumber \quad ~ (\frac{\partial U^l}{\partial x}_{{ij}_\text{truth}} - 
\frac{\partial U^l}{\partial x}_{{ij}_\text{pred}})^2 ~+~(\frac{\partial U^l}{\partial y}_{{ij}_\text{truth}} -\frac{\partial U^l}{\partial y}_{{ij}_\text{pred}})^2 ~+~ \\
&\nonumber \quad ~ (\frac{\partial V^l}{\partial x}_{{ij}_\text{truth}} - \frac{\partial V^l}{\partial x}_{{ij}_\text{pred}})^2 ~+~(\frac{\partial V^l}{\partial y}_{{ij}_\text{truth}} -
\frac{\partial V^l}{\partial y}_{{ij}_\text{pred}})^2],\\
\label{eq:L2}
&\text{L2}_\text{regularization} = \frac{1}{2m}\sum_{l=1}^{L}\sum_{i=1}^{n_l}(\theta^l_{i})^2,
\end{align}
where $U$, and $V$ are the x-component and y-component of the velocity field respectively, and $P$ is the scalar pressure field. $m$ is the batch size, $n_x$ is the number of grid points along the x-direction, $n_y$ is the number of grid points along the y-direction, and $L$ is the number of layers with trainable weights, and $n_l$ represents number of trainable weights in layer $l$.	
MSE is the mean squared error, and GS is gradient sharpening or gradient difference loss (GDL)~\citep{Mathieu:2015, Lee:2018}. In this paper, we use gradient sharpening based on a central difference operator.
The network was trained for $30,000$ epochs with a batch size of $214$ data points, which took 33 GPU hours.
For the separated decoder, the following loss functions are used:
\begin{align}
\label{eq:MSEseparate}
&\text{MSE}_\text{separated} = \frac{1}{m (n_x-2) (n_y-2)}~\sum_{l=1}^{m} \sum_{j=2}^{n_y-1} \sum_{i=2}^{n_x-1}\\
&\nonumber \quad \big[(X^l_{{ij}_\text{truth}}-X^l_{{ij}_\text{pred}})^2],\\
\label{eq:GSseparate}
&\text{GS}_\text{separated} = \frac{1}{2m(n_x-2)(n_y-2)}~\sum_{l=1}^{m} \sum_{j=2}^{n_y-1} \sum_{i=2}^{n_x-1} \\
&\nonumber \quad [(\frac{\partial X^l}{\partial x}_{{ij}_\text{truth}} - \frac{\partial X^l}{\partial x}_{{ij}_\text{pred}})^2 ~+~(\frac{\partial X^l}{\partial y}_{{ij}_\text{truth}} - \frac{\partial X^l}{\partial y}_{{ij}_\text{pred}})^2],
\end{align}
where $X$ stands for $U,~V$ or $P$.

Finding the optimal set of hyper-parameters for the network is an empirical task and is done by performing a grid search consisting of an interval of values of each hyper-parameter, and training many networks with several different combinations of these hyper-parameters. The resulting networks are compared based on generalization tendency and the difference between the truth and prediction.
	
\section{Results and Discussion}
\label{Results}
We first show the capability of the designed network architecture to accurately estimate the velocity and pressure field around different airfoils given only the airfoil shape. 
Then, we quantitatively assess the error measurement followed by a sequence of results which demonstrate usability, accuracy and effectiveness of the network.

\begin{figure}[H]
	\centering
	\resizebox{0.75\textwidth}{!}{%
		\includegraphics{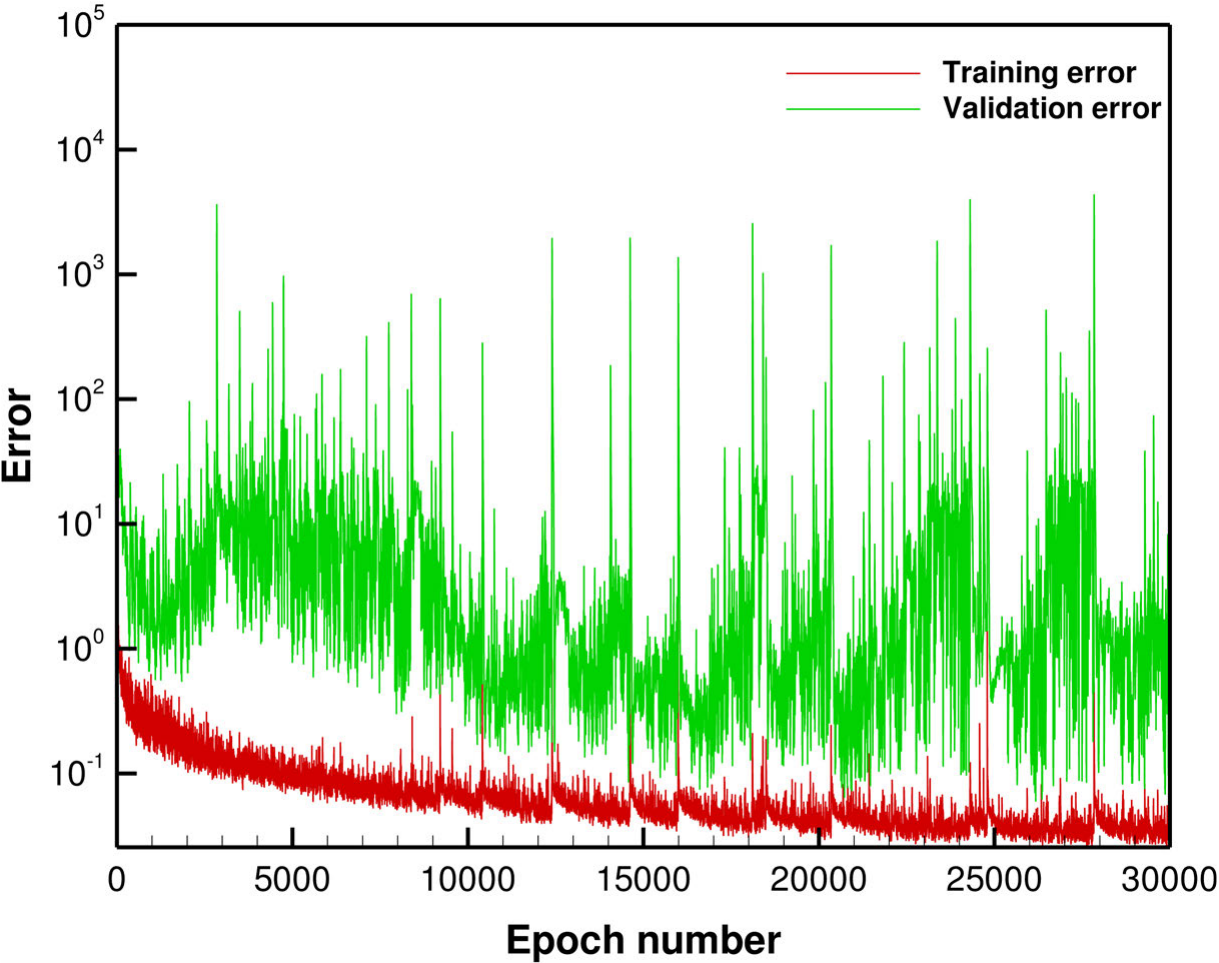}
	}
	\caption{Error and learning curve of the network.}
	\label{fig:fig9}
\end{figure}

Figure~\ref{fig:fig9} illustrates the training and validation results from the network. It shows the working concept of the proposed structure, by incorporating the fluid flow characteristics and airfoil geometry. Results are presented at the epoch number with the lowest validation error.

\subsection{Model validation}
\label{Model}
The Absolute percent error (APE) or the unsigned percentage error is used as a  metric for comparison:
\begin{align}
\text{APE}=\frac{|\text{Prediction} - \text{Truth} |} {|\text{Truth}|} \times 100.
\label{eq:APE}
\end{align}

The mean value of the absolute percent error (MAPE) is standard as a Loss function for regression problems. Here, model evaluation is done using MAPE due to the very intuitive interpretation regarding the relative error and its ease of use.

In this paper, the MAPE between the prediction and the truth is calculated in the wake region of an airfoil and the entire flow field around the airfoil. 
Here, the wake region of the airfoil is an area defined as $\{(x,y)|x\in\left[1.1, 1.5\right],y\in \left[-0.5, 0.5\right]\}$, and $\{(x,y)|x\in\left[-0.5, 1.5\right],y\in \left[-0.5, 0.5\right]\}$ is the entire flow field area around the airfoil. 
The predictions contain $2-3\%$ of points with an error value greater than $100\%$, which are treated as outliers and not included in the reported errors.

\subsection{Numerical simulations}
\label{Experiments}

\subsubsection{Angle of attack variation}
\label{Performance1}
At a fixed Reynolds number ($Re=1\times10^6$) and fixed airfoil shape (S805), we consider simulations with angles of attack of one-degree increments from $\alpha=0^{\circ}$  to $\alpha=20^{\circ}$. 
By using this small set of data (21 data points), we train the network with  50 filters instead of the aforementioned 300 filters in each layer (see Sec.~\ref{Encoder} for more details). The total loss function comprises only an MSE and with no regularization during training. Thus the cost function over the training set is presented as,
\begin{align}
\text{Cost} = \lambda_\text{MSE} \times \text{MSE},
\label{eq:cost1}
\end{align}
where $\lambda_\text{MSE}$ is a user defined parameter (here it is $\lambda_\text{MSE}=1$).

\begin{figure}[H]
	\centering
	\begin{subfigure}{0.45\textwidth}
		\centering
		\resizebox{\textwidth}{!}{%
			\includegraphics{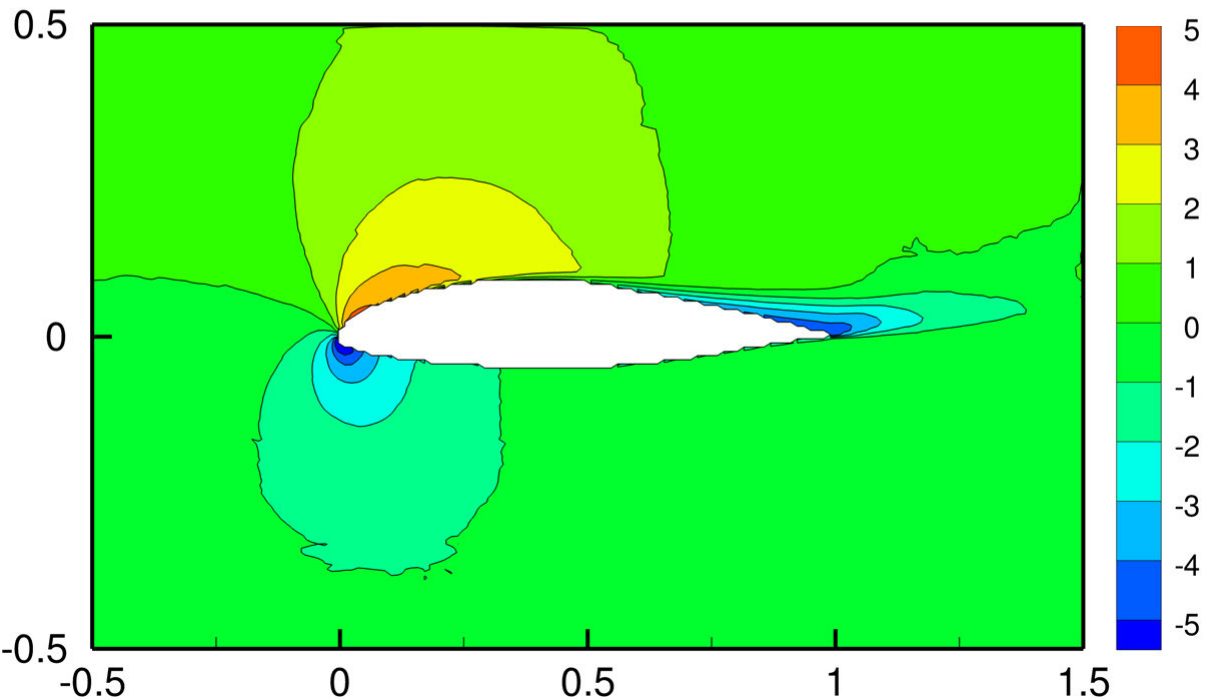}
		}
		\caption{Prediction (separated decoder) \\ without GS.}
		\label{fig:fig10a}
	\end{subfigure}
	\begin{subfigure}{0.45\textwidth}
		\centering
		\resizebox{\textwidth}{!}{%
			\includegraphics{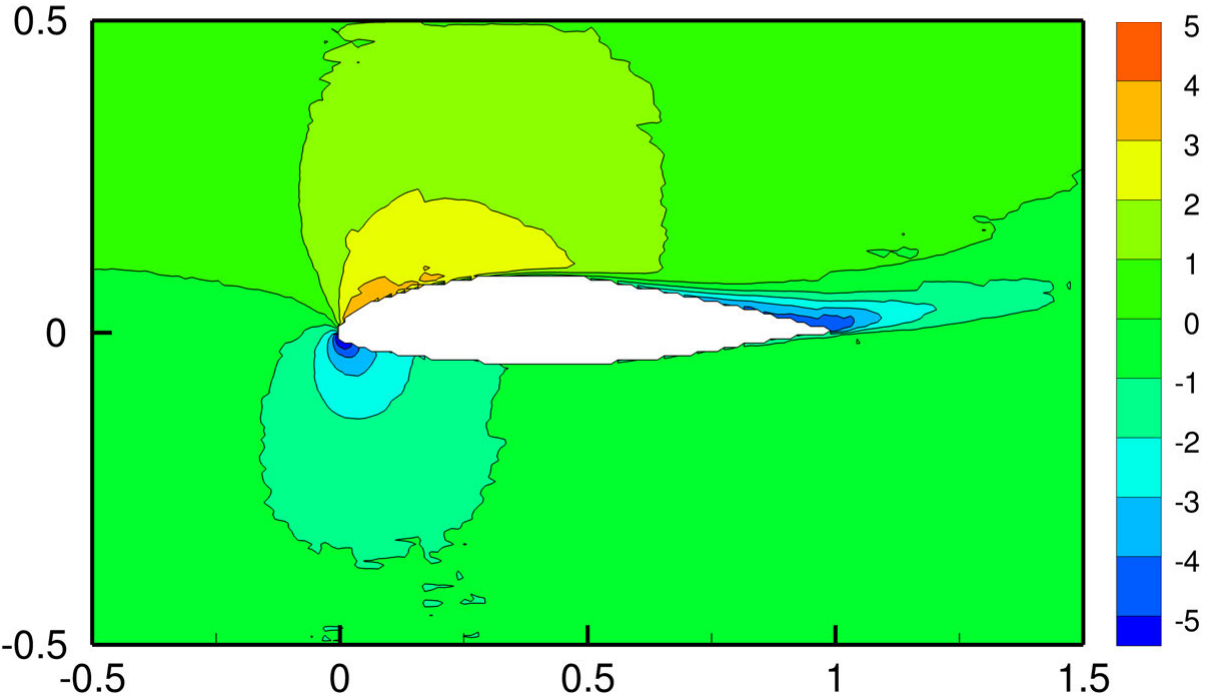}
		}
		\caption{Prediction (shared decoder) \\ without GS.}
		\label{fig:fig10b}
	\end{subfigure}
	\begin{subfigure}{0.45\textwidth}
		\centering
		\resizebox{\textwidth}{!}{%
			\includegraphics{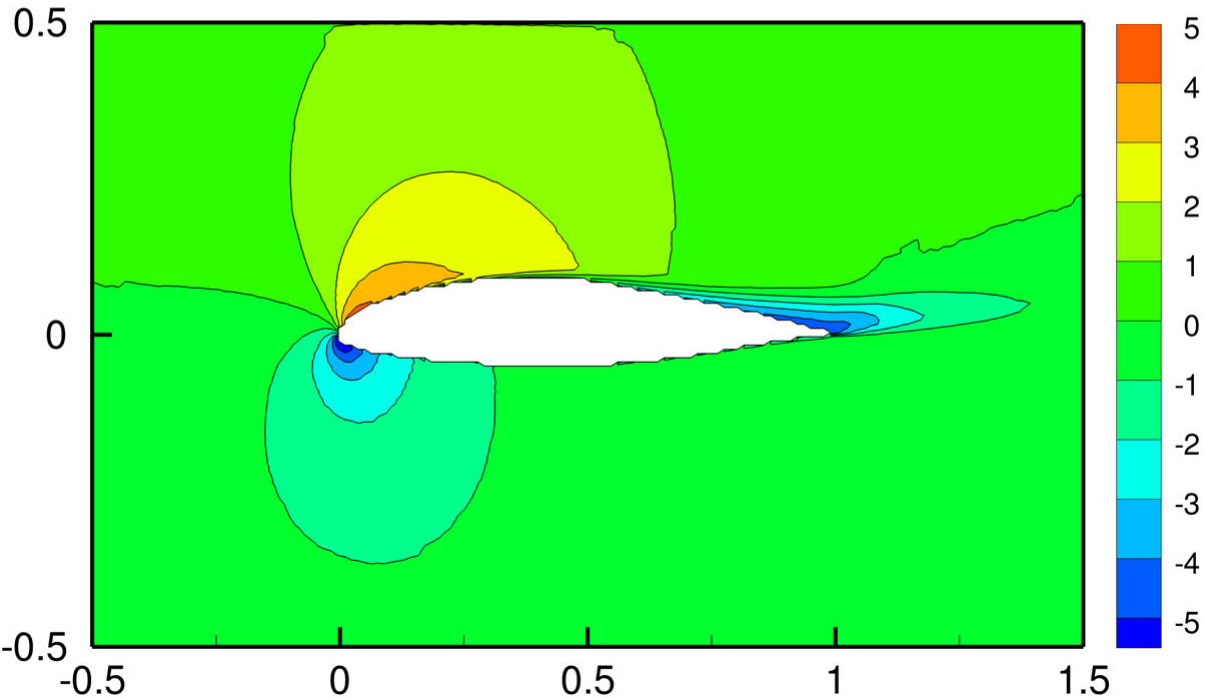}
		}
		\caption{Prediction (shared decoder) \\ with GS.}
		\label{fig:fig10c}
	\end{subfigure}
	\begin{subfigure}{0.45\textwidth}
		\centering
		\resizebox{\textwidth}{!}{%
			\includegraphics{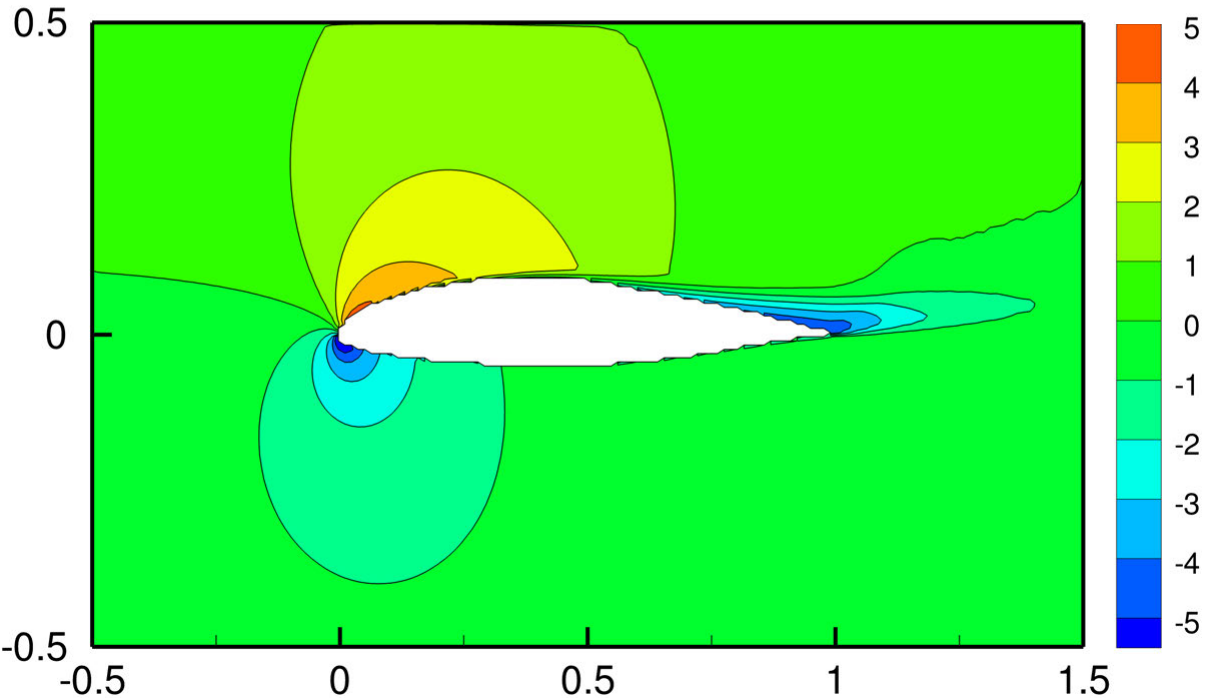}
		}
		\caption{Ground truth. \\ $~$}
		\label{fig:fig10d}
	\end{subfigure}
	\caption{Ground truth (actual observation) vs. Prediction of the x-component of the velocity field around the S805 airfoil with angle of attack of $\alpha = 12.5^\circ$ and the Reynolds number of $1\times10^6$. Ground truth data is interpolated from the structured C-mesh CFD results onto a $150 \times 150$ Cartesian grid and normalized using the standard score normalization.}
	\label{fig:fig10}
\end{figure}

After the network training is complete,  testing is performed on four unseen angles of attacks,  $\alpha=2.5^\circ,~7.5^\circ,~12.5^\circ,~\text{and}~19.5^\circ$ respectively.
Figure~\ref{fig:fig10} shows the comparison between the network prediction and the actual observation from the CFD simulation for the x-component of the velocity field around the S805 airfoil at an angle of attack of $\alpha = 12.5^\circ$. A visual comparison shows that the prediction is in agreement with the truth.

Table.~\ref{tab:tab1separate} and~\ref{tab:tab1shared} present the MAPE calculated in the wake region and the entire flow field around the S805 airfoil (see Fig.\ref{fig:fig10}), where the fluid flow characteristics are the angle of attack of $\alpha = 12.5^\circ$ and the Reynolds number of $1\times10^6$.

\begin{table}[H]
	\begin{center}
		\caption{MAPE for the components of the velocity field ($U$ and $V$ respectively) and pressure in the wake region of the S805 airfoil and the entire flow field around it (separated decoder~Fig.~\ref{fig:fig3a}).}
		\label{tab:tab1separate}
		\begin{tabular}{!{\vrule width 1pt}l!{\vrule width 1pt}c|c|c!{\vrule width 1pt}}
			\hline\noalign{\hrule height 1pt}
			~ & $\mathbf{U}$ & $\mathbf{V}$  & $\mathbf{P}$ \\
			\hline\noalign{\hrule height 1pt}
			\textbf{Error in the wake region} & 24.9\% & 10.15\% & 24.97\% \\
			\textbf{Error in the entire flow} & 13.51\% & 11.92\% & 13.50\% \\ 
			\hline\noalign{\hrule height 1pt}
		\end{tabular}
	\end{center}
\end{table}

\begin{table}[H]
	\begin{center}
		\caption{MAPE for the components of the velocity field ($U$ and $V$ respectively) and pressure in the wake region of the S805 airfoil and the entire flow field around it
			(shared decoder~Fig.~\ref{fig:fig3b}).}
		\label{tab:tab1shared}
		\begin{tabular}{!{\vrule width 1pt}l!{\vrule width 1pt}c|c|c!{\vrule width 1pt}}
			\hline\noalign{\hrule height 1pt}
			~ & $\mathbf{U}$ & $\mathbf{V}$  & $\mathbf{P}$ \\
			\hline\noalign{\hrule height 1pt}
			\textbf{Error in the wake region} & 15.08\% & 7.98\% & 14.82\% \\
			\textbf{Error in the entire flow} & 9.62\% & 8.65\% & 7.31\% \\ 
			\hline\noalign{\hrule height 1pt}
		\end{tabular}
	\end{center}
\end{table}

\begin{figure}[H]
	\centering
	\begin{subfigure}{0.5\textwidth}
		\centering
		\resizebox{\textwidth}{!}{%
			\includegraphics{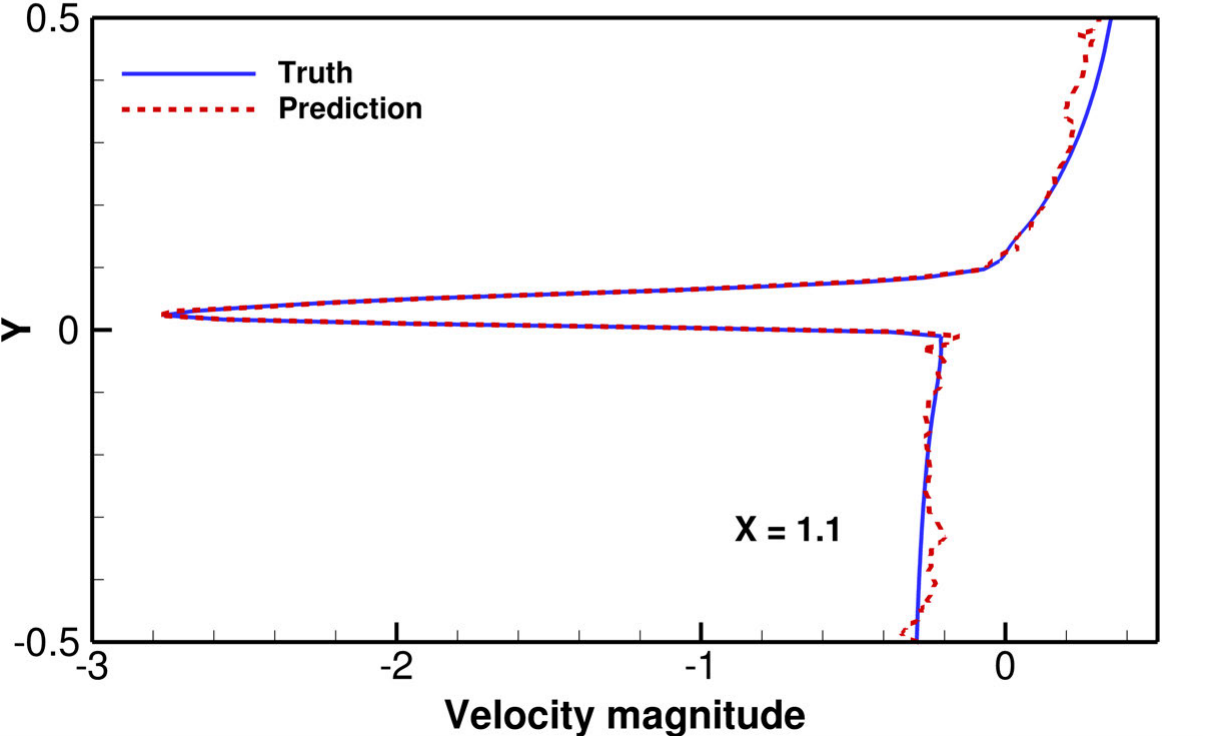}
		}
		\caption{Separated decoder.}
		\label{fig:fig11a}
	\end{subfigure}	
	\begin{subfigure}{0.5\textwidth}
		\centering
		\resizebox{\textwidth}{!}{%
			\includegraphics{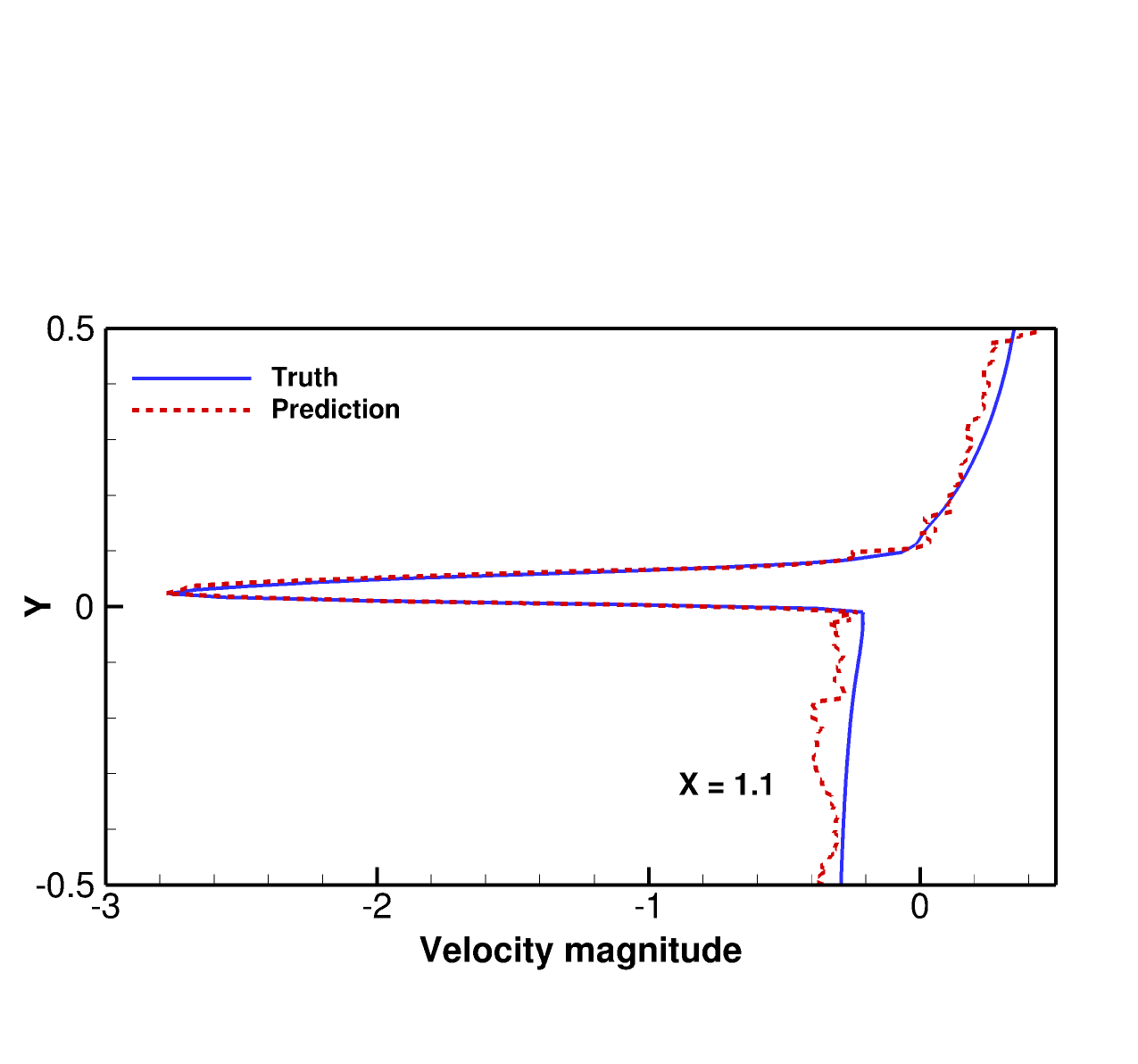}
		}
		\caption{Shared decoder}
		\label{fig:fig11b}
	\end{subfigure}	
	\caption{Prediction of the x-component velocity profile for the S805 airfoil wake at $x=1.1$ with angle of attack of $\alpha = 12.5^\circ$ and the Reynolds number (Re) of $1\times10^6$. Lower plot is for the shared decoder.}
	\label{fig:fig11}
\end{figure}

The results in table.~\ref{tab:tab1separate} and~\ref{tab:tab1shared}, illustrate that the errors in the wake region are generally similar to the errors in the entire flow field. This trend is true not only for this case but also in  subsequent experiments. Figure~\ref{fig:fig11} shows the comparison between the CFD  result and the network prediction of the x-component velocity profile of the airfoil wake at $x=1.1$ (downstream location from the leading edge).


\subsubsection{Shape, angle of attack, and Reynolds number variation}
\label{Shape}
We train the network using $85$ percent of the  252 RANS simulation data-sets, with the variation of the airfoil shape, angle of attack and Reynolds number. Every convolutional layer is composed of 300 convolutional filters (see Sec.\ref{Encoder} for more details).
The total loss function during training comprises an MSE loss function with the L2 regularization. Thus, the cost function over the training set is presented as,
\begin{align}
\text{Cost} = \lambda_\text{MSE} \times \text{MSE} + \lambda_\text{L2} \times \text{L2}_\text{regularization},
\label{eq:cost2}
\end{align}
where $\lambda_\text{MSE}=1$ and $\lambda_\text{L2}=10^{-5}$ are user defined parameters.

\begin{figure}[H]
	\centering
	\begin{subfigure}{0.45\textwidth}
		\centering
		\resizebox{\textwidth}{!}{%
			\includegraphics{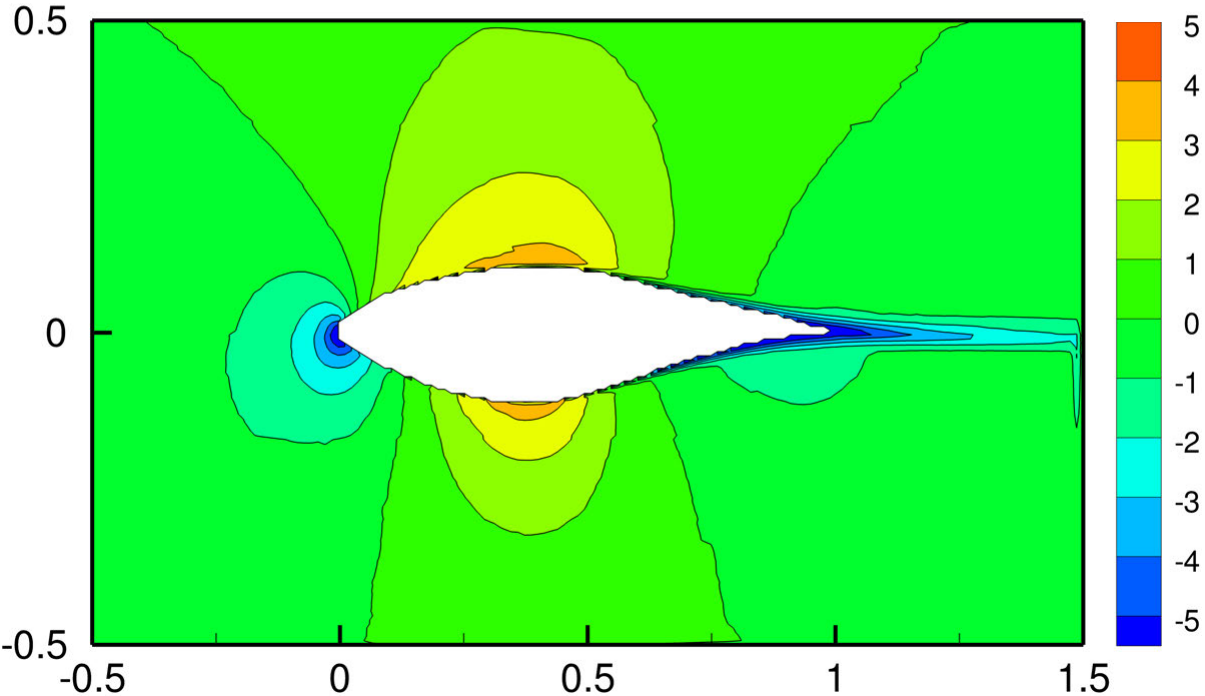}
		}
		\caption{Prediction (separated decoder).}
		\label{fig:fig102a}
	\end{subfigure}
	\begin{subfigure}{0.45\textwidth}
		\centering
		\resizebox{\textwidth}{!}{%
			\includegraphics{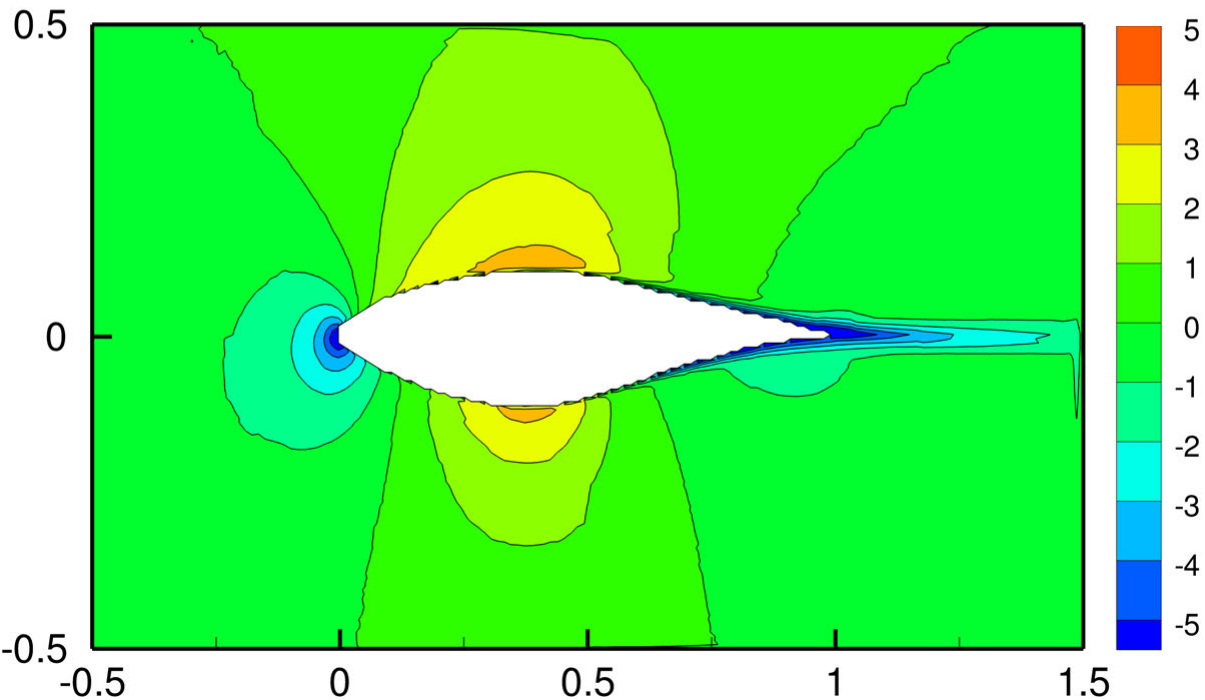}
		}
		\caption{Prediction (shared decoder).}
		\label{fig:fig102b}
	\end{subfigure}		
	\begin{subfigure}{0.45\textwidth}
		\centering
		\resizebox{\textwidth}{!}{%
			\includegraphics{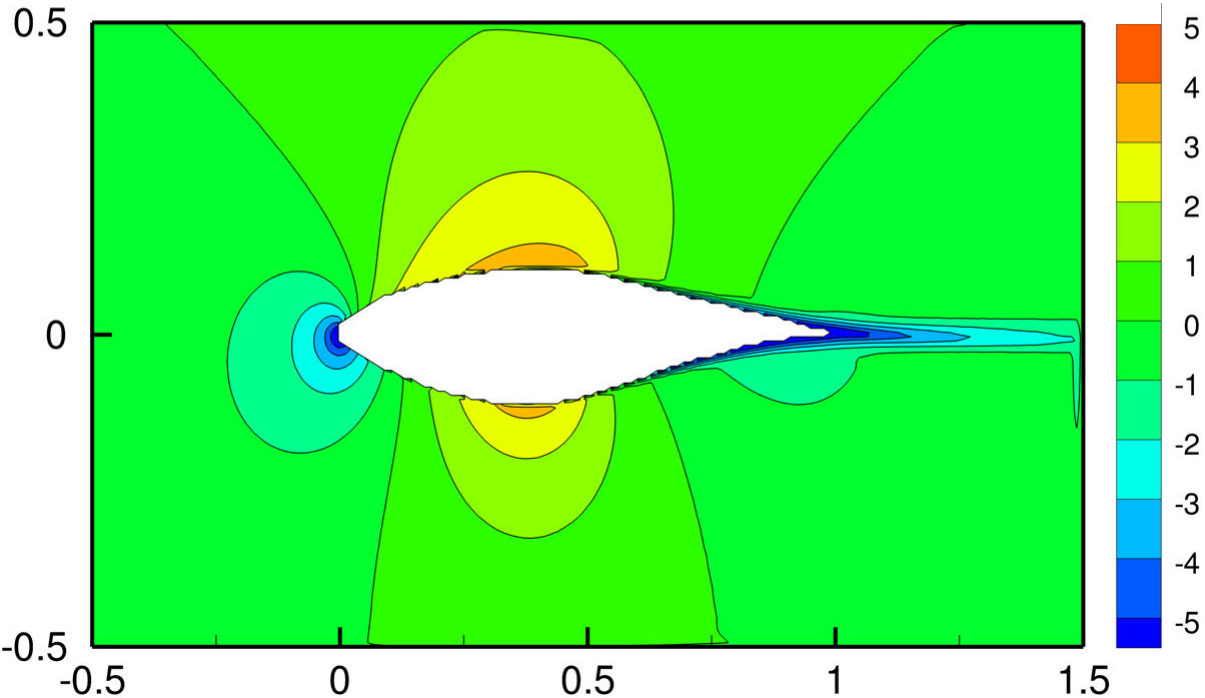}
		}
		\caption{Ground truth.}
		\label{fig:fig102c}
	\end{subfigure}
	\caption{Prediction of the x-component of the velocity field around the S809 airfoil at $\alpha = 1^\circ$ and  $Re = 1\times10^6$. }
	\label{fig:fig102}
\end{figure}
\begin{figure}[H]
	\centering
	\begin{subfigure}{0.45\textwidth}
		\centering
		\resizebox{\textwidth}{!}{%
			\includegraphics{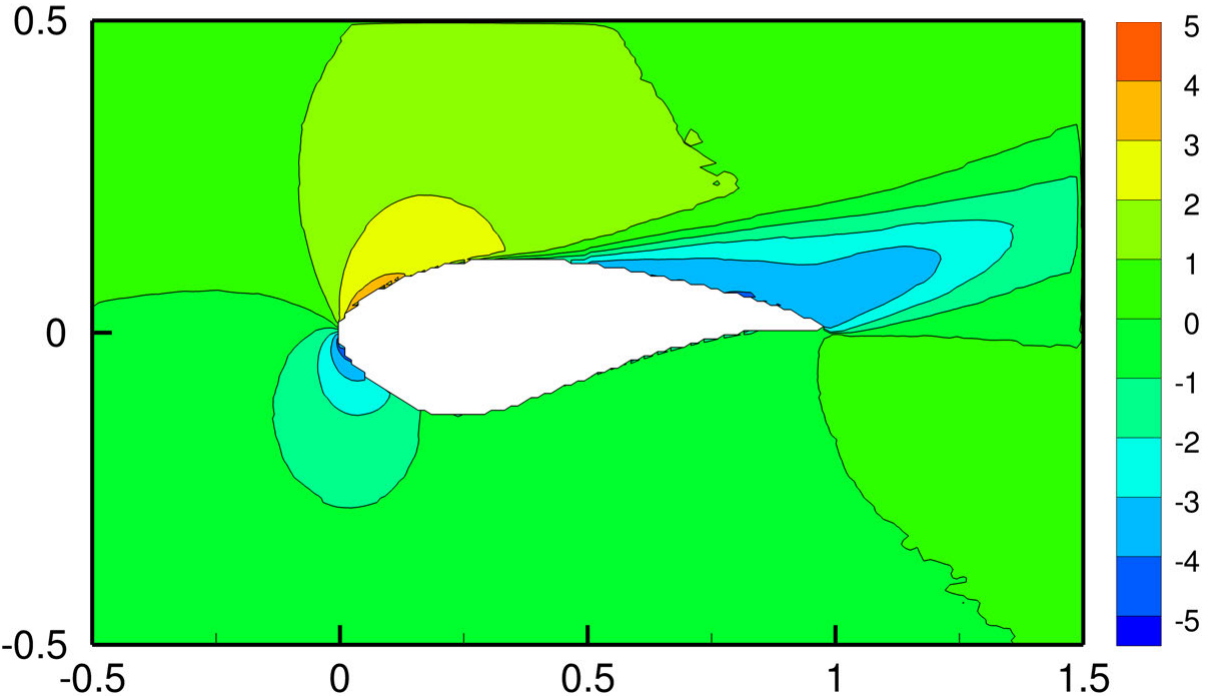}
		}
		\caption{Prediction \\ (separated decoder).}
		\label{fig:fig103a}
	\end{subfigure}
	\begin{subfigure}{0.45\textwidth}
		\centering
		\resizebox{\textwidth}{!}{%
			\includegraphics{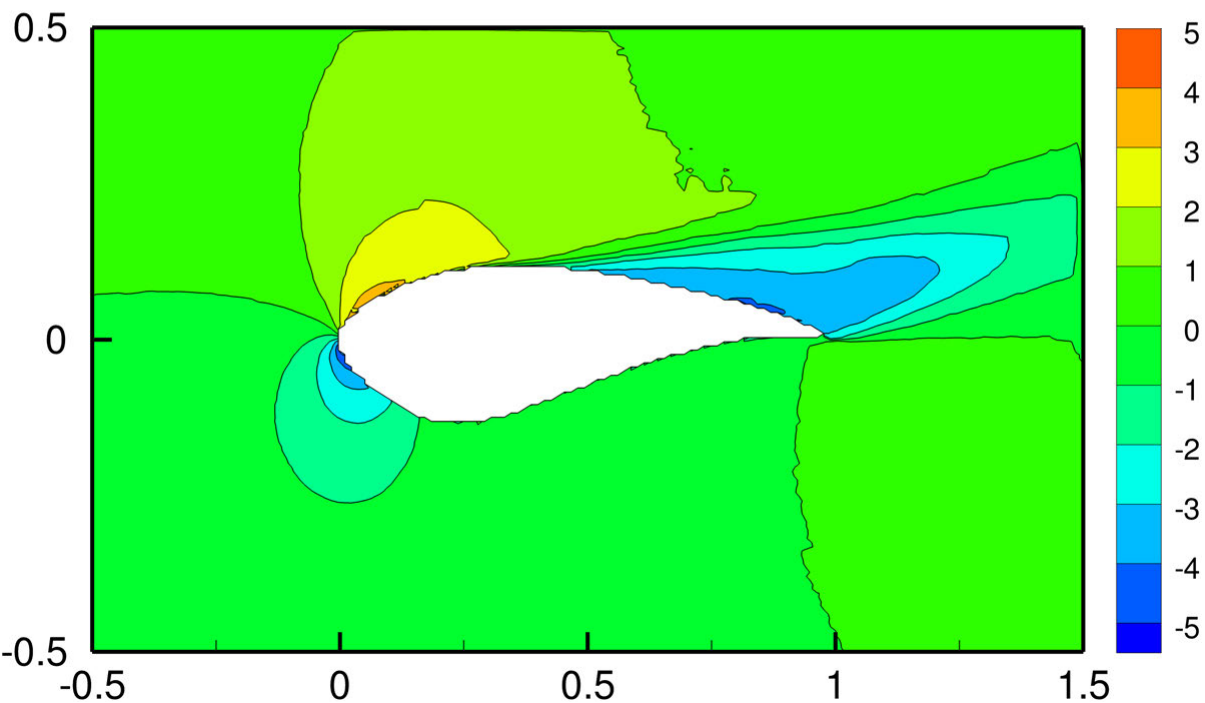}
		}
		\caption{Prediction \\ (shared decoder).}
		\label{fig:fig103b}
	\end{subfigure}
	\begin{subfigure}{0.45\textwidth}
		\centering
		\resizebox{\textwidth}{!}{%
			\includegraphics{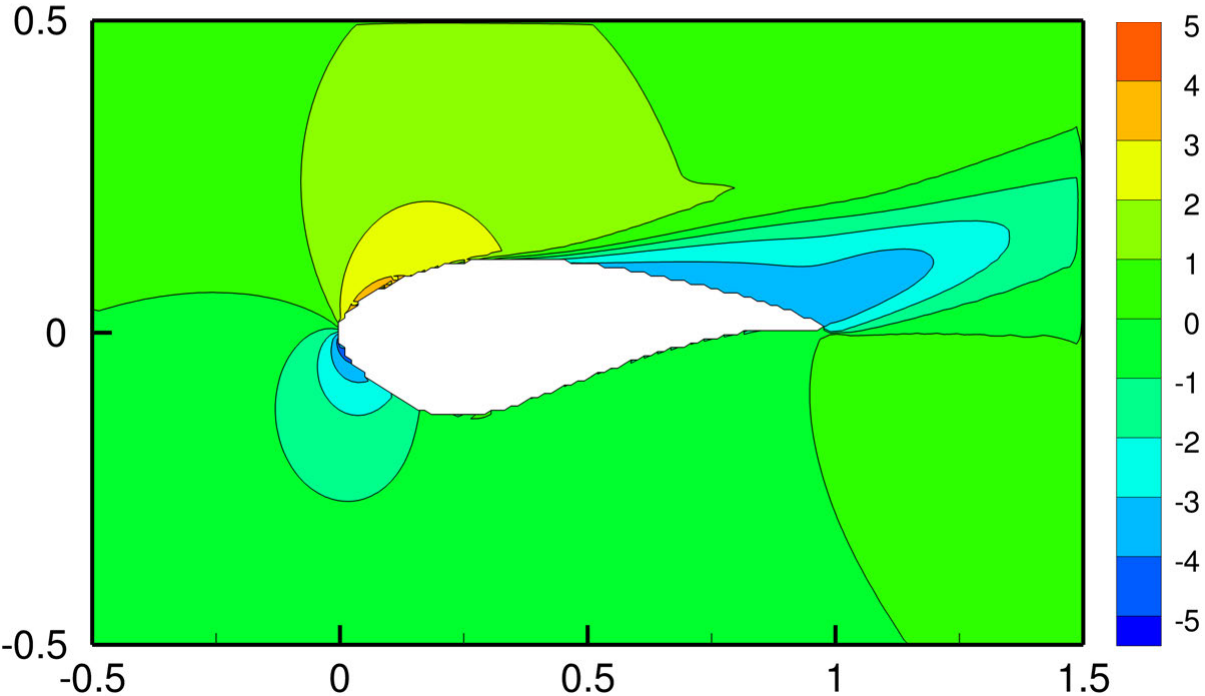}
		}
		\caption{Ground truth.}
		\label{fig:fig103c}
	\end{subfigure}
	\caption{Prediction of the x-component of the velocity field $U$ around the S814 airfoil with $\alpha = 19^\circ$ and $Re = 3\times10^6$.}
	\label{fig:fig103}
\end{figure}

Figures.~\ref{fig:fig102} and~\ref{fig:fig103} present the comparisons between the network predictions and  observations  for the x-component of the velocity field around the S809 and S814 airfoils at $(\alpha = 1^\circ,~Re = 1\times10^6)$ and $(\alpha = 19^\circ,~Re = 3\times10^6)$. 

Quantitative results  are presented in Tables~\ref{tab:tab2separate} and~\ref{tab:tab2shared}.

\begin{table}[H]
	\begin{center}
		\caption{MAPE for the components of the velocity field ($U$ and $V$ respectively) and pressure (shared decoder).}
		\label{tab:tab2separate}
		\begin{tabular}{!{\vrule width 1pt}c|c|c|c|c|c!{\vrule width 1pt}}
			\hline\noalign{\hrule height 1pt}
			\textbf{Airfoil} & \textbf{AOA} & \textbf{Re} & \textbf{Variable} & \textbf{Error in the} & \textbf{Error in the}  \\
			~ & ~ & $\times 10^6$ & ~ & \textbf{wake region} & \textbf{entire flow}  \\
			\hline\noalign{\hrule height 1pt}
			\textbf{S809} & $1^\circ$ & 1 & U & 12.25\% & 10.35\%  \\
			\textbf{S809} & $1^\circ$ & 1 & V & 24.27\% & 11.53\%  \\
			\textbf{S809} & $1^\circ$ & 1 & P & 5.14\% & 8.40\%  \\
			\textbf{S814} & $19^\circ$ & 3 & U & 30.80\% & 13.13\% \\
			\textbf{S814} & $19^\circ$ & 3 & V & 10.43\% & 5.49\%  \\
			\textbf{S814} & $19^\circ$ & 3 & P & 13.84\% & 5.70\%  \\
			\hline\noalign{\hrule height 1pt}
		\end{tabular}
	\end{center}
\end{table}

\begin{table}[H]
	\begin{center}
		\caption{MAPE for the components of the velocity field ($U$ and $V$ respectively) and pressure (separated decoder).}
		\label{tab:tab2shared}
		\begin{tabular}{!{\vrule width 1pt}c|c|c|c|c|c!{\vrule width 1pt}}
			\hline\noalign{\hrule height 1pt}
			\textbf{Airfoil} & \textbf{AOA} & \textbf{Re} & \textbf{Variable} & \textbf{Error in the} & \textbf{Error in the}  \\
			~ & ~ & $\times 10^6$ & ~ & \textbf{wake region} & \textbf{entire flow}  \\
			\hline\noalign{\hrule height 1pt}
			\textbf{S809} & $1^\circ$ & 1 & U & 11.43\% & 7.79\%  \\
			\textbf{S809} & $1^\circ$ & 1 & V & 15.53\% & 8.74\%  \\
			\textbf{S809} & $1^\circ$ & 1 & P & 5.76\% & 7.36\%  \\
			\textbf{S814} & $19^\circ$ & 3 & U & 27.23\% & 13.20\% \\
			\textbf{S814} & $19^\circ$ & 3 & V & 5.57\% & 4.69\%  \\
			\textbf{S814} & $19^\circ$ & 3 & P & 12.93\% & 5.71\%  \\
			\hline\noalign{\hrule height 1pt}
		\end{tabular}
	\end{center}
\end{table}

\subsubsection{Shape, angle of attack, and Reynolds number variation with gradient sharpening}
\label{Gradient}
To penalize the difference of the gradient in the loss function, and to address the lack of sharpness in predictions, we use gradient sharpening (GS)~\citep{Mathieu:2015, Lee:2018} in the loss functions combination and present the cost function over the training set as,
\begin{align}
\text{Cost} = \lambda_\text{MSE} \times \text{MSE} + \lambda_\text{GS} \times \text{GS} +\lambda_\text{L2} \times \text{L2}_\text{regularization},
\label{eq:cost3}
\end{align}
where $\lambda_\text{MSE},~\lambda_\text{GS}~\text{and}~\lambda_\text{L2}$ are the user defined parameters and their values are set via systematic experimentation, as $0.9,~0.1~\text{and}~10^{-5}$ respectively.

\begin{figure}[H]
	\captionsetup{justification=centering}
	\centering
	\begin{subfigure}{0.45\textwidth}
		\centering
		\resizebox{\textwidth}{!}{%
			\includegraphics{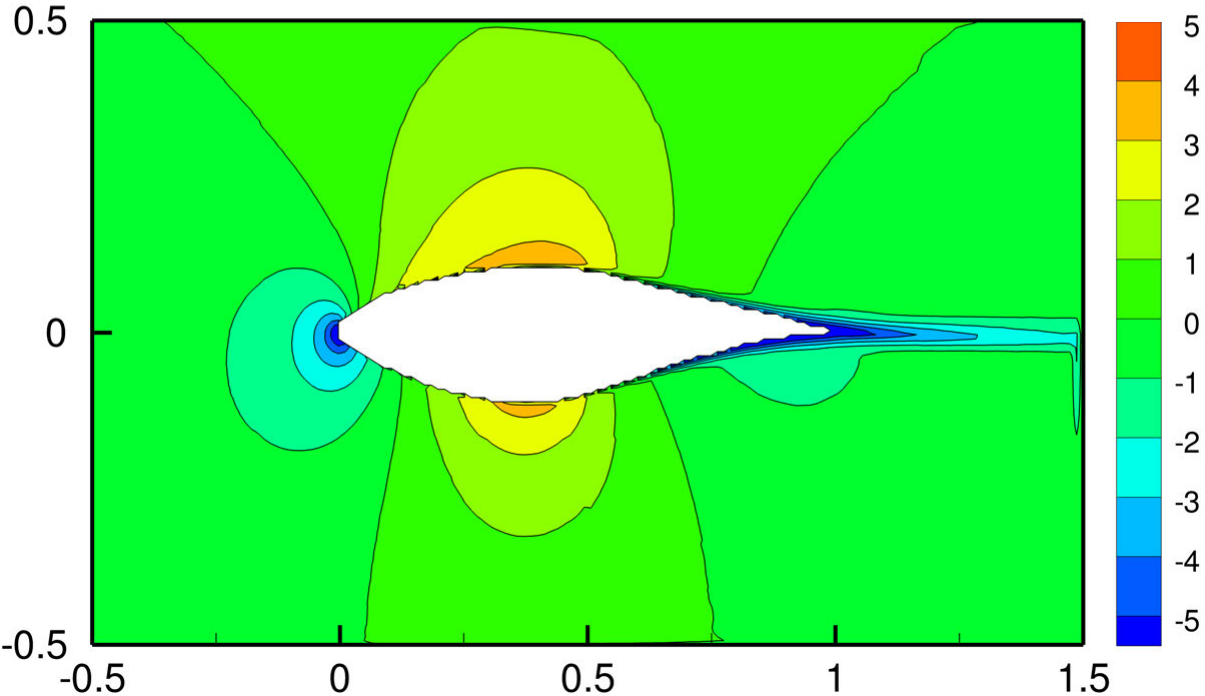}
		}
		\caption{Prediction (shared decoder) \\ with GS.}
		\label{fig:fig104a}
	\end{subfigure}
	\begin{subfigure}{0.45\textwidth}
		\centering
		\resizebox{\textwidth}{!}{%
			\includegraphics{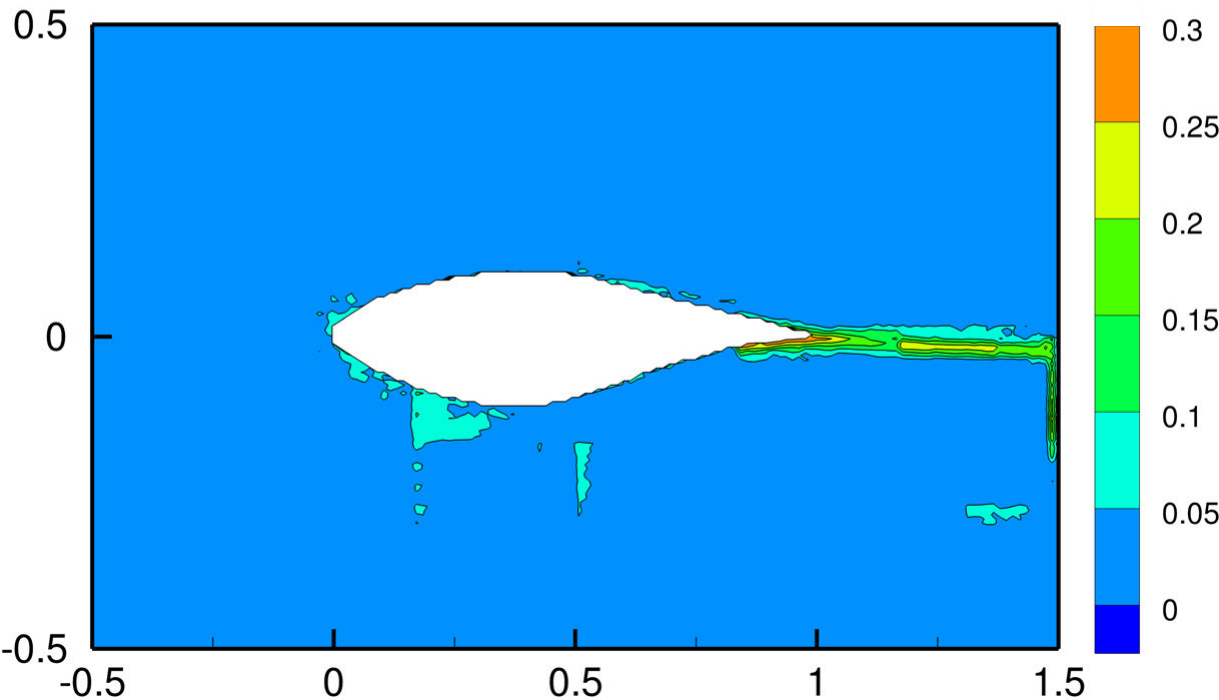}
		}
		\caption{Absolute difference \\ (w.r.t ground truth)}
		\label{fig:fig104c}
	\end{subfigure}
	\begin{subfigure}{0.45\textwidth}
		\centering
		\resizebox{\textwidth}{!}{%
			\includegraphics{u_1_1_s809_NOGS}
		}
		\caption{Prediction (separated decoder) \\ without GS.}
		\label{fig:fig104b}
	\end{subfigure}
	\begin{subfigure}{0.45\textwidth}
		\centering
		\resizebox{\textwidth}{!}{%
			\includegraphics{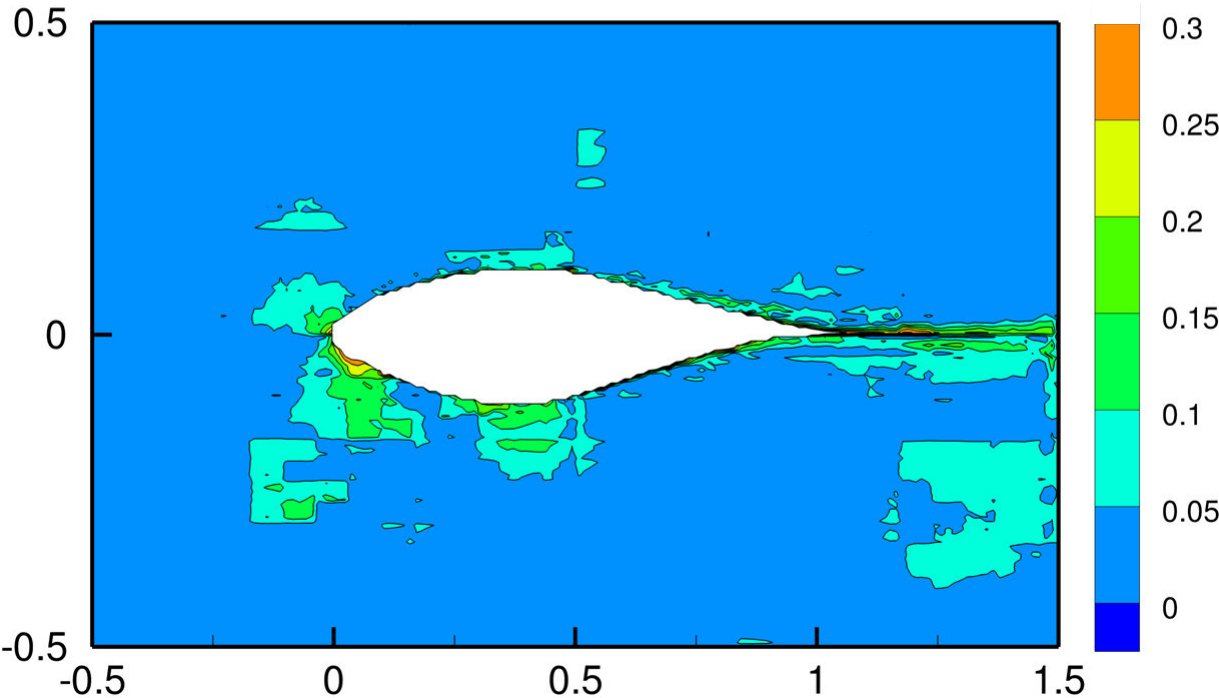}
		}
		\caption{Absolute difference \\ (w.r.t ground truth)}
		\label{fig:fig104d}
	\end{subfigure}
	\begin{subfigure}{0.45\textwidth}
		\centering
		\resizebox{\textwidth}{!}{%
			\includegraphics{u_1_1_s809_NOGS_shared}
		}
		\caption{Prediction (shared decoder) \\ without GS.}
		\label{fig:fig104e}
	\end{subfigure}
	\begin{subfigure}{0.45\textwidth}
		\centering
		\resizebox{\textwidth}{!}{%
			\includegraphics{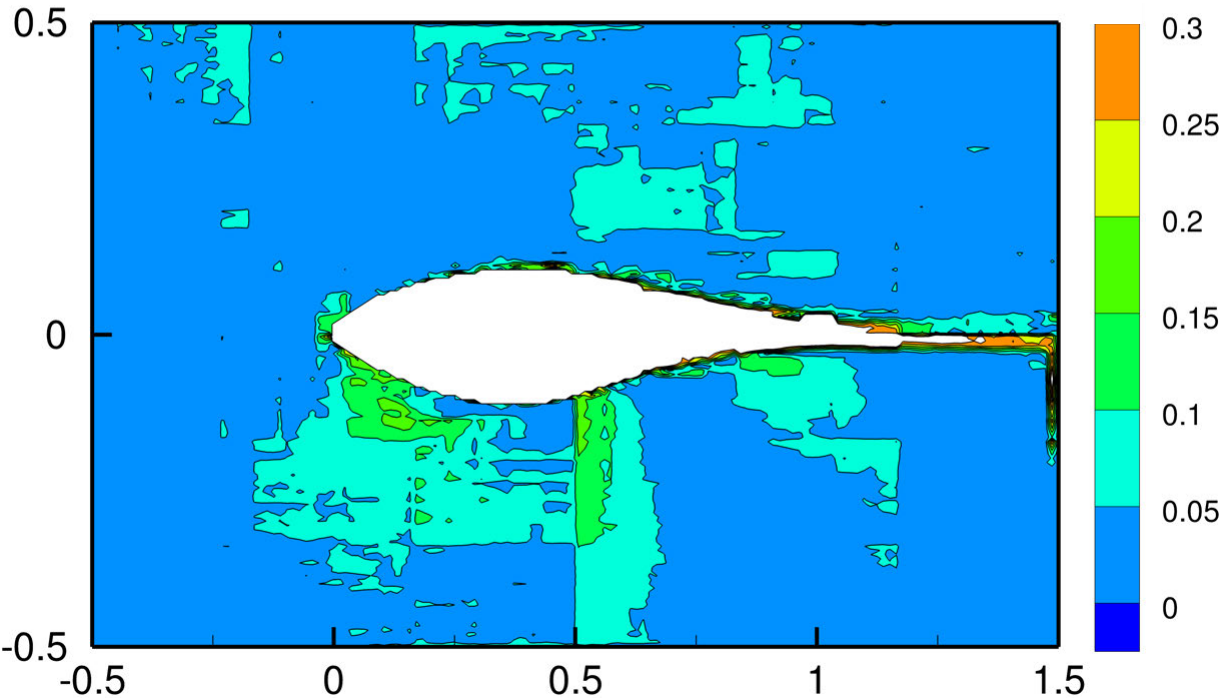}
		}
		\caption{Absolute difference \\ (w.r.t ground truth)}
		\label{fig:fig104f}
	\end{subfigure}		
	\caption{Prediction of the x-component of the velocity field $U_p$ around the S809 airfoil at $\alpha = 1^\circ$ and  $Re = 1\times10^6$, with and without gradient sharpening.}
	\label{fig:fig104}
\end{figure}

\begin{figure}[H]
	\captionsetup{justification=centering}
	\centering
	\begin{subfigure}{0.45\textwidth}
		\centering
		\resizebox{\textwidth}{!}{%
			\includegraphics{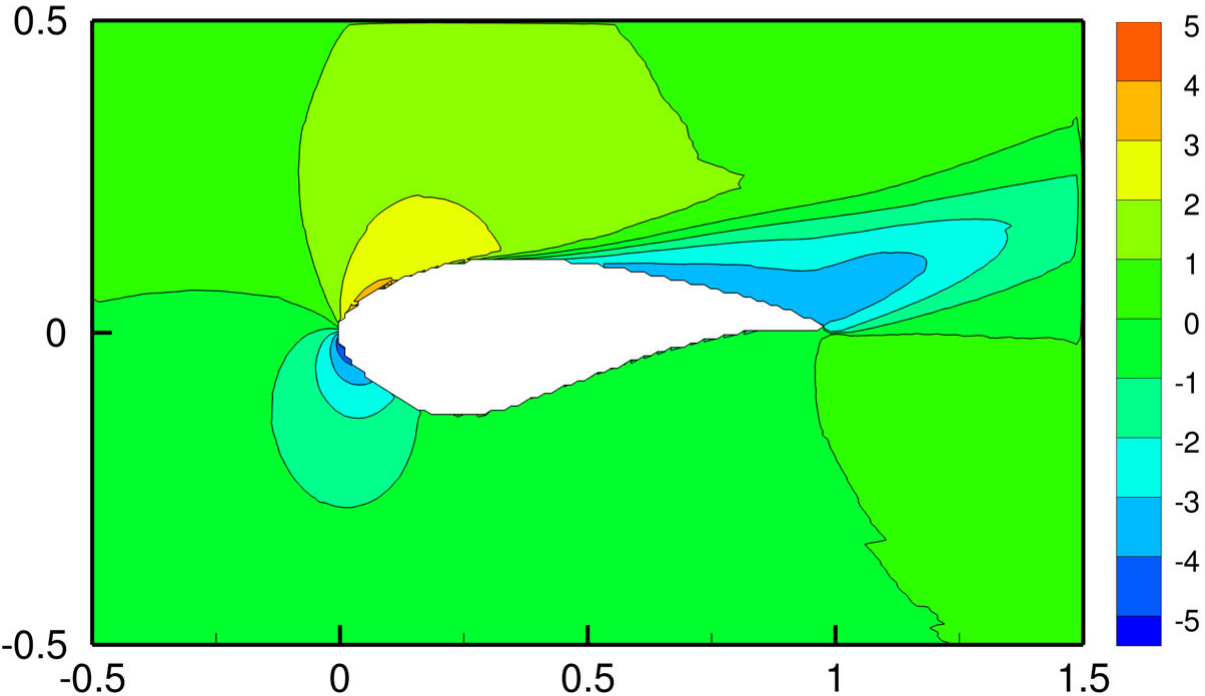}
		}
		\caption{Prediction (shared decoder) \\ with GS.}
		\label{fig:fig105a}
	\end{subfigure}
	\begin{subfigure}{0.45\textwidth}
		\centering
		\resizebox{\textwidth}{!}{%
			\includegraphics{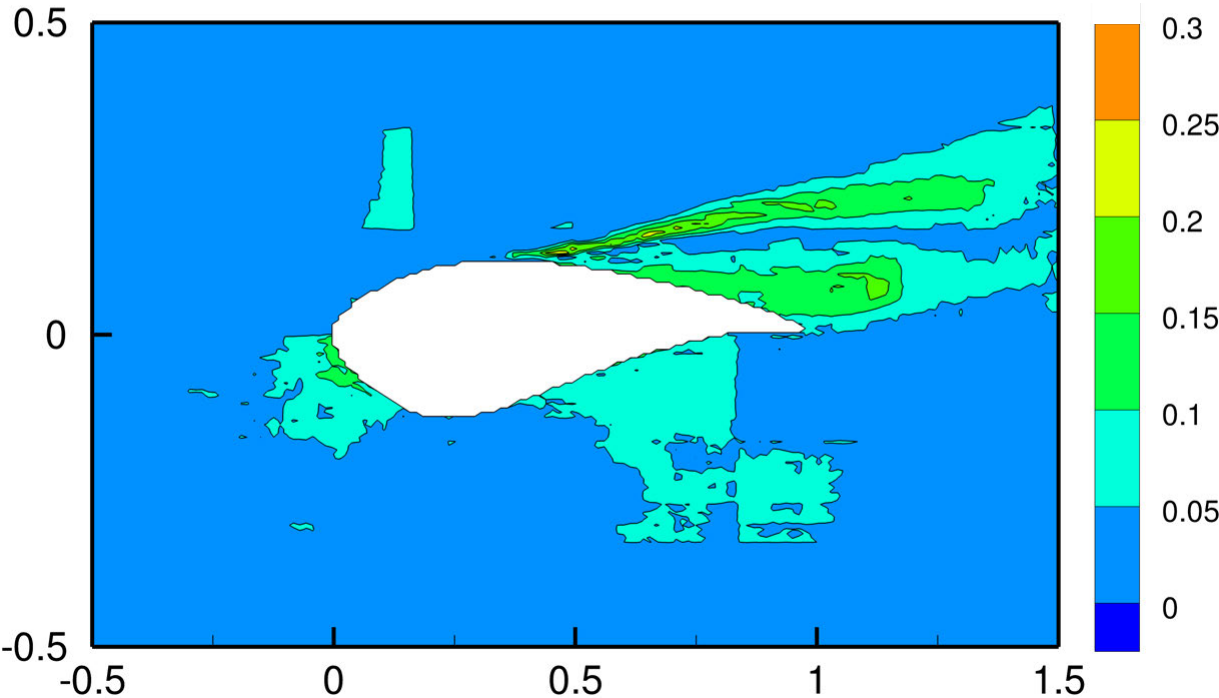}
		}
		\caption{Absolute difference \\ (w.r.t ground truth)}
		\label{fig:fig105c}
	\end{subfigure}
	\begin{subfigure}{0.45\textwidth}
		\centering
		\resizebox{\textwidth}{!}{%
			\includegraphics{u_3_19_s814_NOGS}
		}
		\caption{Prediction (separated decoder) \\ without GS.}
		\label{fig:fig105b}
	\end{subfigure}
	\begin{subfigure}{0.45\textwidth}
		\centering
		\resizebox{\textwidth}{!}{%
			\includegraphics{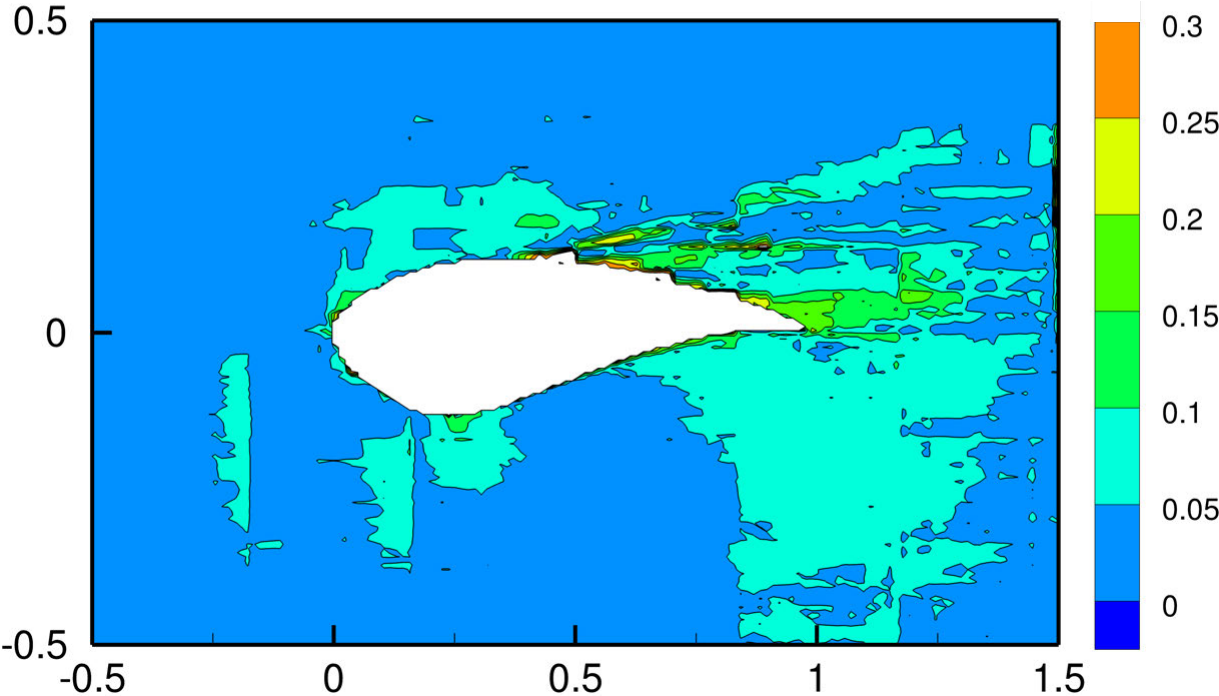}
		}
		\caption{Absolute difference \\ (w.r.t ground truth)}
		\label{fig:fig105d}
	\end{subfigure}
	
	\begin{subfigure}{0.45\textwidth}
		\centering
		\resizebox{\textwidth}{!}{%
			\includegraphics{u_3_19_s814_NOGS_shared}
		}
		\caption{Prediction (shared decoder) \\ without GS.}
		\label{fig:fig105e}
	\end{subfigure}	
	\begin{subfigure}{0.45\textwidth}
		\centering
		\resizebox{\textwidth}{!}{%
			\includegraphics{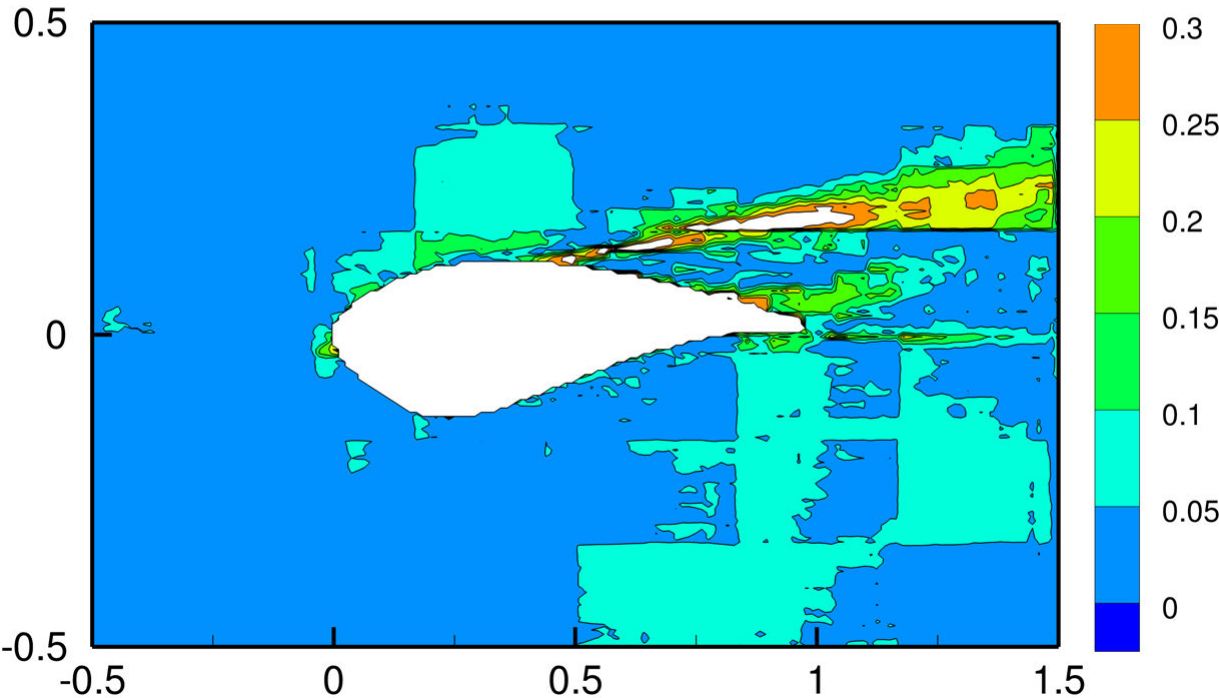}
		}
		\caption{Absolute difference \\ (w.r.t ground truth)}
		\label{fig:fig105f}
	\end{subfigure}	
	\caption{Prediction of the x-component of the velocity field $U_p$ around the S814 airfoil at $\alpha = 19^\circ$ and  $Re = 3\times10^6$, with and without gradient sharpening.}
	\label{fig:fig105}
\end{figure}

Figures.~\ref{fig:fig104} and~\ref{fig:fig105} present the comparisons between the network predictions with and without GS loss for the x-component of the velocity field around S809 and S814 airfoils respectively.

Visual comparisons of the predictions and the absolute difference with and without GS as illustrated in Figs.~\ref{fig:fig104} and~\ref{fig:fig105} are proofs of further gains and sharpness in the network predictions. 
The ``absolute difference''\label{absolute} between the prediction and ground truth, for example, is defined as the absolute difference in the subtraction of each element in prediction from the corresponding element in ground truth.
The MAPE for the components of the velocity field and pressure of the airfoils (S809 and S814 discussed above) are presented in table.~\ref{tab:tab3separate} and~\ref{tab:tab3shared}. The errors are reported in the wake region and the entire flow field around the airfoils with and without GS.
\begin{table}[H]
	\begin{center}
		\caption{ MAPE for the components of the velocity field ($U$ and $V$ respectively) and pressure (separated decoder).}
		\label{tab:tab3separate}
		\resizebox{\textwidth}{!}{
			\begin{tabular}{!{\vrule width 1pt}c!{\vrule width 1pt}c|c|c|c|c|c|c!{\vrule width 1pt}}
				\hline\noalign{\hrule height 1pt}
				\textbf{Airfoil} & \textbf{AOA} & \textbf{Re} & \textbf{Variable} & \textbf{Error in the} & \textbf{Error in the} & \textbf{Error in the} & \textbf{Error in the}  \\
				~ & ~ & ~ & ~ & \textbf{wake region} & \textbf{wake region} & \textbf{entire flow} & \textbf{entire flow} \\ 
				~ & ~ & $\times 10^6$ & ~ & \textbf{without GS} & \textbf{with GS} & \textbf{without GS} & \textbf{with GS} \\
				\hline\noalign{\hrule height 1pt}
				\textbf{S809} & $1^\circ$ & 1 & U & 11.43\% & 11.30 \% & 7.79\% & 9.92\% \\
				\textbf{S809} & $1^\circ$ & 1 & V & 15.53\% & 19.43 \% & 8.74\% & 8.88 \%  \\
				\textbf{S809} & $1^\circ$ & 1 & P & 5.76\% & 6.82 \% & 7.36\%  & 8.09 \% \\
				\textbf{S814} & $19^\circ$ & 3 & U & 27.23\% & 9.49 \% & 13.20\% & 7.38 \% \\
				\textbf{S814} & $19^\circ$ & 3 & V & 5.57\% & 3.09\% & 4.69\% & 2.94 \% \\
				\textbf{S814} & $19^\circ$ & 3 & P & 12.93\% & 4.71\% & 5.71\% & 3.11 \%  \\
				\textbf{S805} & $4^\circ$ & 3 & U & 13.40\% & 4.47\% & 6.44\% & 2.99 \%  \\
				\textbf{S805} & $4^\circ$ & 3 & V & 3.96\% & 1.42\% & 5.60\% & 4.85 \%  \\
				\textbf{S805} & $4^\circ$ & 3 & P & 8.48\% & 3.72\% & 5.37\% & 3.22 \%  \\		    
				\hline\noalign{\hrule height 1pt}
		\end{tabular}}
	\end{center}
\end{table}

\begin{table}[H]
	\begin{center}
		\caption{MAPE  for the components of the velocity field ($U$ and $V$ respectively) and pressure with and without GS (shared decoder).}
		\label{tab:tab3shared}
		\resizebox{\textwidth}{!}{
			\begin{tabular}{!{\vrule width 1pt}c!{\vrule width 1pt}c|c|c|c|c|c|c!{\vrule width 1pt}}
				\hline\noalign{\hrule height 1pt}
				\textbf{Airfoil} & \textbf{AOA} & \textbf{Re} & \textbf{Variable} & \textbf{Error in the} & \textbf{Error in the} & \textbf{Error in the} & \textbf{Error in the}  \\
				~ & ~ & ~ & ~ & \textbf{wake region} & \textbf{wake region} & \textbf{entire flow} & \textbf{entire flow} \\ 
				~ & ~ & $\times 10^6$ & ~ & \textbf{without GS} & \textbf{with GS} & \textbf{without GS} & \textbf{with GS} \\
				\hline\noalign{\hrule height 1pt}
				\textbf{S809} & $1^\circ$ & 1 & U & 12.25\% & 8.35 \% & 10.35\% & 5.57\% \\
				\textbf{S809} & $1^\circ$ & 1 & V & 24.27\% & 17.43 \% & 11.53\% & 6.52 \%  \\
				\textbf{S809} & $1^\circ$ & 1 & P & 5.14\% & 3.18 \% & 8.40\%  & 3.86 \% \\
				\textbf{S814} & $19^\circ$ & 3 & U & 30.80\% & 19.89 \% & 13.13\% & 11.37 \% \\
				\textbf{S814} & $19^\circ$ & 3 & V & 10.43\% & 3.52\% & 5.49\% & 3.04 \% \\
				\textbf{S814} & $19^\circ$ & 3 & P & 13.84\% & 7.95\% & 5.70\% & 3.57 \%  \\
				\textbf{S805} & $4^\circ$ & 3 & U & 15.68\% & 15.15\% & 6.50\% & 9.59 \%  \\
				\textbf{S805} & $4^\circ$ & 3 & V & 4.20\% & 1.29\% & 7.63\% & 10.70 \%  \\
				\textbf{S805} & $4^\circ$ & 3 & P & 8.58\% & 32.77\% & 4.03\% & 15.06 \%  \\
				\hline\noalign{\hrule height 1pt}
		\end{tabular}}
	\end{center}
\end{table}

The predictions with GS in the loss function compared to not having it show significantly reduced errors in the wake region of the airfoil (twenty percent or more in the x-component of the velocity and pressure predictions) and obvious gains and sharpness in the entire flow field around the airfoil.

\begin{figure}[H]
	\centering
	\begin{subfigure}{0.45\textwidth}
		\centering
		\resizebox{\textwidth}{!}{%
			\includegraphics{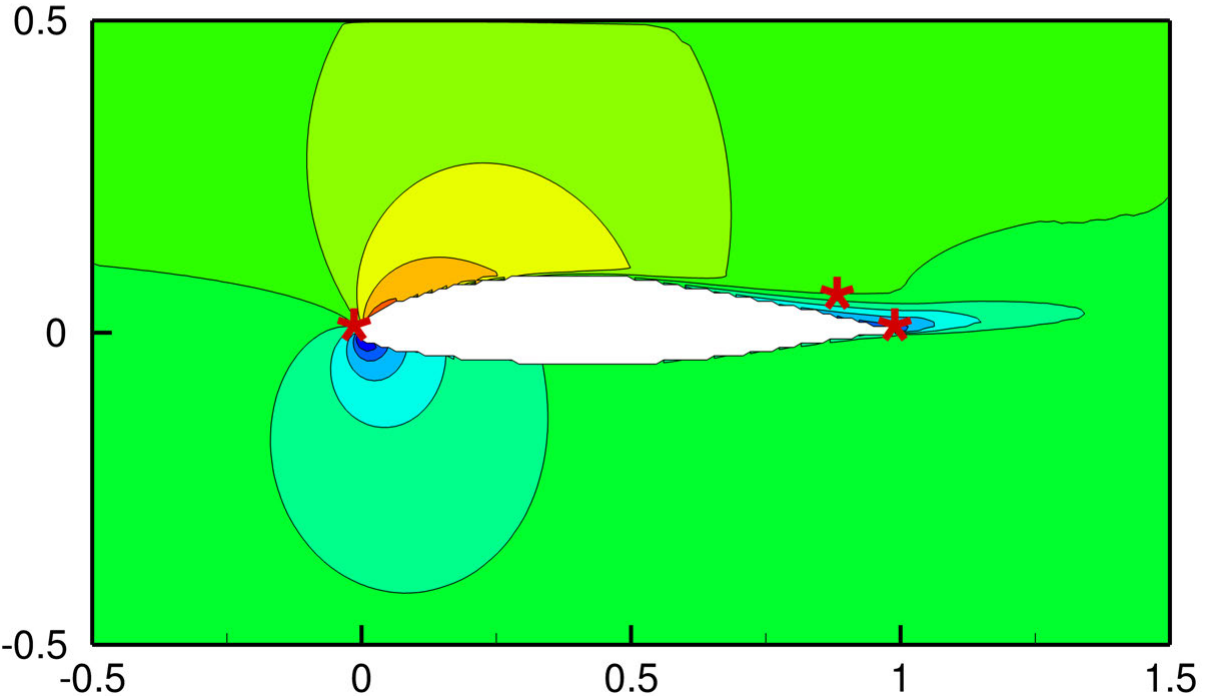}
		}
		\caption{S805 airfoil.}
	\end{subfigure}
	\begin{subfigure}{0.45\textwidth}
		\centering
		\resizebox{\textwidth}{!}{%
			\includegraphics{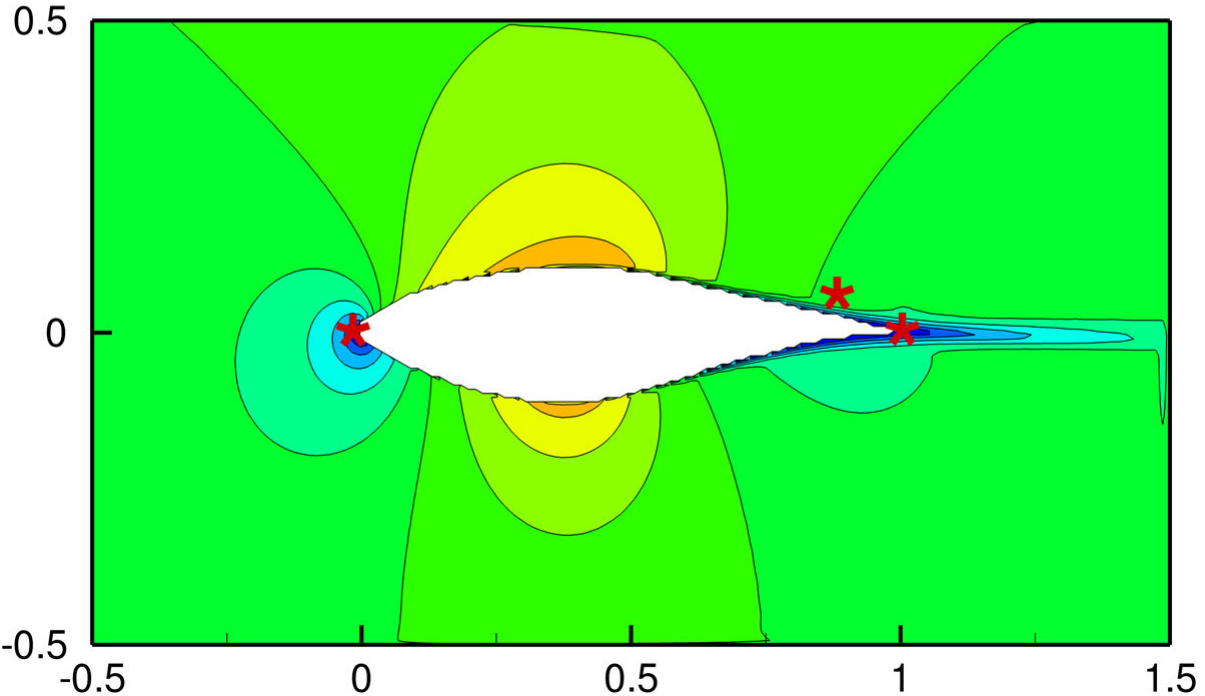}
		}
		\caption{S809 airfoil.}
	\end{subfigure}
	\begin{subfigure}{0.45\textwidth}
		\centering
		\resizebox{\textwidth}{!}{%
			\includegraphics{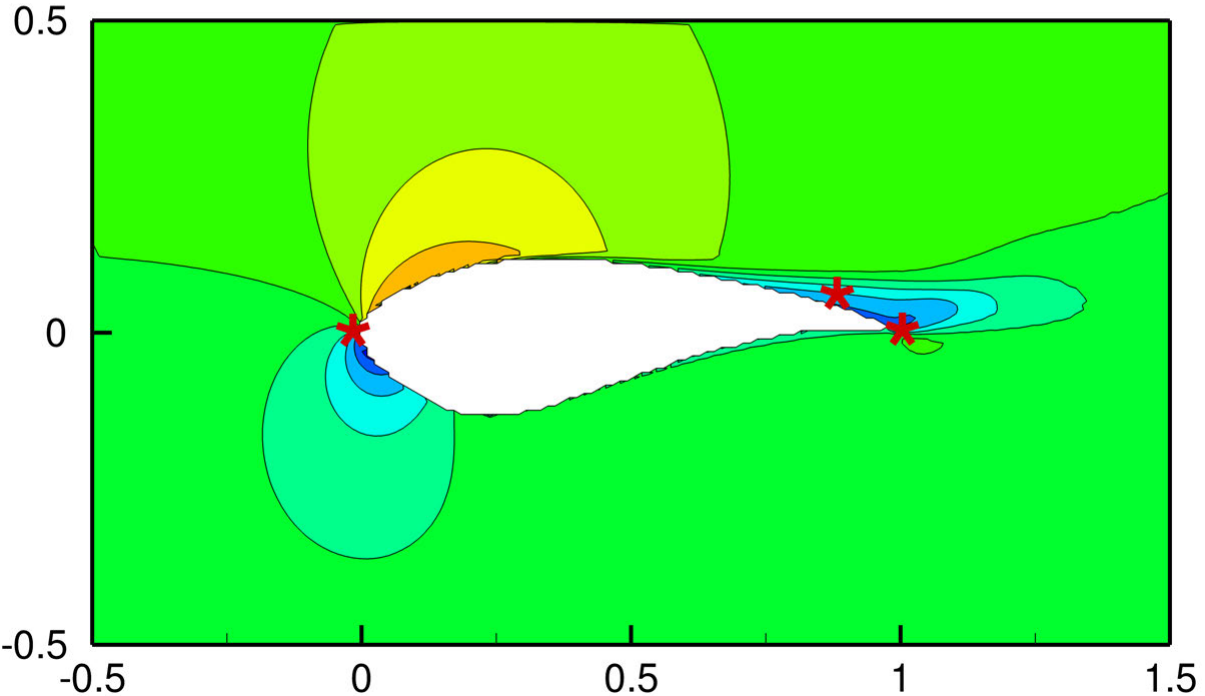}
		}
		\caption{S814 airfoil.}
	\end{subfigure}
	\caption{Three probes around each airfoil which are leading edge probe (LE), trailing-edge probe (TE), and the probe at the wake region of the airfoil. Each red star symbol, depicts a probe.}
	\label{fig:fig200}
\end{figure}

To further compare the accuracy of the network predictions, we use three probes around different airfoils in different flow conditions. 
These probes are leading edge probe (LE), trailing-edge probe (TE), and the probe at the wake region of an airfoil. 
Figure~\ref{fig:fig200} illustrates these three probes  around different airfoils, S805, S809, and S814, respectively.

Table.~\ref{tab:tab4} presents the APE (Eq.~\ref{eq:APE}) at the probe locations (LE, TE, and wake region probe).
\begin{table}[H]
	\begin{center}
		\caption{APE at probe locations (shared decoder).}
		\label{tab:tab4}
		\resizebox{\textwidth}{!}{
			\begin{tabular}{!{\vrule width 1pt}c|c|c|c|c|c|c!{\vrule width 1pt}}
				\hline\noalign{\hrule height 1pt}
				\textbf{Airfoil} & \textbf{AOA} & \textbf{Re} & \textbf{Variable} & \textbf{Error at} & \textbf{Error at} & \textbf{Error at} \\ 
				~ & ~ & $\times 10^6$ & ~ & \textbf{LE probe} & \textbf{TE probe} & \textbf{wake probe} \\
				\hline\noalign{\hrule height 1pt}
				\textbf{S805} & $12^\circ$ & 3 & U  & 2.36\% & 1.67\% & 31.66\% \\ 
				\textbf{S805} & $12^\circ$ & 3 & V  & 0.96\% & 0.74\% & 1.50\% \\
				\textbf{S805} & $12^\circ$ & 3 & P  &  9.35\% & 6.56\% & 22.44\% \\
				\textbf{S809} & $1^\circ$ & 3 & U & 0.08\% & 2.38\% & 5.45\% \\
				\textbf{S809} & $1^\circ$ & 3 & V & 0.73\% & 2.22\% & 1.93\% \\
				\textbf{S809} & $1^\circ$ & 3 & P & 0.23\% & 3.16\% & 1.07\% \\
				\textbf{S814} & $15^\circ$ & 2 & U & 0.75\% & 1.98\% & 1.64\% \\
				\textbf{S814} & $15^\circ$ & 2 & V & 0.76\% & 1.17\% & 0.18\% \\
				\textbf{S814} & $15^\circ$ & 2 & P & 5.67\% & 6.56\% & 8.02\% \\
				\hline\noalign{\hrule height 1pt}
		\end{tabular}}
	\end{center}
\end{table}

\begin{figure}[H]
	\centering
	\begin{subfigure}{0.32\textwidth}
		\centering
		\resizebox{\textwidth}{!}{%
			\includegraphics{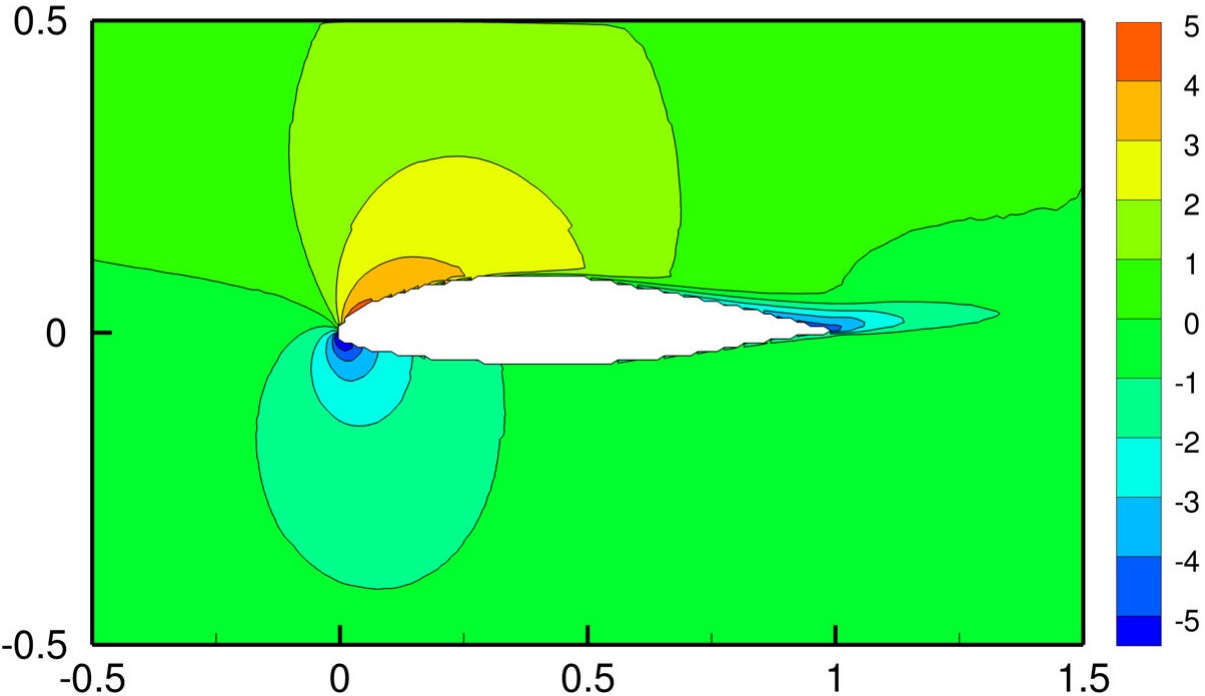}
		}
		\caption{Prediction ($U_p$).}
		\label{fig:fig17a}
	\end{subfigure}
	\begin{subfigure}{0.32\textwidth}
		\centering
		\resizebox{\textwidth}{!}{%
			\includegraphics{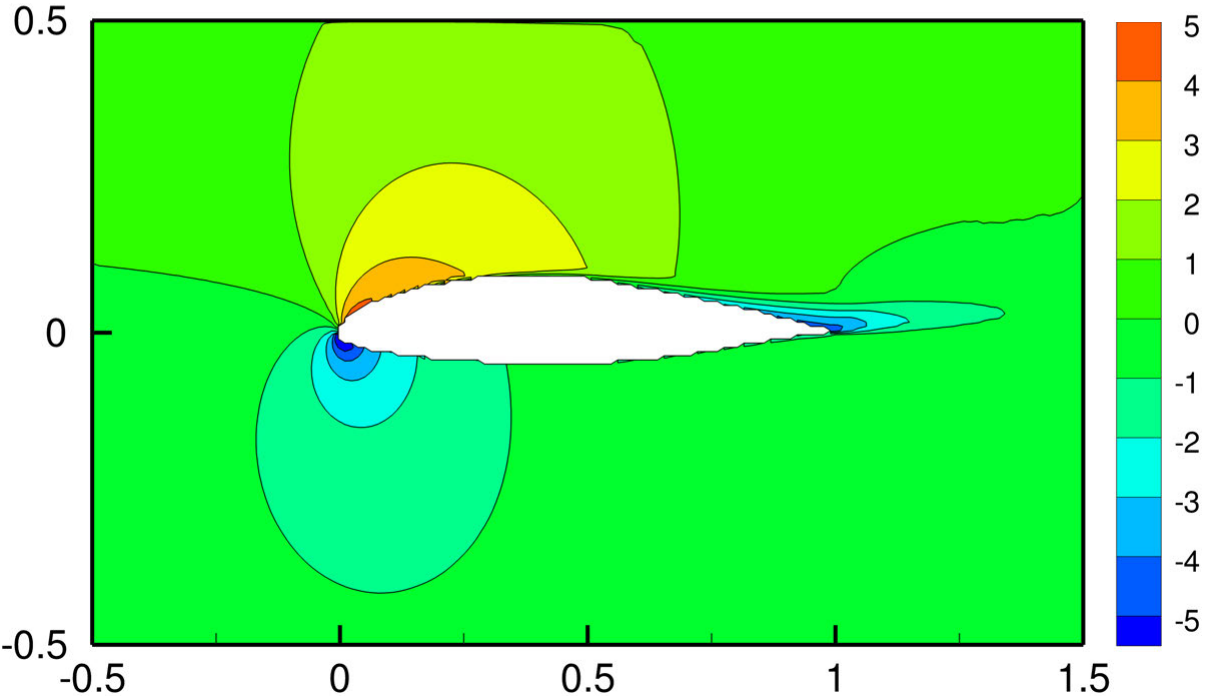}
		}
		\caption{Ground truth ($U_t$).}
		\label{fig:fig17b}
	\end{subfigure}
	\begin{subfigure}{0.32\textwidth}
		\centering
		\resizebox{\textwidth}{!}{%
			\includegraphics{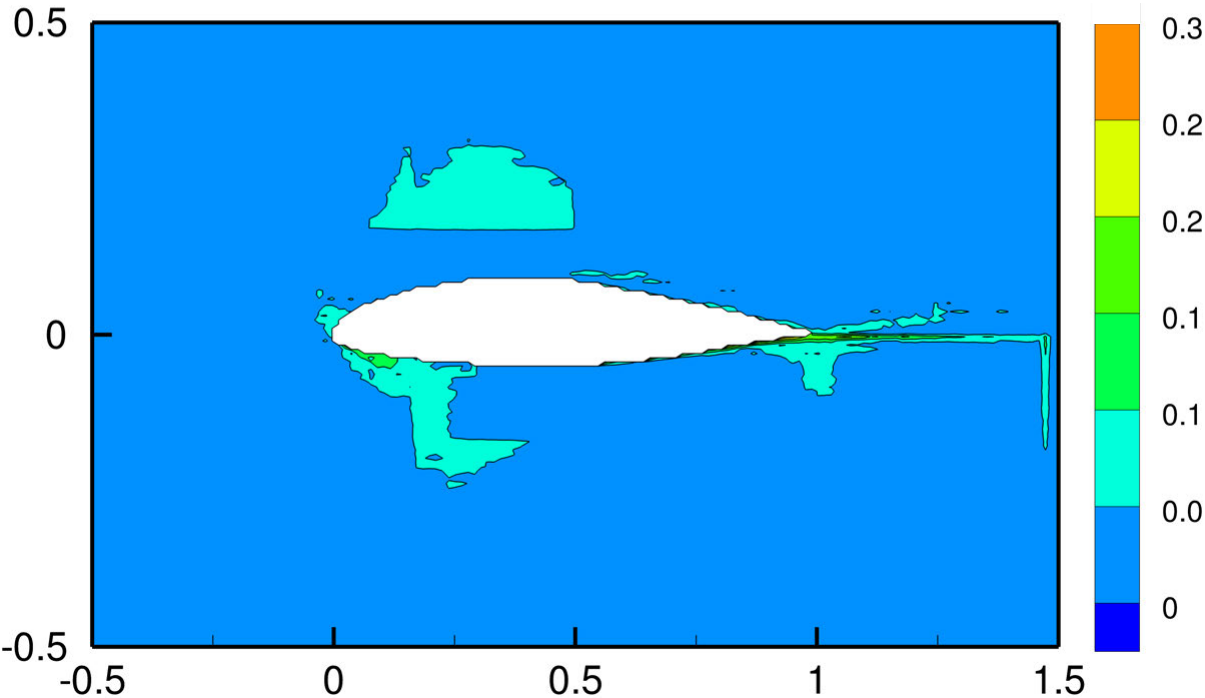}
		}
		\caption{Absolute difference.} 
		\label{fig:fig17c}
	\end{subfigure}
	\begin{subfigure}{0.32\textwidth}
		\centering
		\resizebox{\textwidth}{!}{%
			\includegraphics{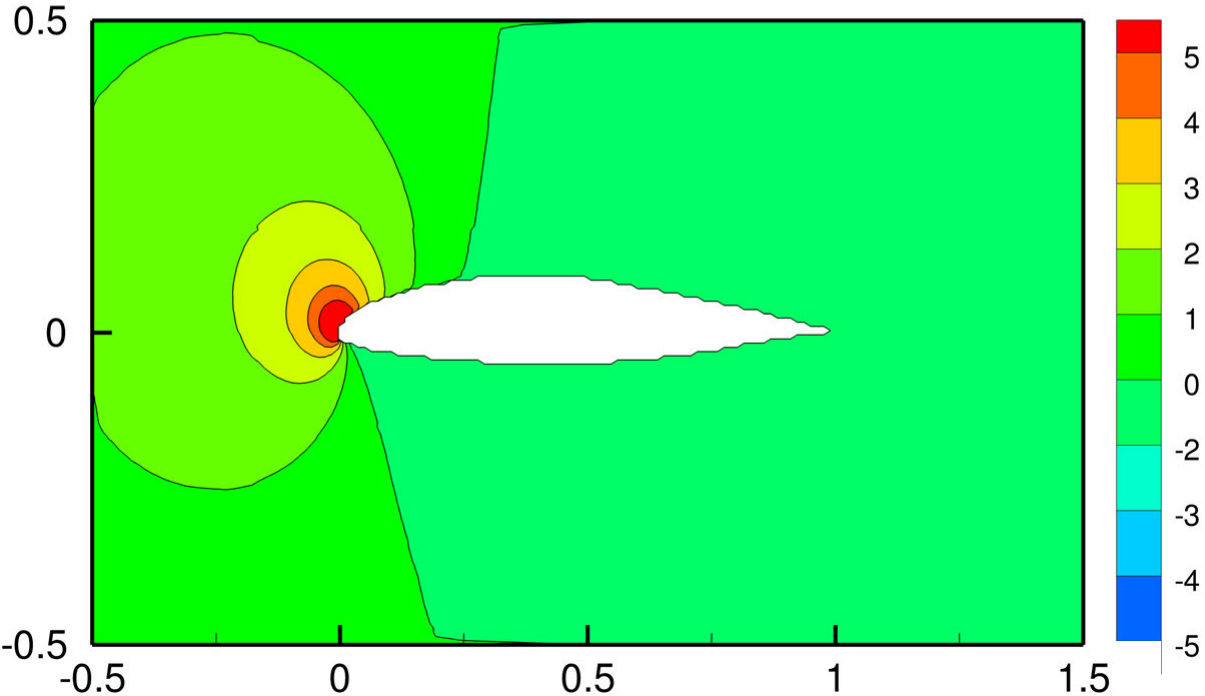}
		}
		\caption{Prediction ($V_p$).}
		\label{fig:fig17d}
	\end{subfigure}
	\begin{subfigure}{0.32\textwidth}
		\centering
		\resizebox{\textwidth}{!}{%
			\includegraphics{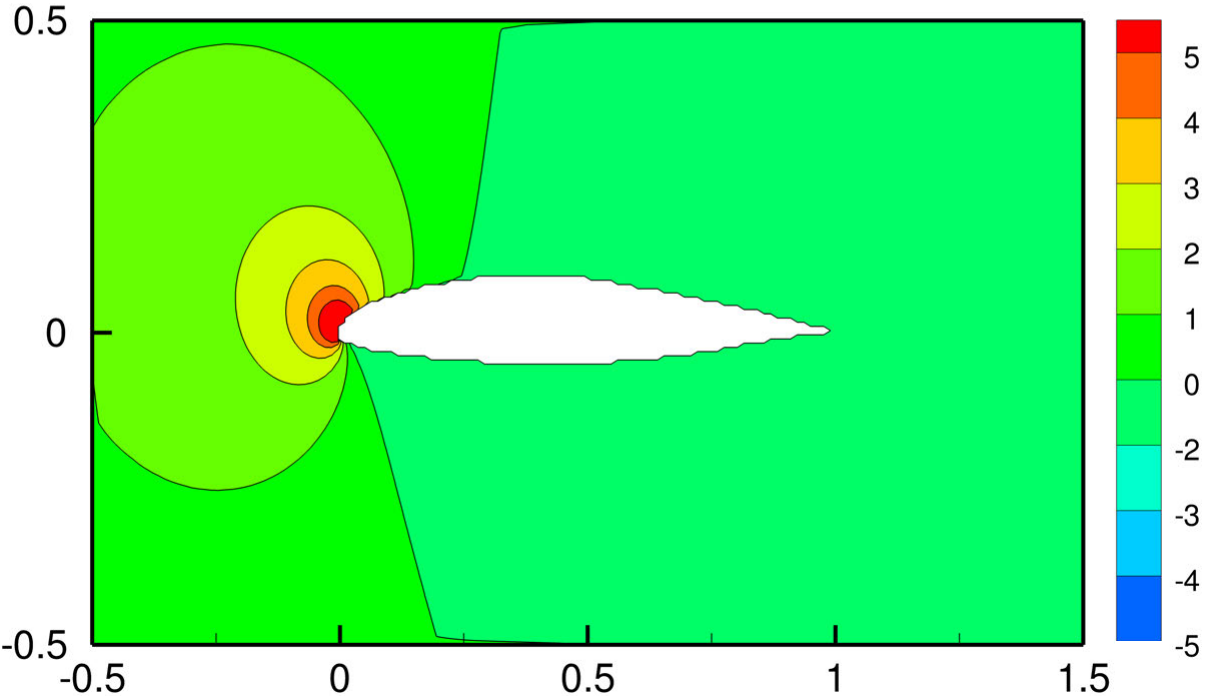}
		}
		\caption{Ground truth ($V_t$).}
		\label{fig:fig17e}
	\end{subfigure}
	\begin{subfigure}{0.32\textwidth}
		\centering
		\resizebox{\textwidth}{!}{%
			\includegraphics{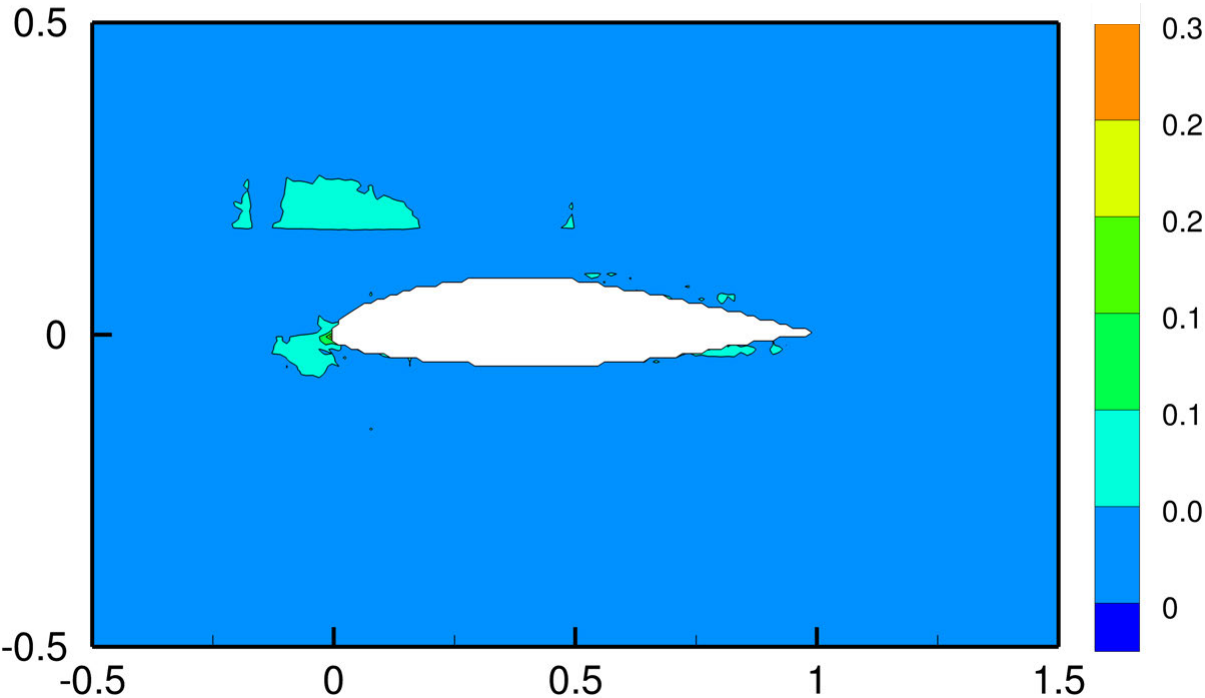}
		}
		\caption{Absolute difference.} 
		\label{fig:fig17f}
	\end{subfigure}
	\begin{subfigure}{0.32\textwidth}
		\centering
		\resizebox{\textwidth}{!}{%
			\includegraphics{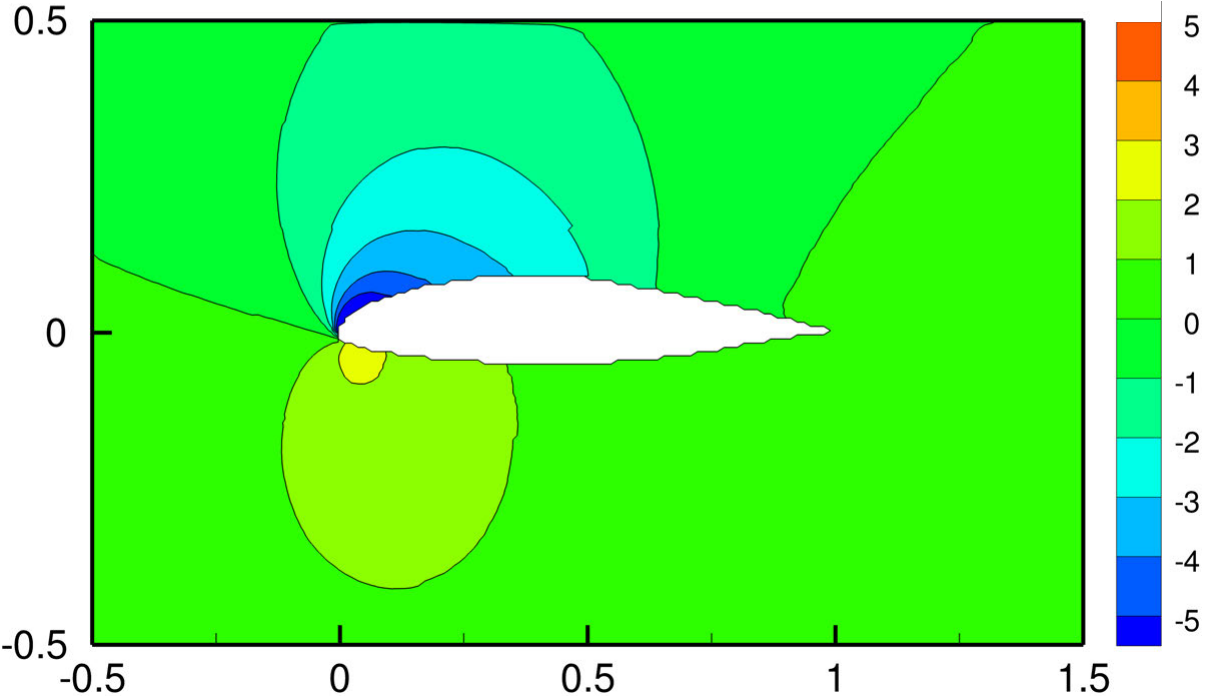}
		}
		\caption{Prediction ($P_p$).}
		\label{fig:fig17g}
	\end{subfigure}
	\begin{subfigure}{0.32\textwidth}
		\centering
		\resizebox{\textwidth}{!}{%
			\includegraphics{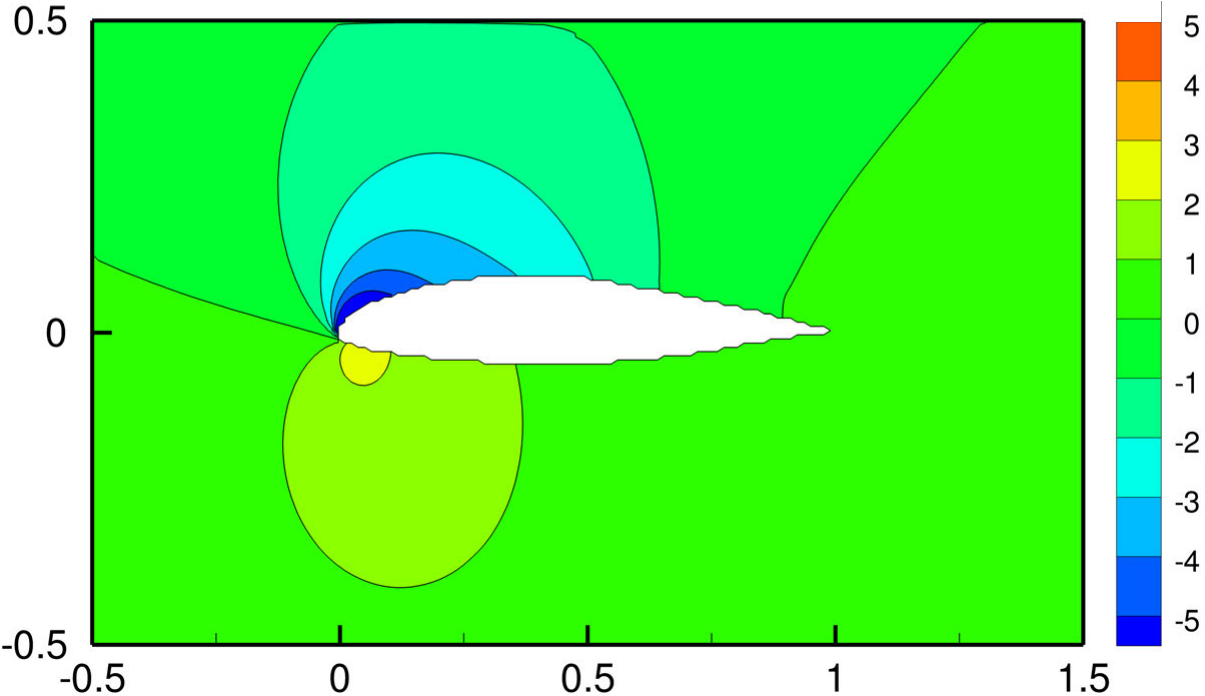}
		}
		\caption{Ground truth ($P_t$).}
		\label{fig:fig17h}
	\end{subfigure}
	\begin{subfigure}{0.32\textwidth}
		\centering
		\resizebox{\textwidth}{!}{%
			\includegraphics{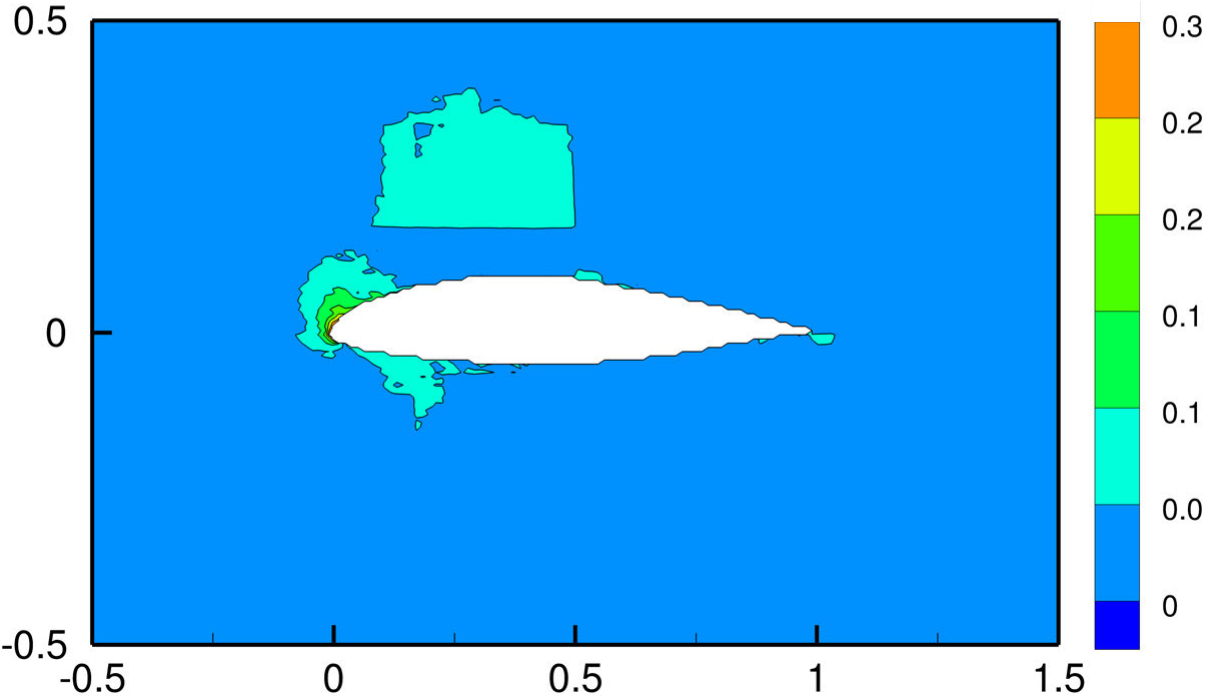}
		}
		\caption{Absolute difference.} 
		\label{fig:fig17i}
	\end{subfigure}
	\caption{Ground truth vs. Prediction of the velocity field components and pressure around the S805 airfoil at $\alpha = 12^\circ$ and $Re = 3\times10^6$.}
	\label{fig:fig17}
\end{figure}

\begin{figure}[H]
	\centering
	\begin{subfigure}{0.32\textwidth}
		\centering
		\resizebox{\textwidth}{!}{%
			\includegraphics{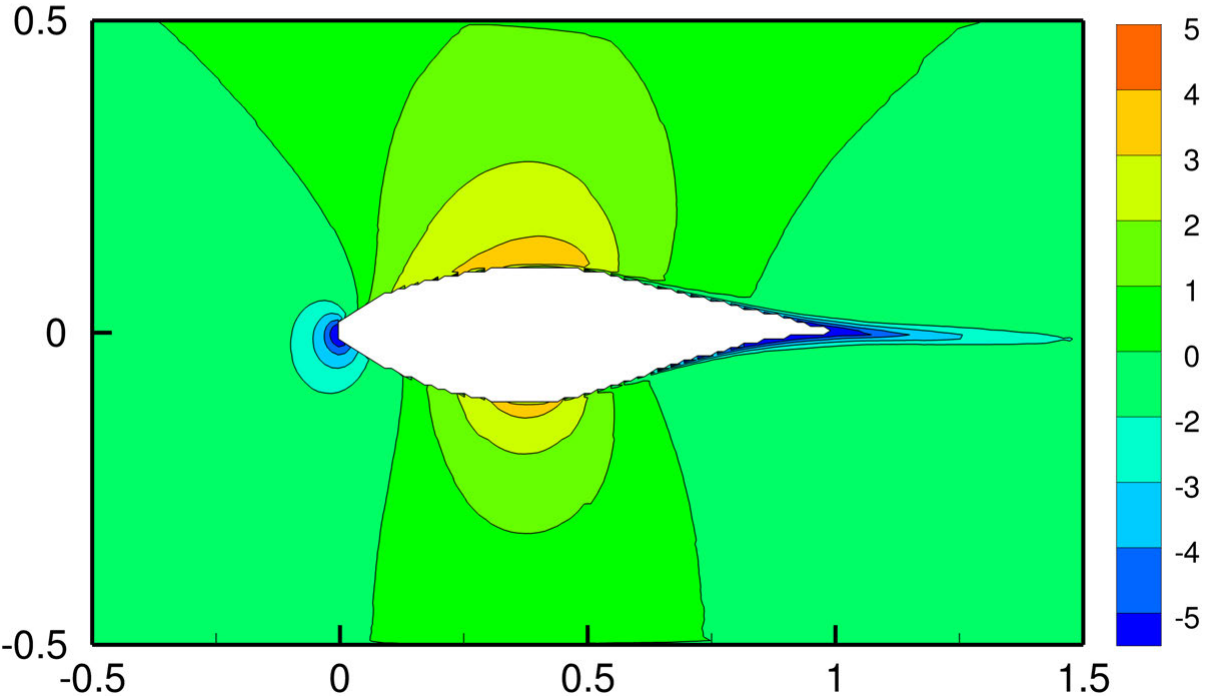}
		}
		\caption{Prediction ($U_p$).}
		\label{fig:fig18a}
	\end{subfigure}
	\begin{subfigure}{0.32\textwidth}
		\centering
		\resizebox{\textwidth}{!}{%
			\includegraphics{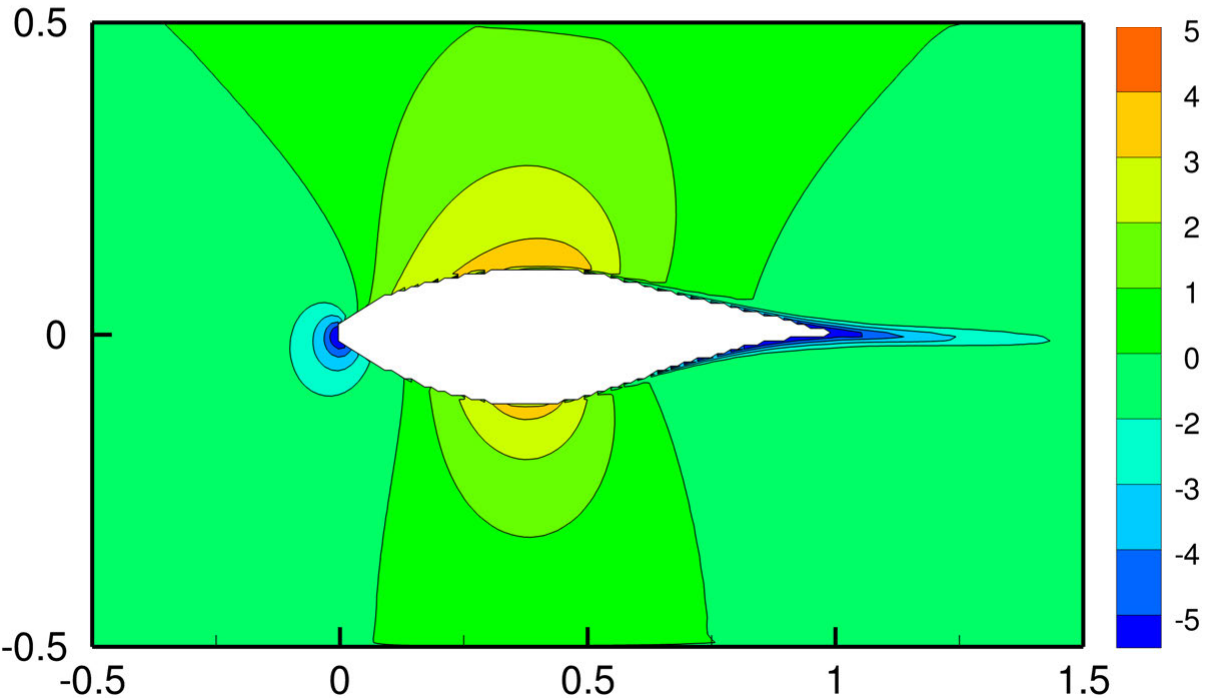}
		}
		\caption{Ground truth ($U_t$).}
		\label{fig:fig18b}
	\end{subfigure}
	\begin{subfigure}{0.32\textwidth}
		\centering
		\resizebox{\textwidth}{!}{%
			\includegraphics{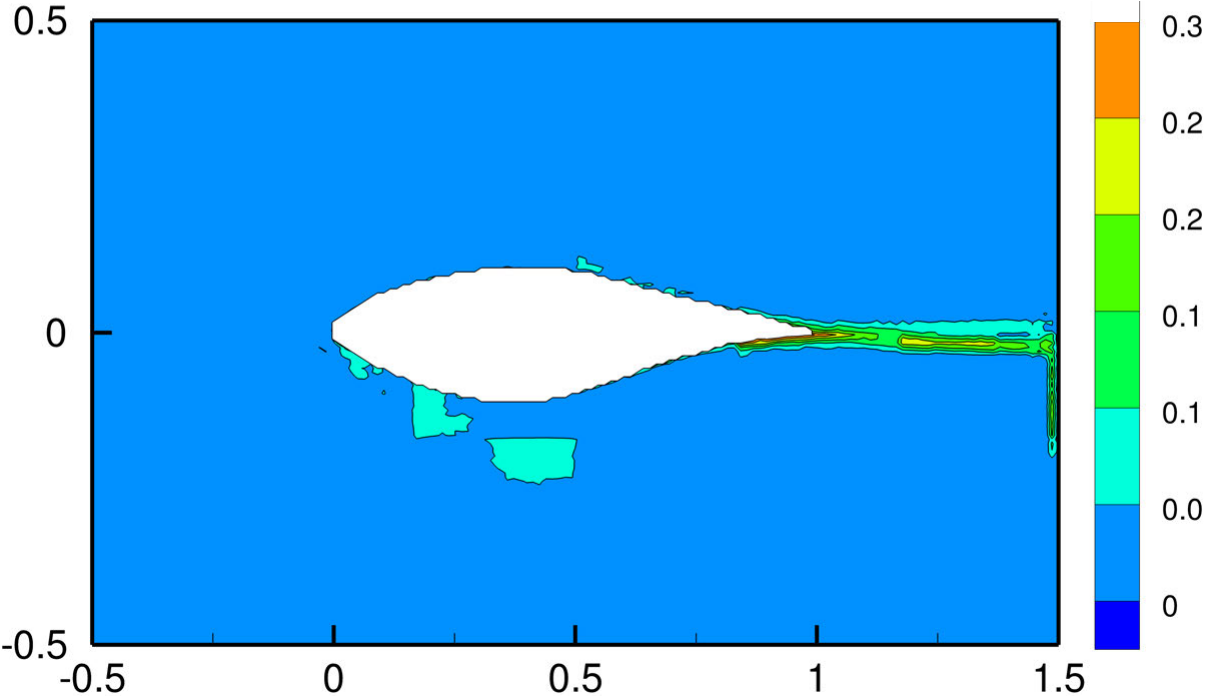}
		}
		\caption{Absolute difference.}
		\label{fig:fig18c}
	\end{subfigure}
	\begin{subfigure}{0.32\textwidth}
		\centering
		\resizebox{\textwidth}{!}{%
			\includegraphics{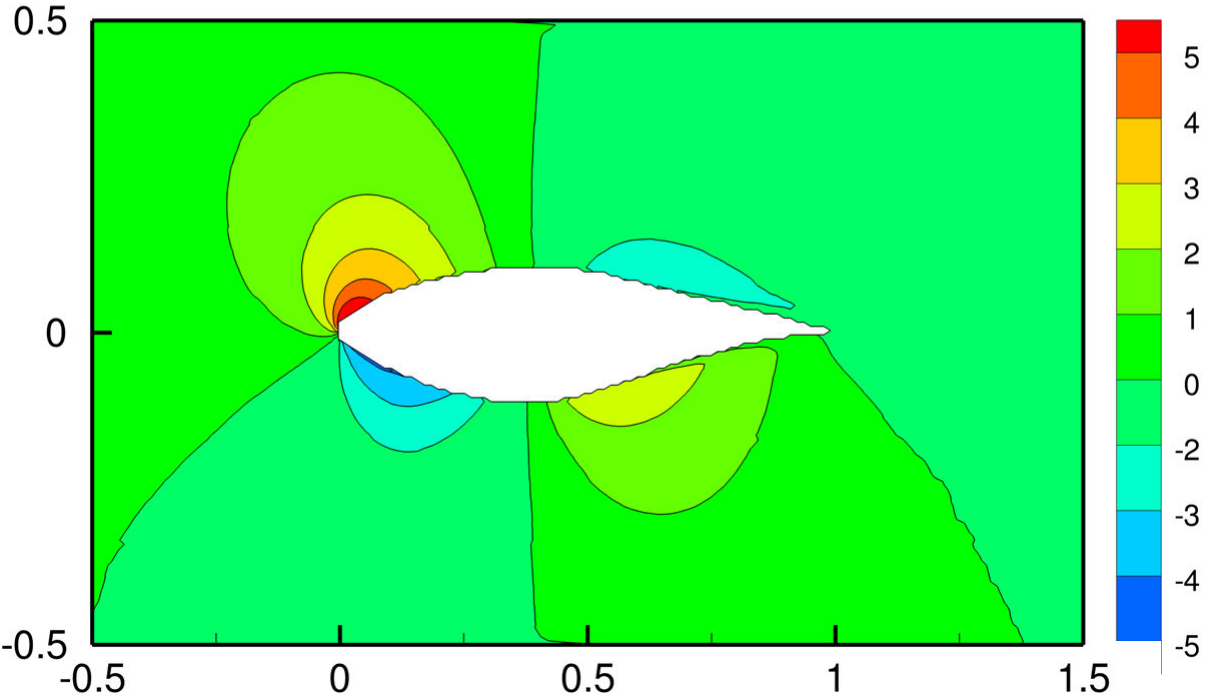}
		}
		\caption{Prediction ($V_p$).}
		\label{fig:fig18d}
	\end{subfigure}
	\begin{subfigure}{0.32\textwidth}
		\centering
		\resizebox{\textwidth}{!}{%
			\includegraphics{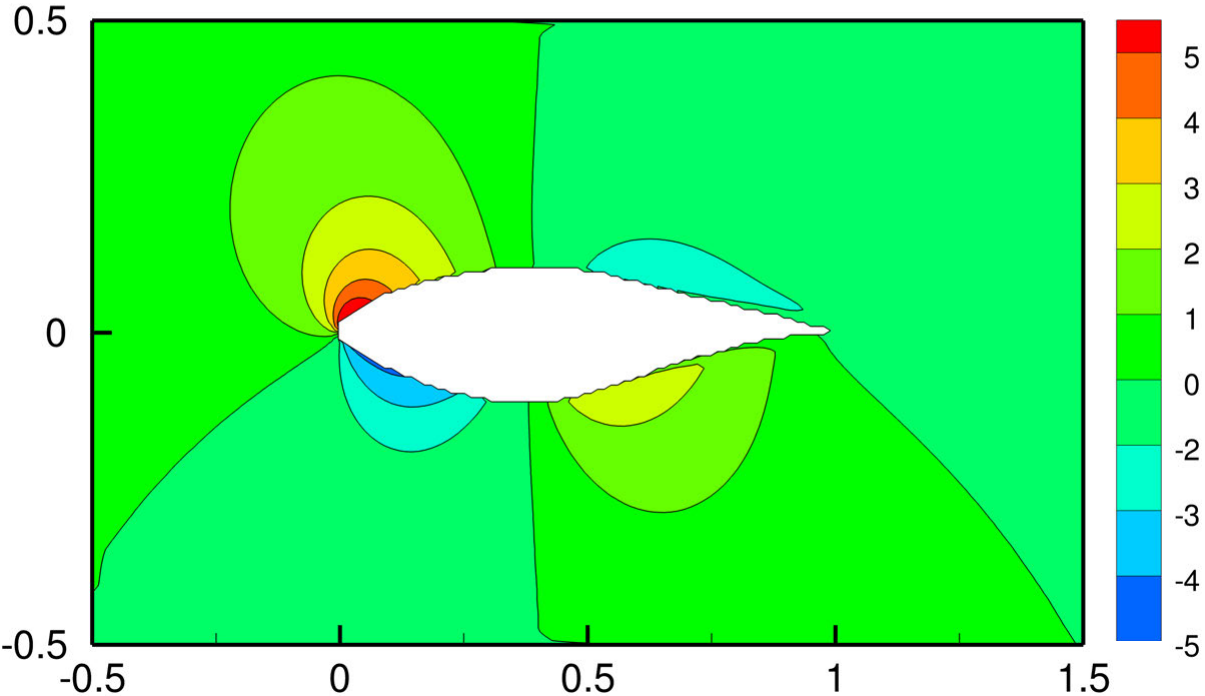}
		}
		\caption{Ground truth ($V_t$).}
		\label{fig:fig18e}
	\end{subfigure}
	\begin{subfigure}{0.32\textwidth}
		\centering
		\resizebox{\textwidth}{!}{%
			\includegraphics{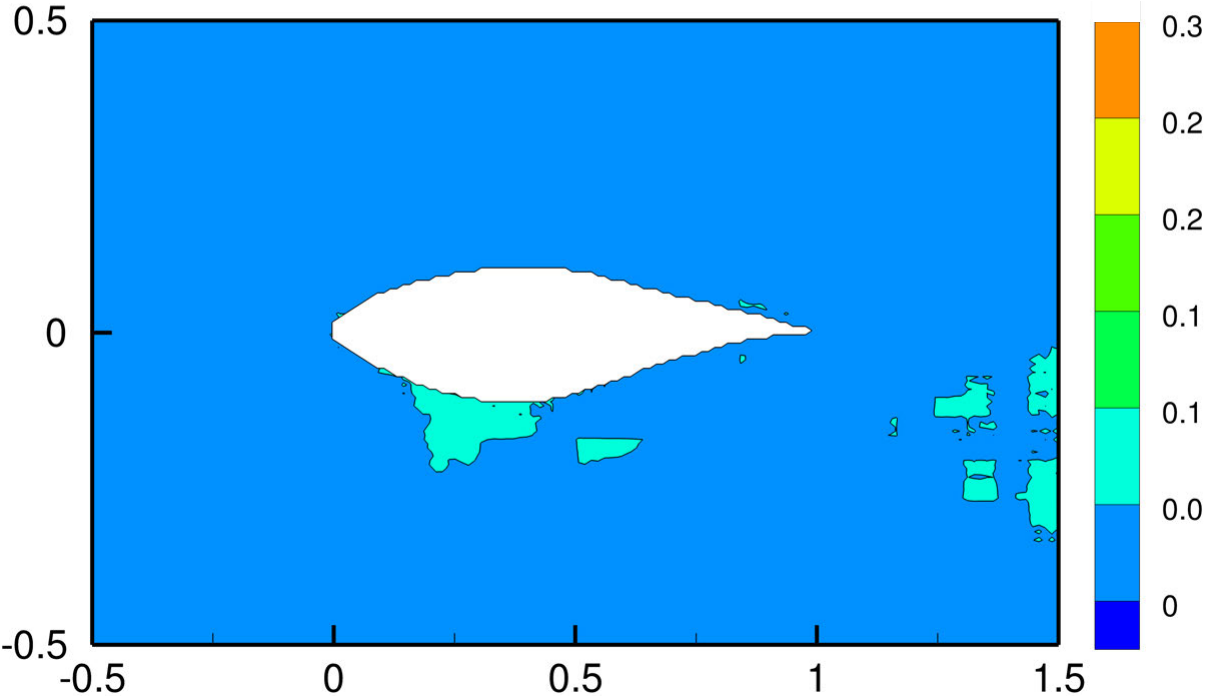}
		}
		\caption{Absolute difference.}
		\label{fig:fig18f}
	\end{subfigure}
	\begin{subfigure}{0.32\textwidth}
		\centering
		\resizebox{\textwidth}{!}{%
			\includegraphics{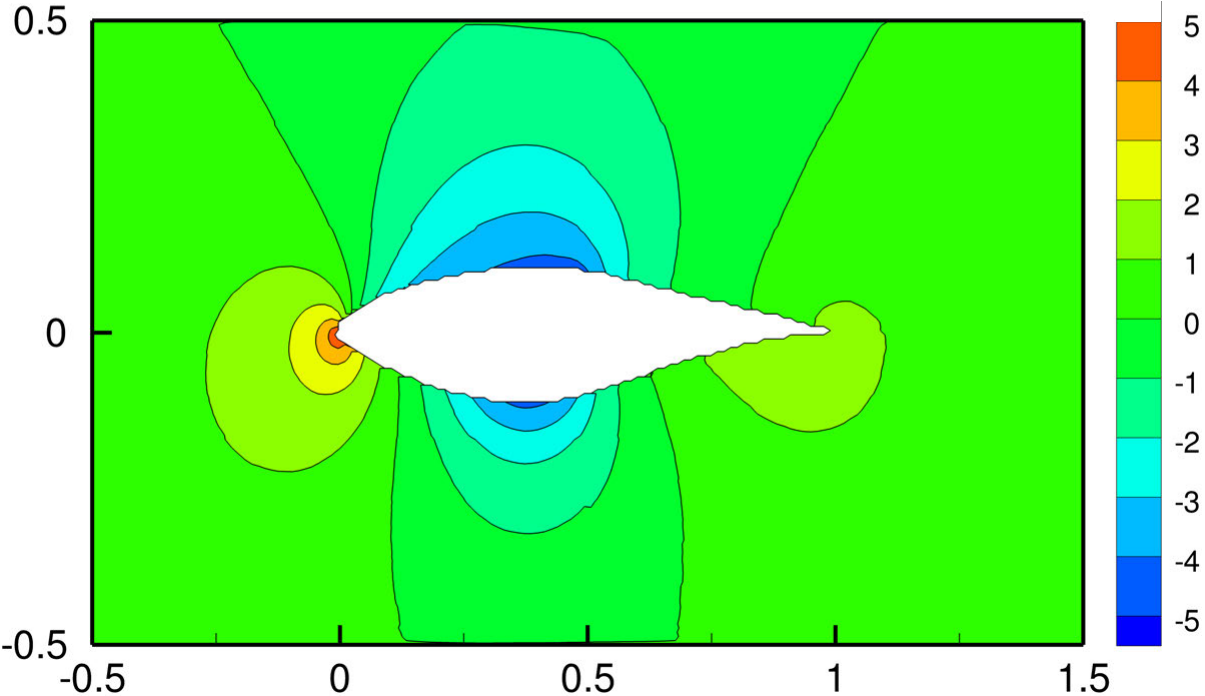}
		}
		\caption{Prediction ($P_p$).}
		\label{fig:fig18g}
	\end{subfigure}
	\begin{subfigure}{0.32\textwidth}
		\centering
		\resizebox{\textwidth}{!}{%
			\includegraphics{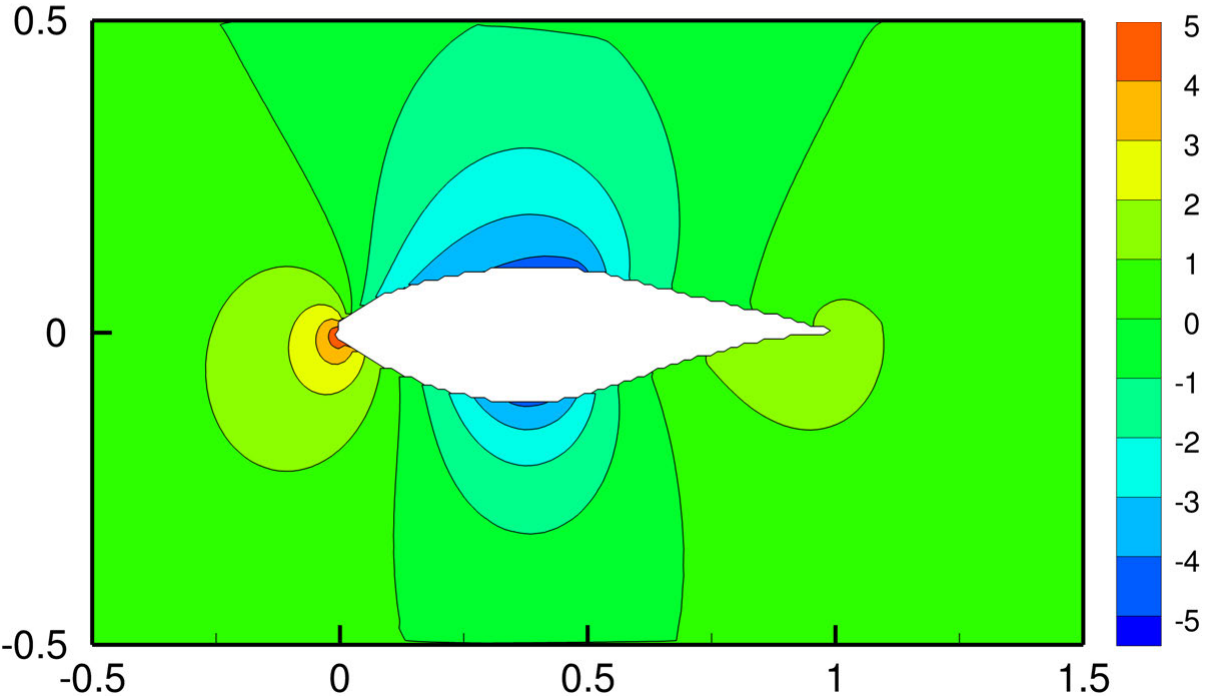}
		}
		\caption{Ground truth ($P_t$).}
		\label{fig:fig18h}
	\end{subfigure}
	\begin{subfigure}{0.32\textwidth}
		\centering
		\resizebox{\textwidth}{!}{%
			\includegraphics{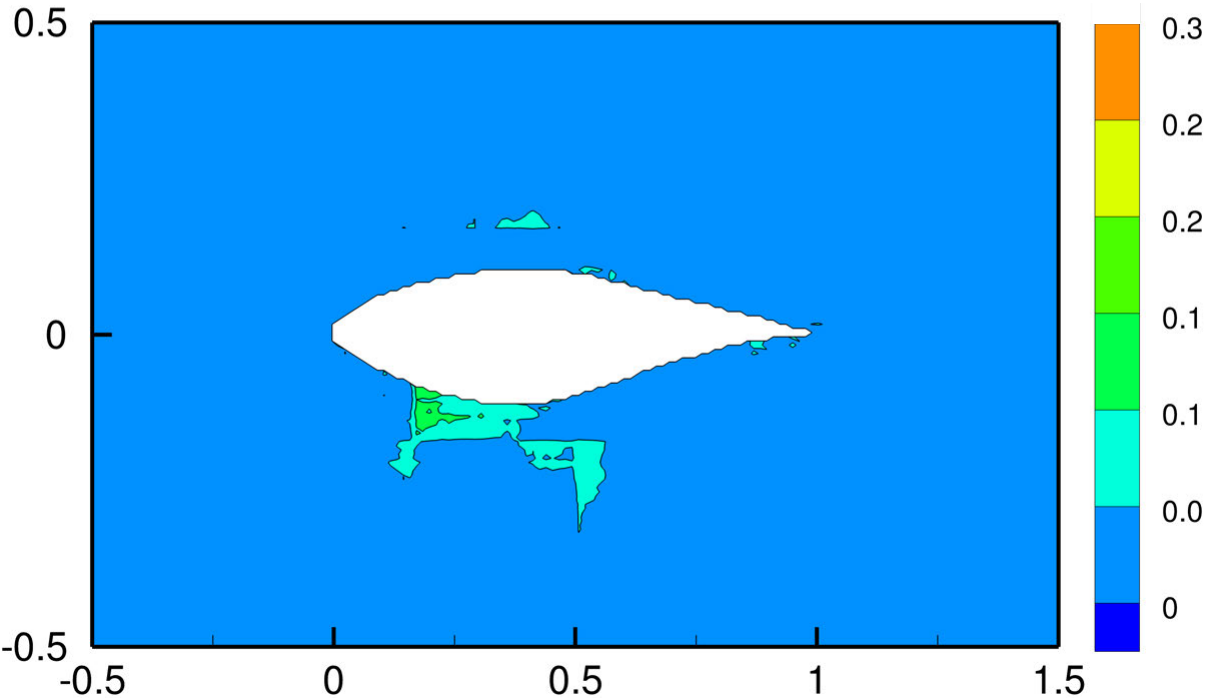}
		}
		\caption{Absolute difference.}
		\label{fig:fig18i}
	\end{subfigure}
	\caption{Ground truth vs. Prediction of the velocity field components and pressure around the S809 airfoil at  $\alpha =1^\circ$ and $Re = 3\times10^6$.}
	\label{fig:fig18}
\end{figure}

\begin{figure}[H]
	\centering
	\begin{subfigure}{0.32\textwidth}
		\centering
		\resizebox{\textwidth}{!}{%
			\includegraphics{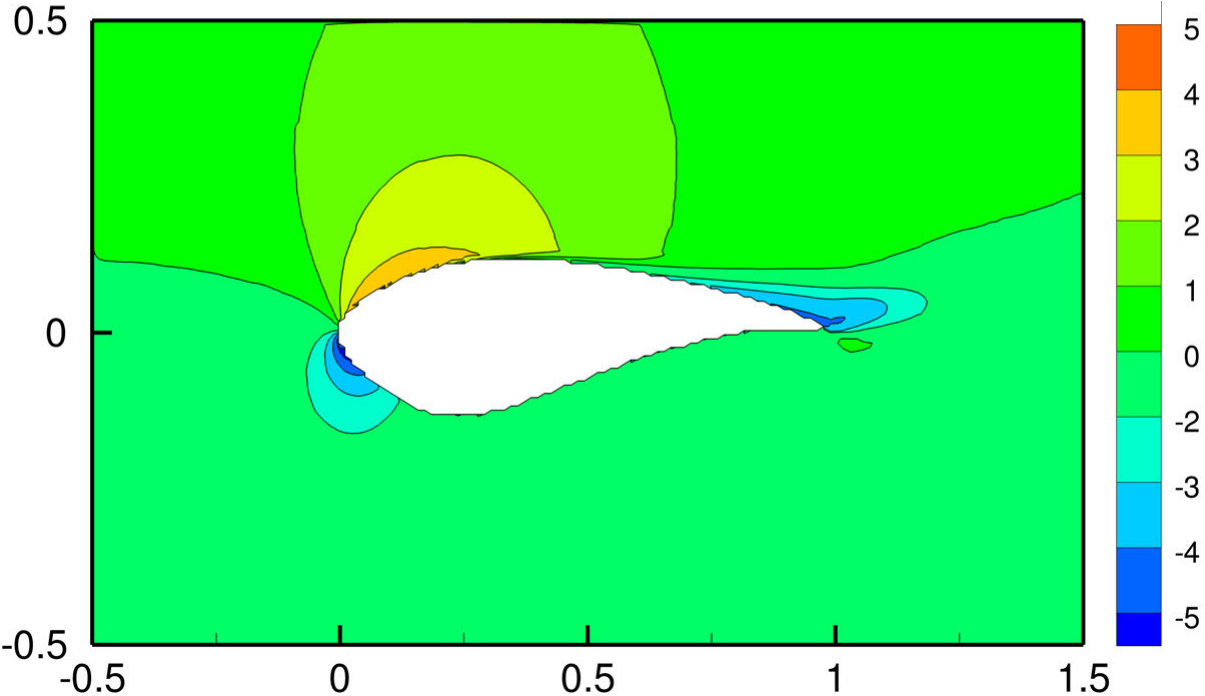}
		}
		\caption{Prediction ($U_p$).}
		\label{fig:fig19a}
	\end{subfigure}
	\begin{subfigure}{0.32\textwidth}
		\centering
		\resizebox{\textwidth}{!}{%
			\includegraphics{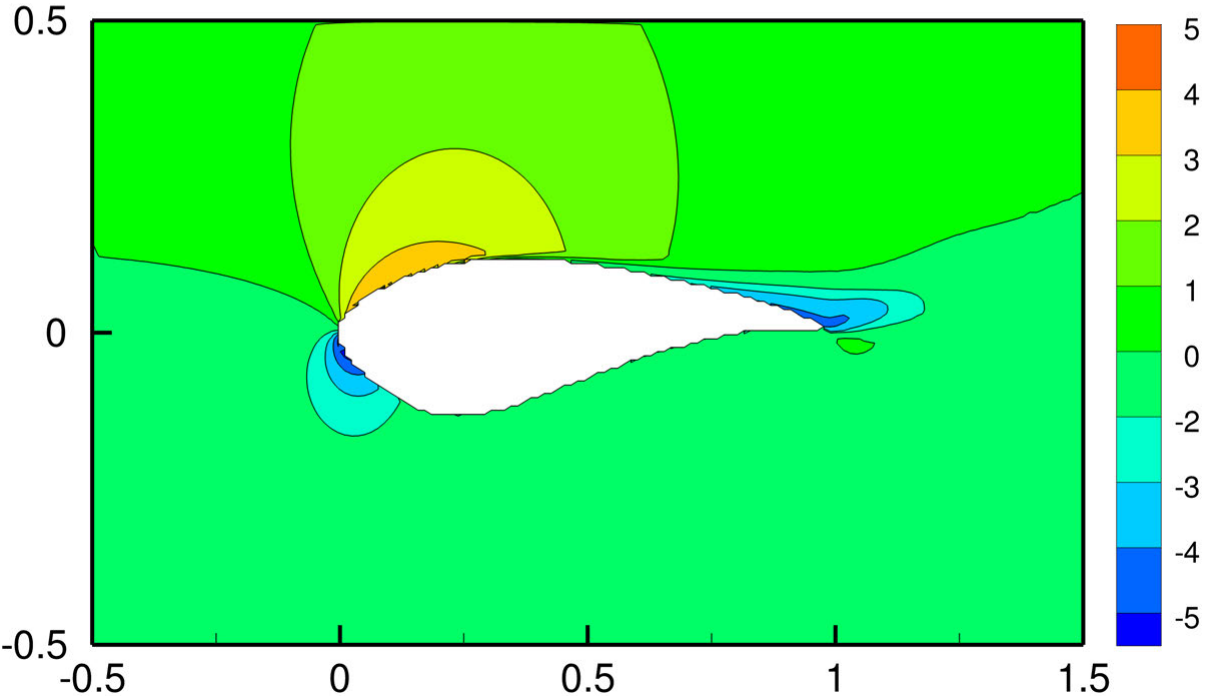}
		}
		\caption{Ground truth ($U_t$).}
		\label{fig:fig19b}
	\end{subfigure}
	\begin{subfigure}{0.32\textwidth}
		\centering
		\resizebox{\textwidth}{!}{%
			\includegraphics{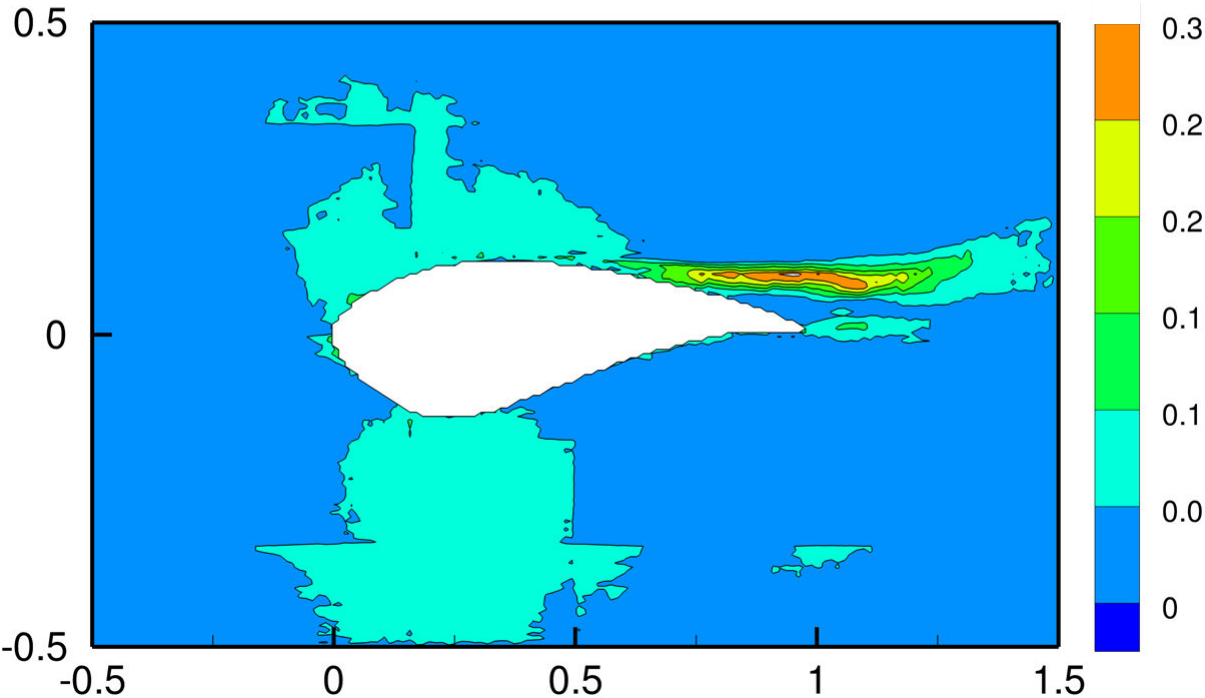}
		}
		\caption{Absolute difference.}
		\label{fig:fig19c}
	\end{subfigure}
	\begin{subfigure}{0.32\textwidth}
		\centering
		\resizebox{\textwidth}{!}{%
			\includegraphics{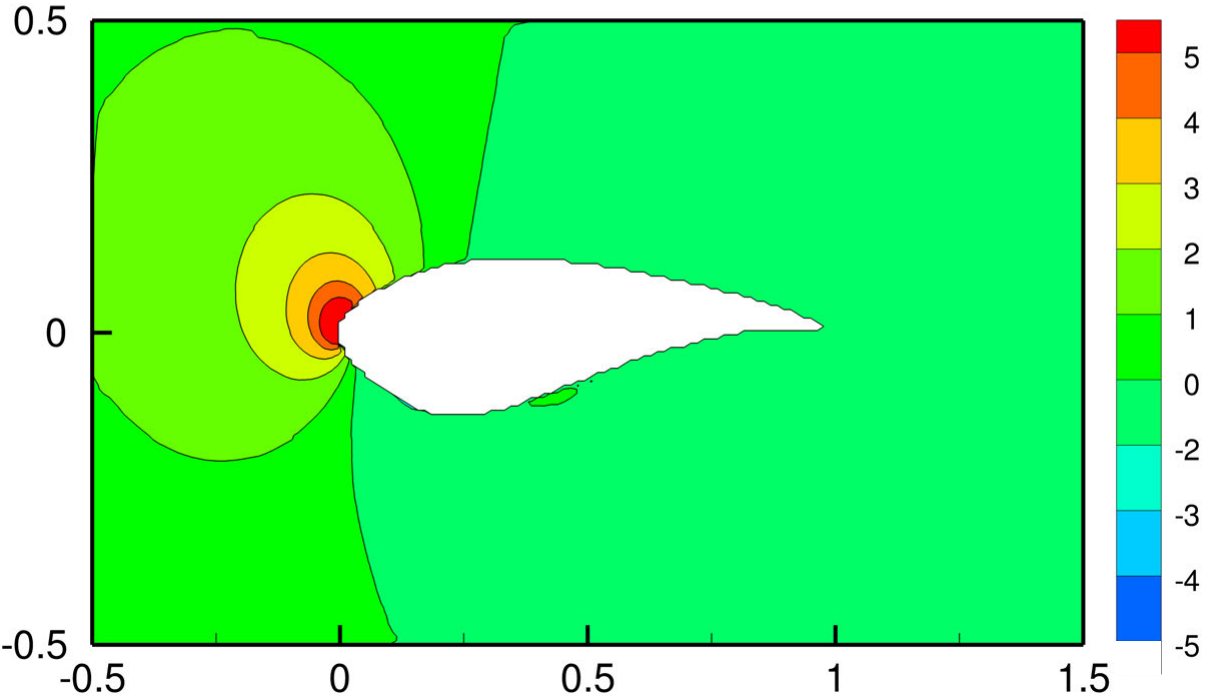}
		}
		\caption{Prediction ($V_p$).}
		\label{fig:fig19d}
	\end{subfigure}
	\begin{subfigure}{0.32\textwidth}
		\centering
		\resizebox{\textwidth}{!}{%
			\includegraphics{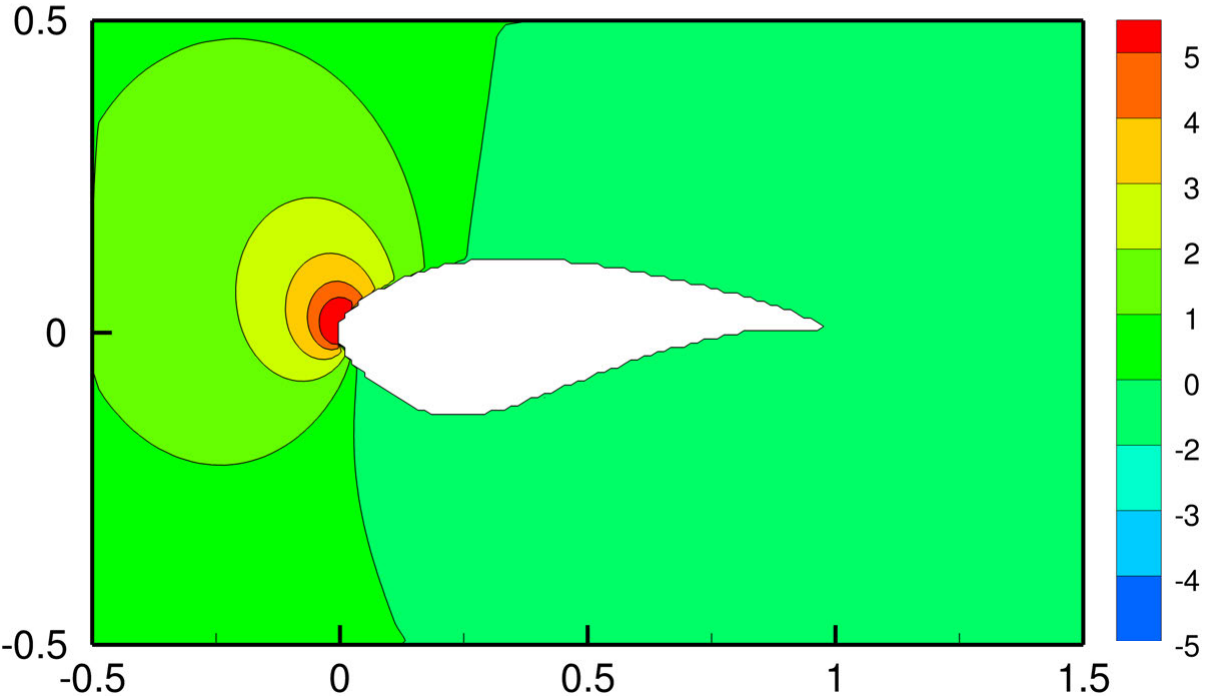}
		}
		\caption{Ground truth ($V_t$).}
		\label{fig:fig19e}
	\end{subfigure}
	\begin{subfigure}{0.32\textwidth}
		\centering
		\resizebox{\textwidth}{!}{%
			\includegraphics{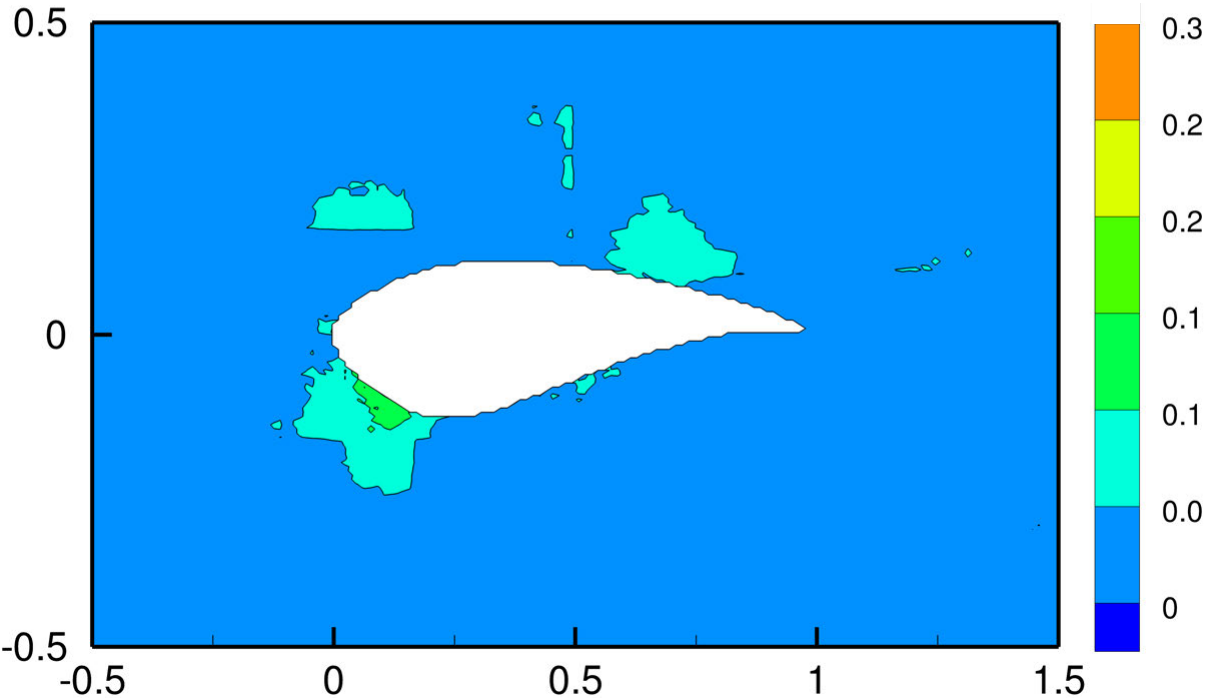}
		}
		\caption{Absolute difference.}
		\label{fig:fig19f}
	\end{subfigure}
	\begin{subfigure}{0.32\textwidth}
		\centering
		\resizebox{\textwidth}{!}{%
			\includegraphics{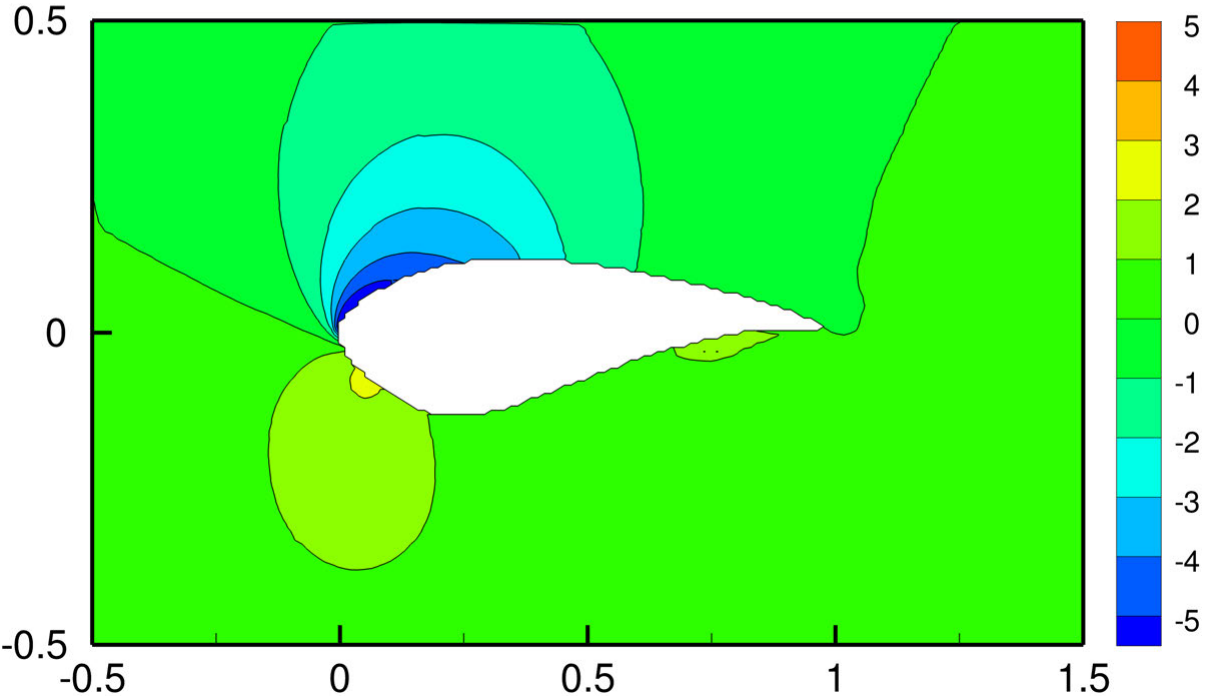}
		}
		\caption{Prediction ($P_p$).}
		\label{fig:fig19g}
	\end{subfigure}
	\begin{subfigure}{0.32\textwidth}
		\centering
		\resizebox{\textwidth}{!}{%
			\includegraphics{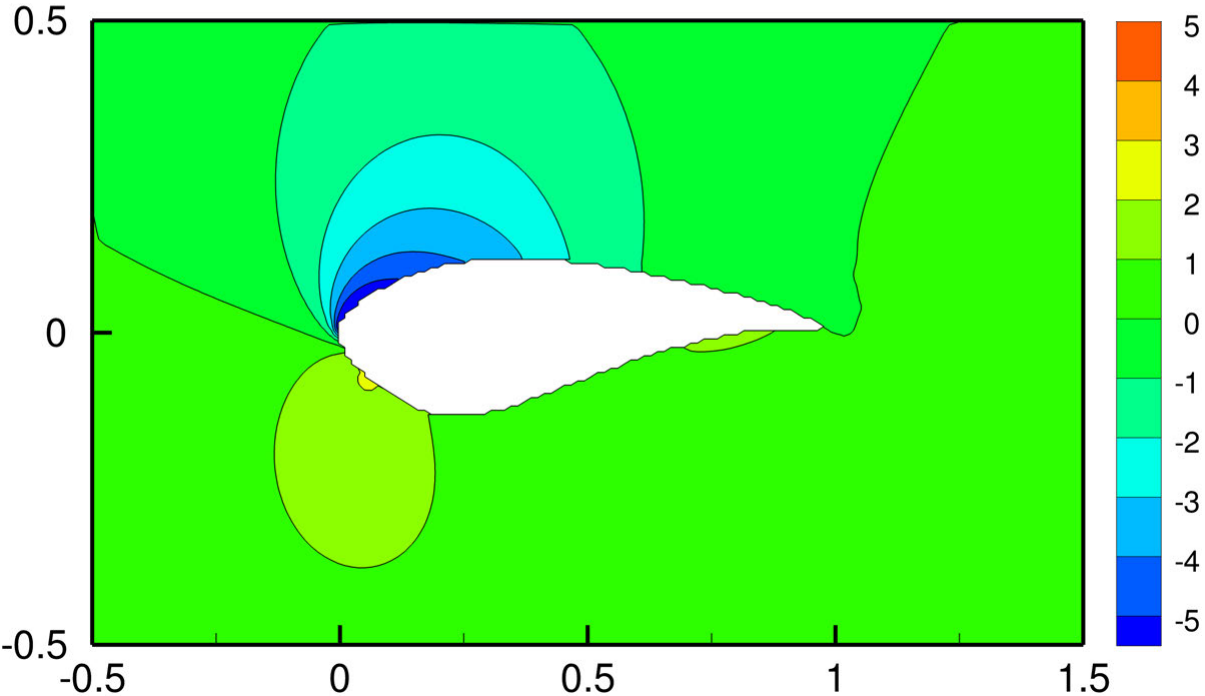}
		}
		\caption{Ground truth ($P_t$).}
		\label{fig:fig19h}
	\end{subfigure}
	\begin{subfigure}{0.32\textwidth}
		\centering
		\resizebox{\textwidth}{!}{%
			\includegraphics{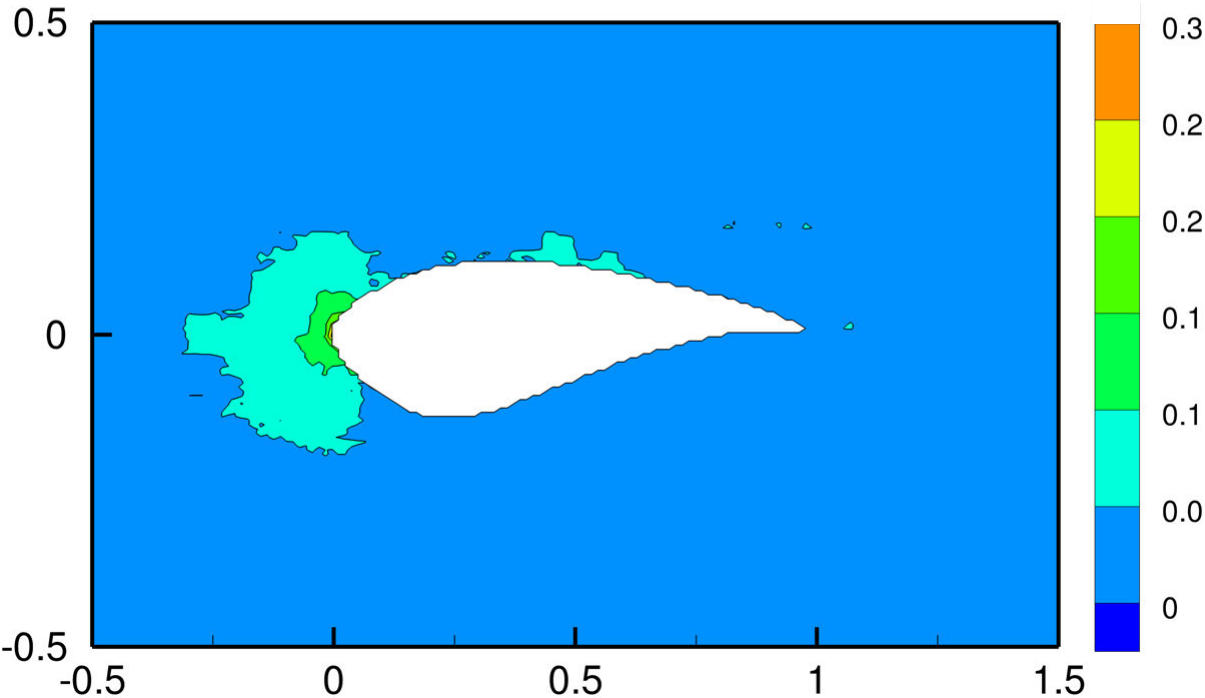}
		}
		\caption{Absolute difference.}
		\label{fig:fig19i}
	\end{subfigure}
	\caption{Ground truth vs. Prediction of the velocity field components and pressure around the S814 airfoil at  $\alpha = 15^\circ$ and $Re = 2\times10^6$.}
	\label{fig:fig19}
\end{figure}

Figures.~\ref{fig:fig17},~\ref{fig:fig18} and~\ref{fig:fig19} illustrate the flow-field predictions with gradient sharpening in the loss function and in comparison with the reference results from the OVERTURNS CFD code.

\begin{figure}[H]
	\centering
	\resizebox{0.75\textwidth}{!}{%
		\includegraphics{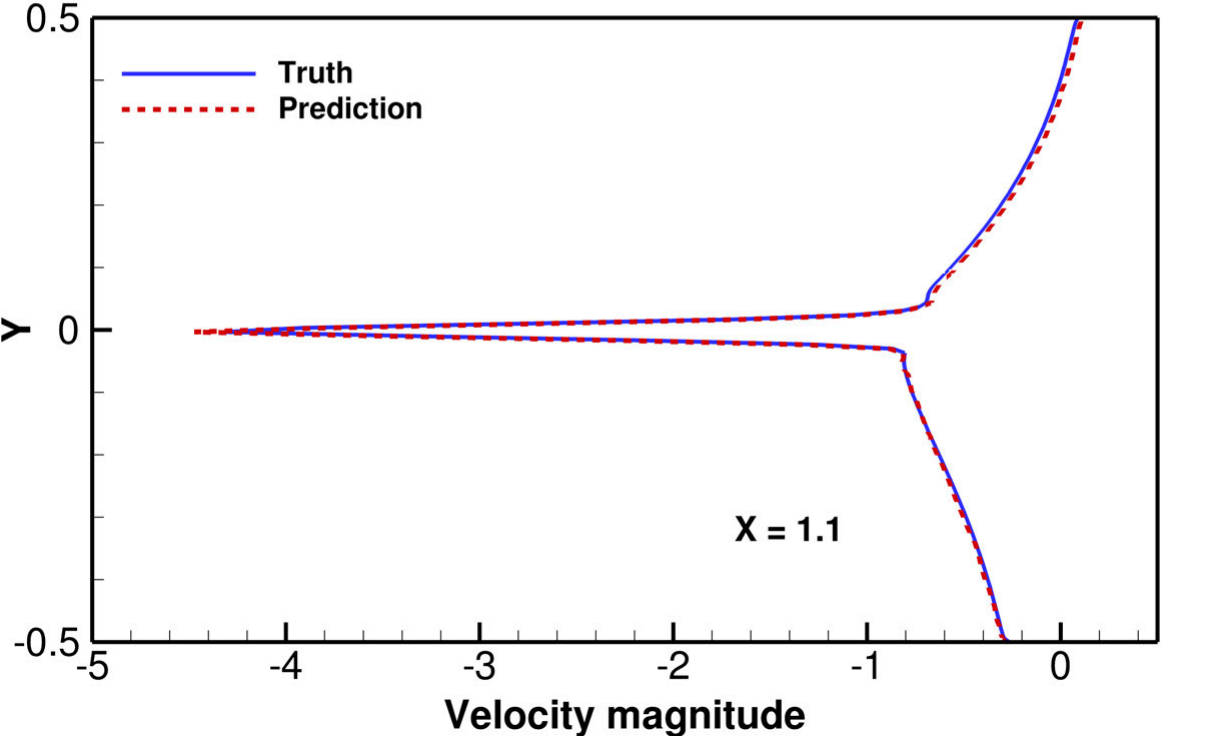}
	}
	\caption{Wake velocity (x-component) prediction at $x=1.1$ downstream of the trailing edge for the S809 airfoil at $\alpha = 1^\circ$ and $Re = 1\times10^6$.}
	\label{fig:fig500}
\end{figure}

\begin{figure}[H]
	\centering
	\resizebox{0.75\textwidth}{!}{%
		\includegraphics{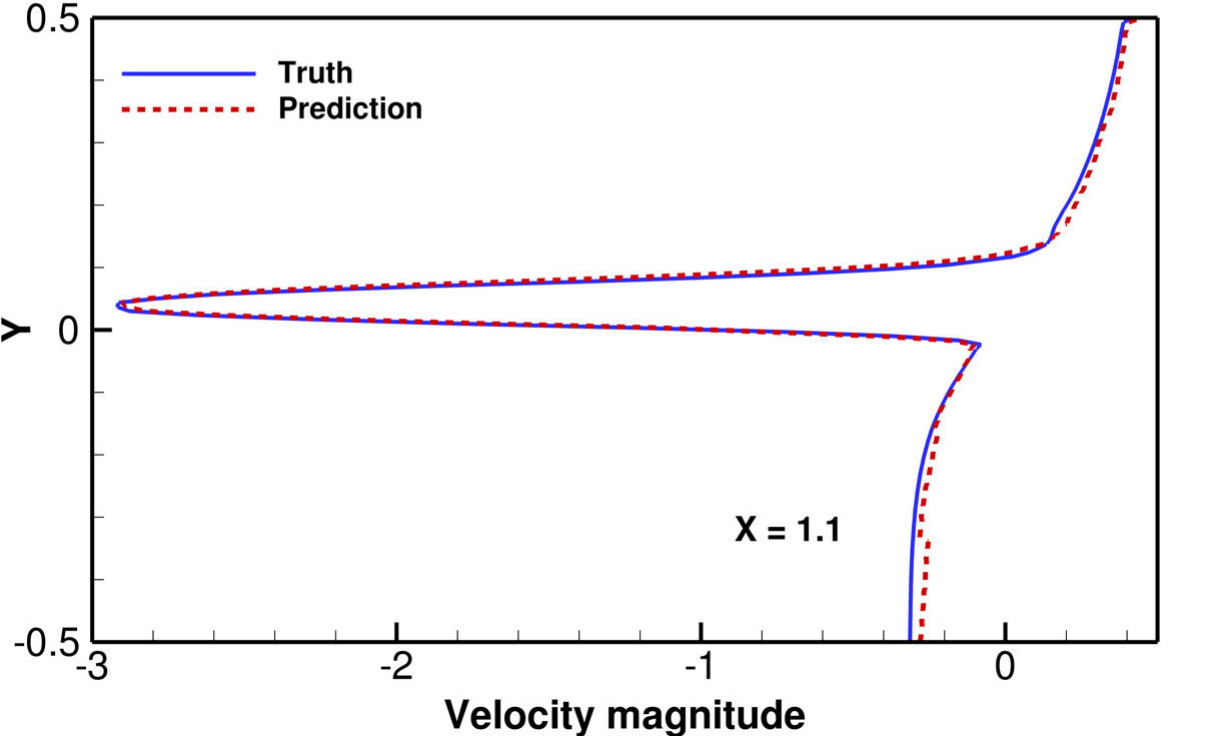}
	}
	\caption{Wake velocity (x-component) prediction at $x=1.1$ downstream of the trailing edge for the S814 airfoil at $\alpha = 15^\circ$ and $Re = 2\times10^6$.}
	\label{fig:fig501}
\end{figure}

Figures.~\ref{fig:fig500} and~\ref{fig:fig501} illustrate the x-component velocity profile of the airfoil wake at $x=1.1$ (downstream location from the leading edge). These predictions include  GS in the loss function. 

\begin{figure}[H]
	\centering
	\resizebox{0.6\textwidth}{!}{%
		\includegraphics{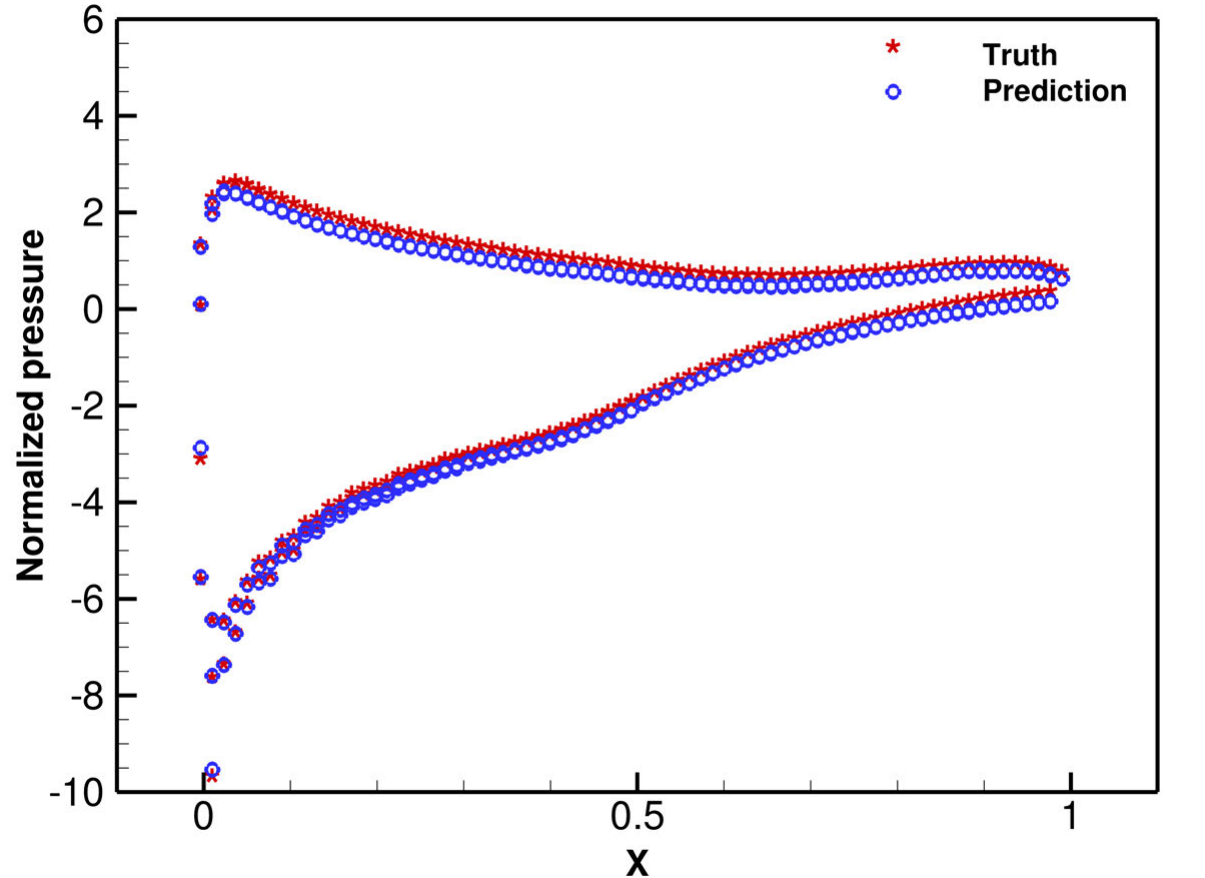}
	}
	\caption{Pressure prediction (standard score normalization) along the surface of the S805 airfoil at $\alpha = 12^\circ$ and $Re = 3\times10^6$.}
	\label{fig:fig22}
\end{figure}

\begin{figure}[H]
	\centering
	\resizebox{0.6\textwidth}{!}{%
		\includegraphics{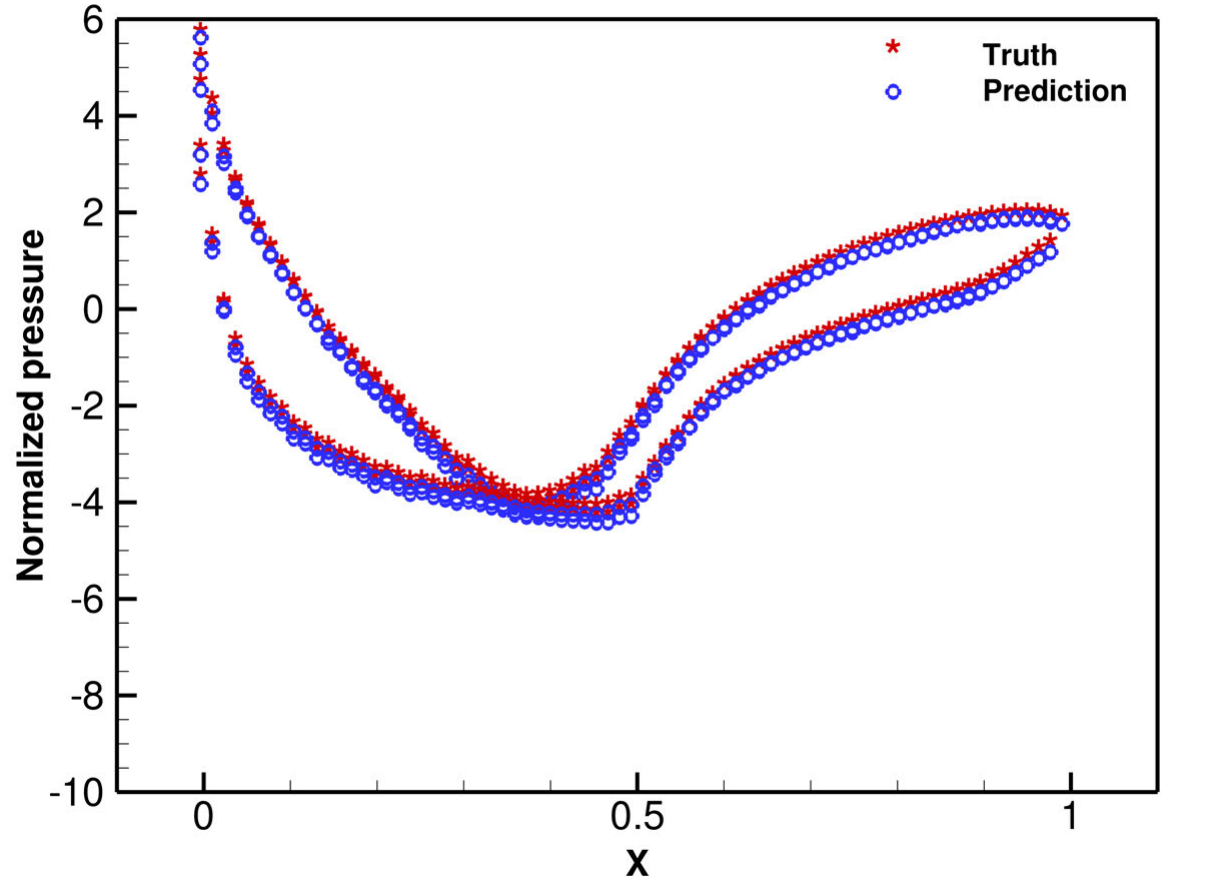}
	}
	\caption{Pressure prediction (standard score normalization) along the surface of the S809 airfoil at $\alpha = 1^\circ$ and $Re = 3\times10^6$.}
	\label{fig:fig23}
\end{figure}

\begin{figure}[H]
	\centering
	\resizebox{0.6\textwidth}{!}{%
		\includegraphics{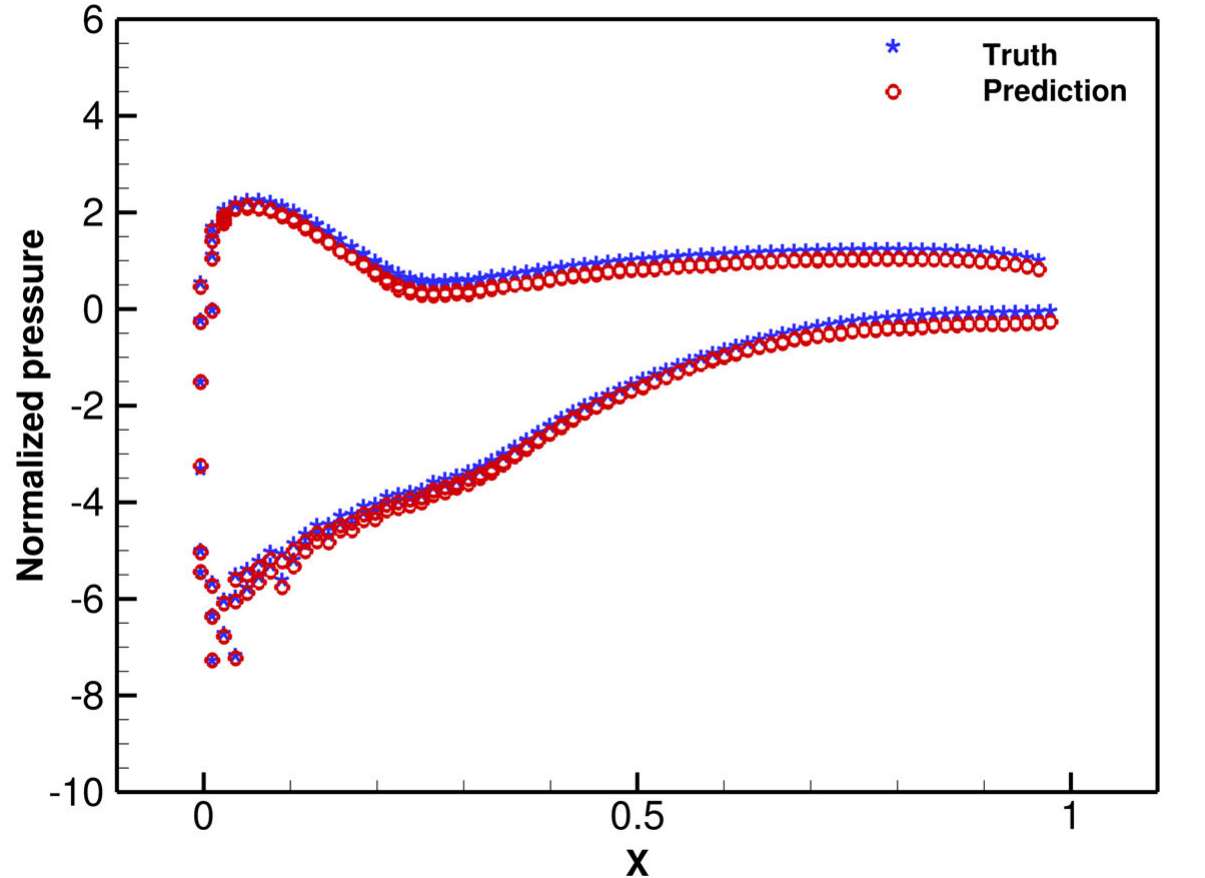}
	}
	\caption{Pressure  prediction (standard score normalization) along the surface of the S814 airfoil at $\alpha = 15^\circ$ and $Re = 2\times10^6$.}
	\label{fig:fig24}
\end{figure}

As a further comparison of the network prediction accuracy, we consider the pressure distribution on the upper and lower boundaries.
Figures.~\ref{fig:fig22},~\ref{fig:fig23}, and~\ref{fig:fig24} depicts the Ground truth vs. Predictions of the normalized pressure using the standard score normalization along the surface of the S805, S809, and S814 airfoils respectively. 
It is noteworthy that the surface with a one-pixel gap adjacent to the airfoil surface is used to obtain the pressure values. This change is due to the masking of the airfoil as an input during the training.

Overall, results are in good agreement with the ground truth simulation results in the entire range of angles of attacks and Reynolds numbers for the three different airfoils.

\subsubsection{Prediction for unseen airfoil shapes}
\label{Unseen}
\begin{figure}[H]
	\centering
	\resizebox{0.75\textwidth}{!}{%
		\includegraphics{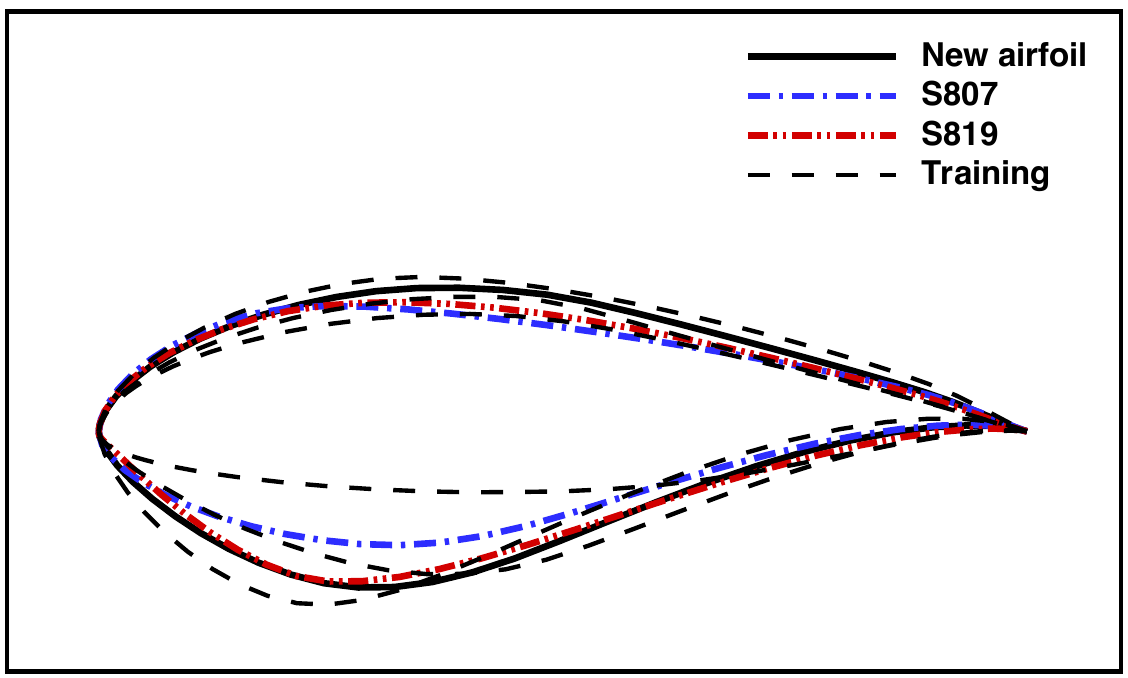}
	}
	\caption{Illustration of the newly created airfoil in black which is an averaged shape between the S809 and S814 airfoils respectively.  The S807 and S819 airfoils are also illustrated in blue and red respectively.}
	\label{fig:fig25}
\end{figure}

To further explore the predictive ability and accuracy of the trained network, three unseen geometries are considered as shown in Figure~\ref{fig:fig25}). The first one, denoted by "new airfoil" is an averaged shape of S809 and S814 airfoils. In addition, the S807 and S819 airfoils are also considered.  

\begin{figure}[H]
	\centering
	\begin{subfigure}{0.45\textwidth}
		\centering
		\resizebox{\textwidth}{!}{%
			\includegraphics{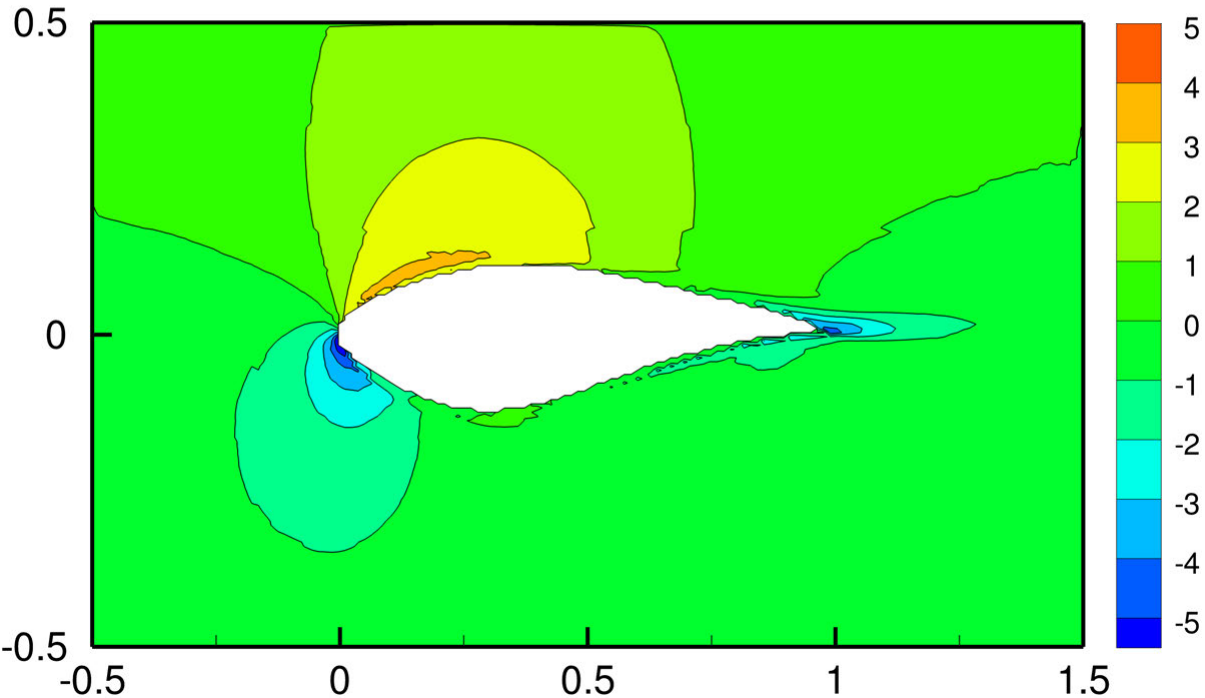}
		}
		\caption{Prediction ($U_p$).}
		\label{fig:fig26a}
	\end{subfigure}
	\begin{subfigure}{0.45\textwidth}
		\centering
		\resizebox{\textwidth}{!}{%
			\includegraphics{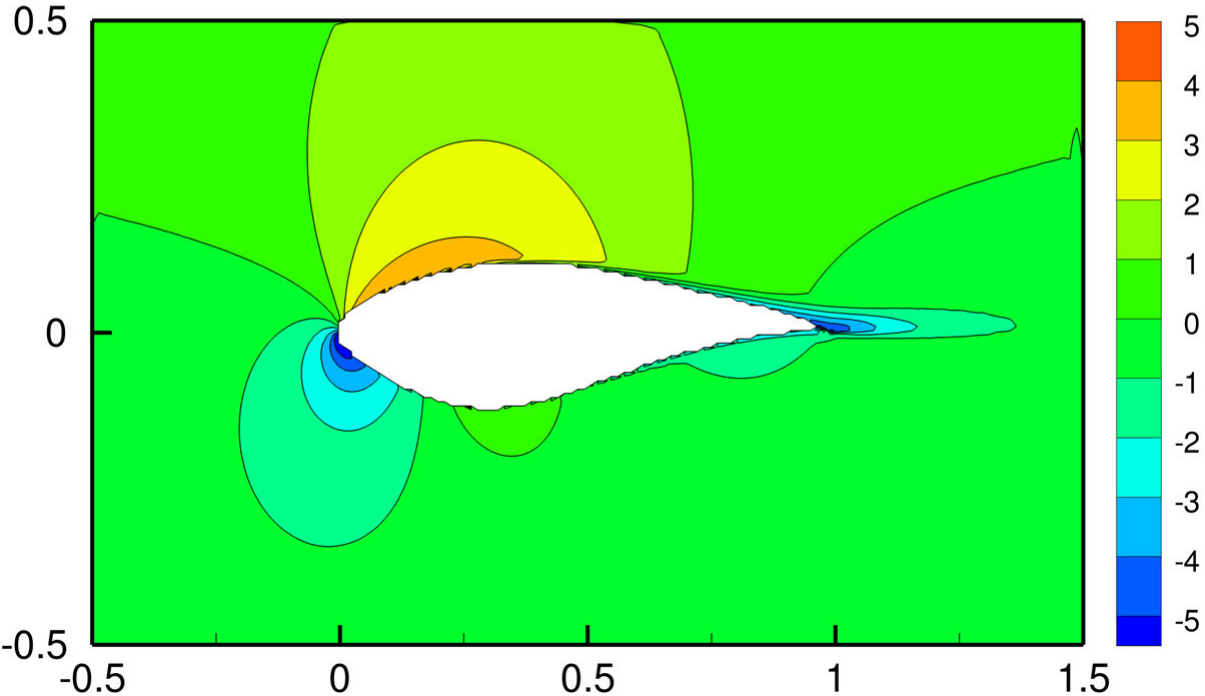}
		}
		\caption{Ground truth ($U_t$).}
		\label{fig:fig26b}
	\end{subfigure}
	\begin{subfigure}{0.45\textwidth}
		\centering
		\resizebox{\textwidth}{!}{%
			\includegraphics{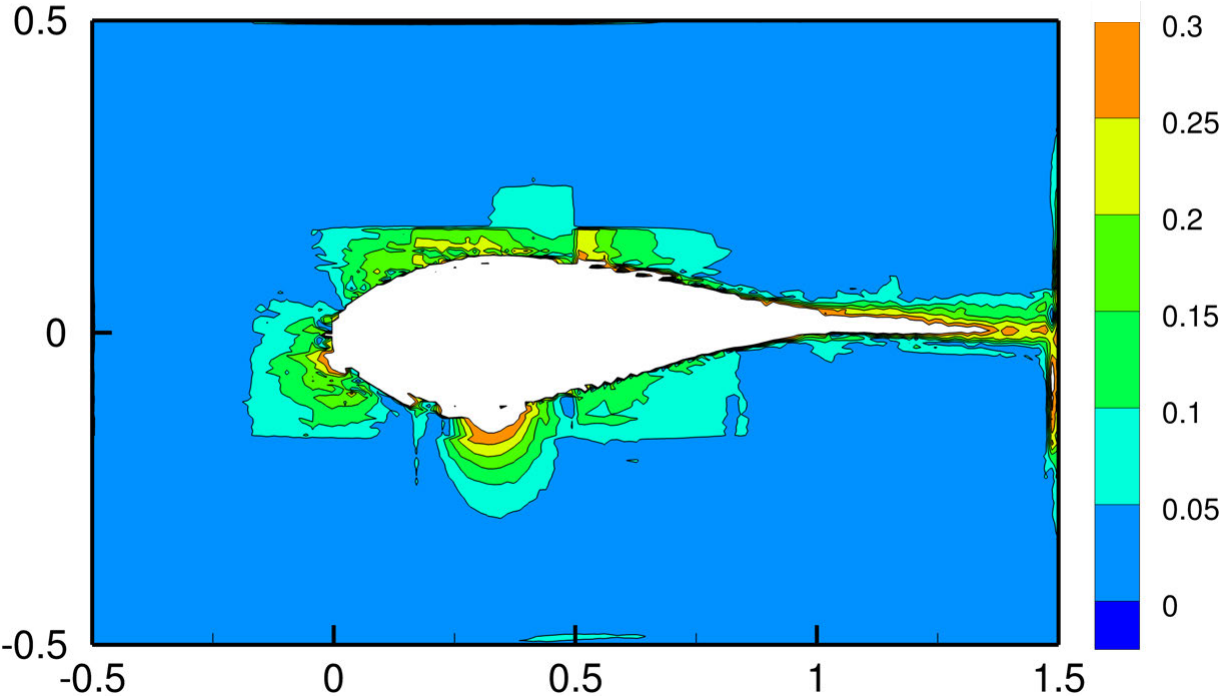}
		}
		\caption{Absolute difference.}
		\label{fig:fig26c}
	\end{subfigure}
	\caption{Ground truth vs. Prediction of the x-component of the velocity field ($U_t$ vs. $U_p$) and absolute difference ($|U_t-U_p|$) around the unseen airfoil at $\alpha = 9^\circ$ and $Re = 3\times10^6$.}
	\label{fig:fig26}
\end{figure}

\begin{figure}[H]
	\centering
	\begin{subfigure}{0.45\textwidth}
		\centering
		\resizebox{\textwidth}{!}{%
			\includegraphics{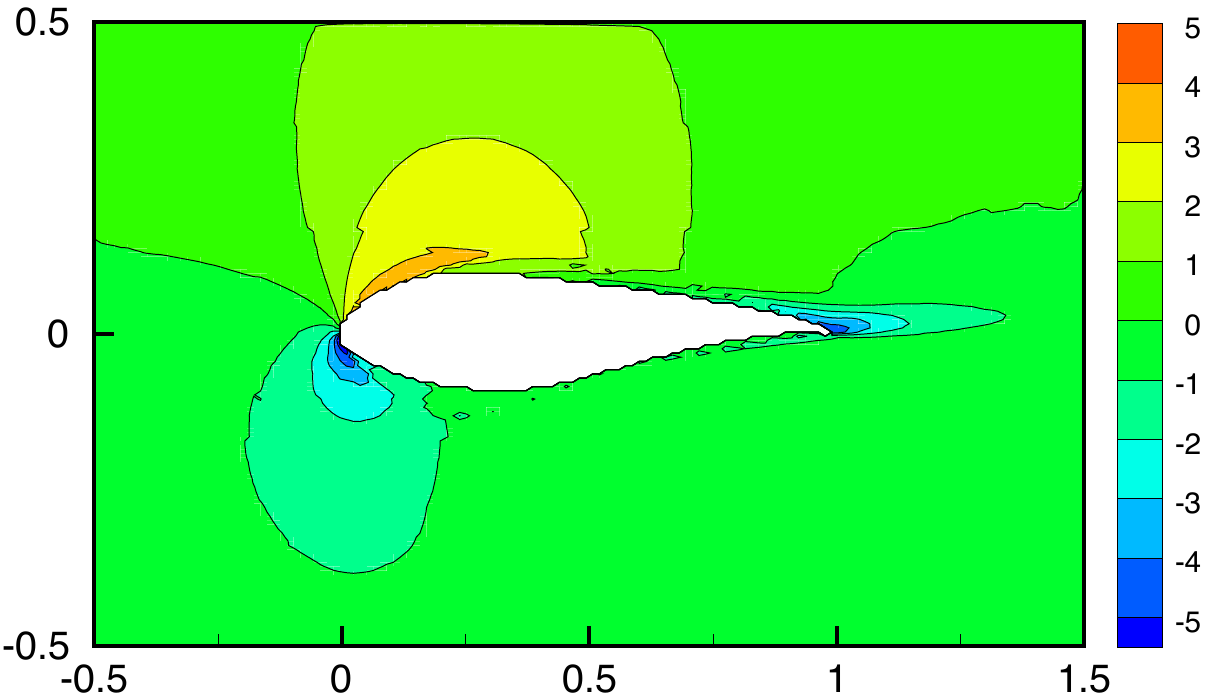}
		}
		\caption{Prediction ($U_p$).}
		\label{fig:fig27a}
	\end{subfigure}
	\begin{subfigure}{0.45\textwidth}
		\centering
		\resizebox{\textwidth}{!}{%
			\includegraphics{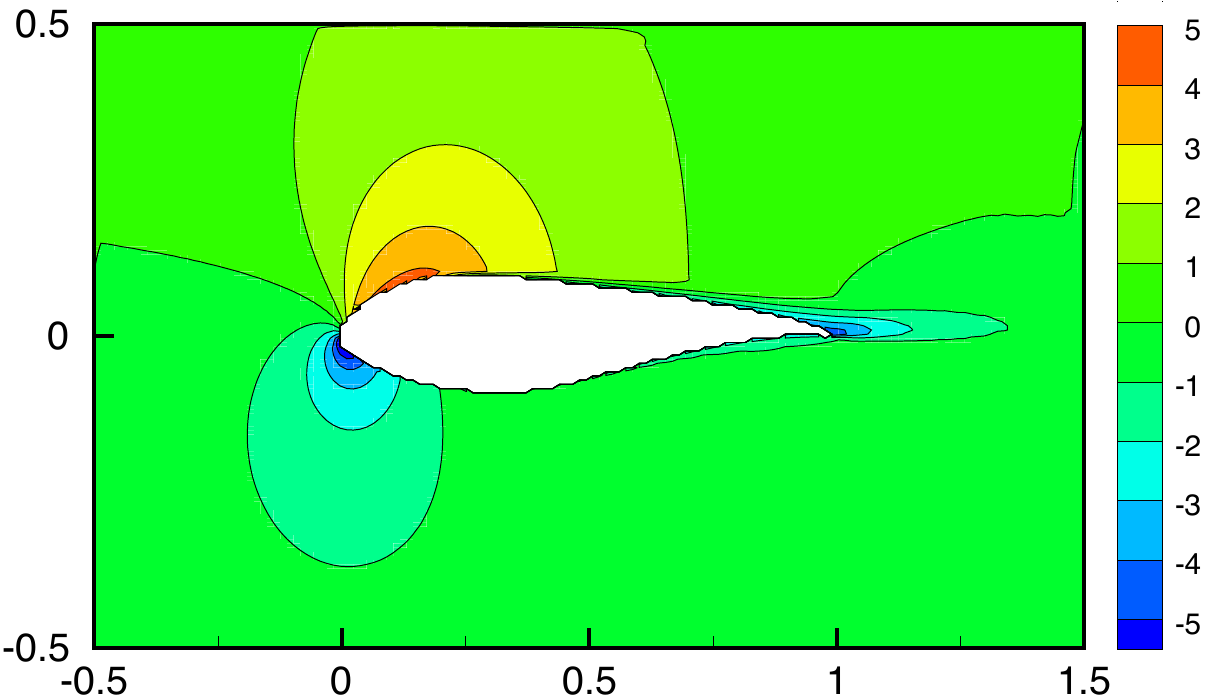}
		}
		\caption{Ground truth ($U_t$).}
		\label{fig:fig27b}
	\end{subfigure}

	\caption{Ground truth vs. Prediction of the x-component of the velocity field ($U_t$ vs. $U_p$) around the S807 airfoil at $\alpha = 7^\circ$ and $Re = 1\times10^6$.}
	\label{fig:fig27}
\end{figure}

\begin{figure}[H]
	\centering
	\begin{subfigure}{0.45\textwidth}
		\centering
		\resizebox{\textwidth}{!}{%
			\includegraphics{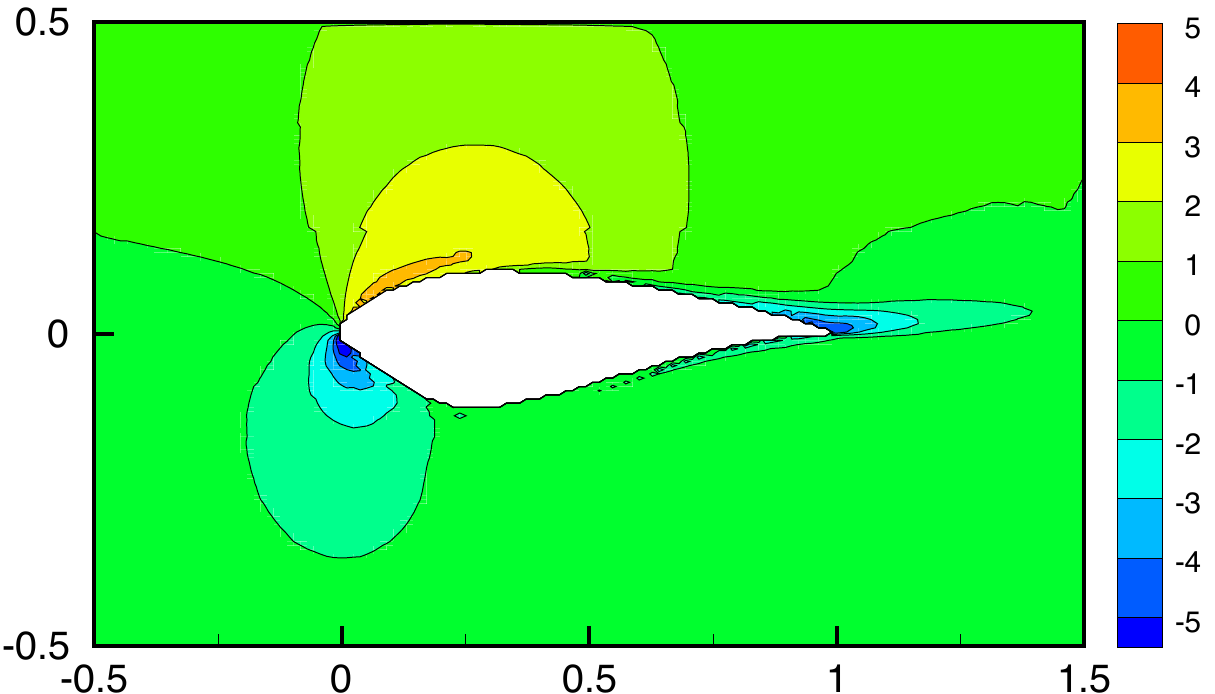}
		}
		\caption{Prediction ($U_p$).}
		\label{fig:fig28a}
	\end{subfigure}
	\begin{subfigure}{0.45\textwidth}
		\centering
		\resizebox{\textwidth}{!}{%
			\includegraphics{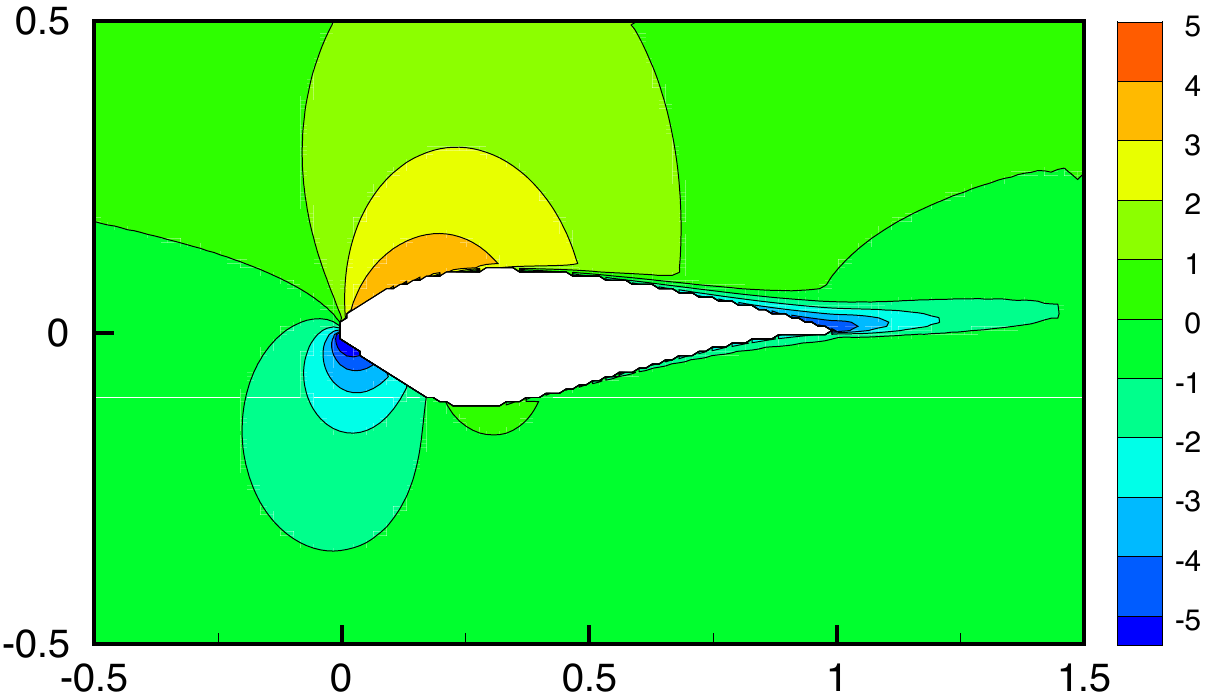}
		}
		\caption{Ground truth ($U_t$).}
		\label{fig:fig28b}
	\end{subfigure}
	
	\caption{Ground truth vs. Prediction of the x-component of the velocity field ($U_t$ vs. $U_p$) around the S819 airfoil at $\alpha = 7^\circ$ and $Re = 1\times10^6$.}
	\label{fig:fig28}
\end{figure}

Figures~\ref{fig:fig26},~\ref{fig:fig27},~\ref{fig:fig28} illustrate the prediction of the network on the unseen  airfoils in comparison to  CFD simulations. 

\begin{table}[H]
	\begin{center}
		\caption{MAPE for the components of the velocity field ($U$ and $V$ respectively) and pressure in the wake region of S807 and S819 airfoils respectively and the entire flow field around them.}
		\label{tab:tab6}
		\begin{tabular}{!{\vrule width 1pt}c|c|c|c!{\vrule width 1pt}}
		\hline\noalign{\hrule height 1pt}
		\textbf{Airfoil} & \textbf{Variable} & \textbf{Error in the} & \textbf{Error in the} \\ 
		~  & ~ & \textbf{wake region} & \textbf{entire flow} \\
		\hline\noalign{\hrule height 1pt}
		\textbf{New} &  U  & 10.55\% & 6.41\%  \\ 
		\textbf{New} &  V  & 5.71\% & 8.59\%  \\
		\textbf{New} &  P  &  4.41\% & 5.27\%  \\ \hline
		\textbf{S807} &  U  & 11.35\% & 8.4\%  \\ 
		\textbf{S807} &  V  & 2.2\% & 8.8\%  \\
		\textbf{S807} &  P  &  6.0\% & 7.6\%  \\ \hline
		\textbf{S819} &  U & 13.3\% & 10.3\%  \\
		\textbf{S819} & V & 2.4\% & 8.6\%  \\
		\textbf{S819} & P & 5.5\% & 7.5\%  \\
		\hline\noalign{\hrule height 1pt}
		\end{tabular}
	\end{center}
\end{table}

Table.~\ref{tab:tab6} provides a quantification of the results, and suggests good generalization properties of the network to an unseen shape. 

\section{Conclusions and Future Work}
\label{Conclusion}
A flexible approximation model based on convolutional neural networks was developed for efficient prediction of aerodynamic flow fields.
 Shared-encoding and decoding was used and found to be computationally more efficient compared to separated alternatives. The use of convolution operations, parameter sharing and robustness to noise using the gradient sharpening were shown to enhance  predictive capabilities. The Reynolds number,  angle of attack, and the shape of the airfoil in the form of a signed distance function are used as inputs to the network and the outputs are the velocity and pressure fields.
	
The  framework was utilized to predict the Reynolds Averaged Navier--Stokes flow field around different airfoil geometries under variable flow conditions. The network predictions on a single GPU were four orders of magnitude faster compared to the  RANS solver, at mean square error levels of less than 10\% over the entire flow field. Predictions were possible  with a small number of training simulations, and accuracy improvements were demonstrated by employing  gradient sharpening. Furthermore, the capability of the  network was evaluated for unseen airfoil shapes.

The results illustrate that the CNNs can enable near real-time simulation-based design and optimization, opening  avenues for an efficient design process. It is noteworthy that using three airfoil shapes in training, is a data limitation and reduces the general prediction behavior for unseen airfoil geometries from other families. Future work will seek to use a rich data set including multiple airfoil families in training and to augment the training data-sets to convert a set of input data into a broader set of slightly altered data~\citep{Shijie:2017} using operations such as translation and rotation.
This augmentation would effectively help the network from learning irrelevant patterns, and substantially boost the performance. 
Furthermore, exploring physical loss functions can be helpful in explicitly imposing physical constraints  such as the conservation of mass and momentum to the networks. 

\section*{Acknowledgements}
This work was supported by General Motors Corporation under a contract titled ``Deep Learning and Reduced Order Modeling for Automotive Aerodynamics.'' Computing resources were provided by the NSF via grant 1531752 MRI: Acquisition of Conflux, A Novel Platform for Data-Driven Computational Physics (Tech. Monitor: Stefan Robila). 

\section*{Appendix: Governing equations}
The RANS equations are derived by ensemble-averaging the conservation equations of mass, momentum and energy. These equations, for compressible flow are given by: 
\begin{align}
&\frac{\partial\bar\rho}{\partial t}+\frac{\partial\left(\bar\rho\hat u_i\right)}{\partial x_i} =0 \\
&\frac{\partial\left(\bar\rho\hat u_i\right)}{\partial t}+\frac{\partial\left(\bar\rho\hat u_i\hat u_j\right)}{\partial x_j} =-\frac{\partial\bar p}{\partial x_i}+\frac{\partial\bar\sigma_{ij}}{\partial x_j}+\frac{\partial\tau_{ij}}{\partial x_j} \\
&\frac{\partial\left(\bar\rho\hat E\right)}{\partial t}+\frac{\partial\left(\bar\rho\hat H\hat u_j\right)}{\partial x_j} = \frac{\partial}{\partial x_j}\left(\bar\sigma_{ij}\hat u_i+\overline{\sigma_{ij}u_i''}\right) - \\
&\nonumber \quad\quad \frac{\partial}{\partial x_j}\left(-\hat\kappa\frac{\partial\hat T}{\partial x_j}+c_P\overline{\rho u_j'' T''}-\hat u_i\tau_{ij}+\frac{1}{2}\overline{\rho u_i'' u_i'' u_j''}\right),
\end{align}
where the overbar indicates conventional time-average mean,	$u_i$ is the fluid velocity, $\rho$ is the density, $p$ is the pressure, $\tau_{ij}$ is the Reynolds stress term, $c_P$ is the heat capacity at constant pressure, and $\kappa$ is the kinetic energy of the fluctuating field (local turbulent kinetic energy). 
The density weighted time averaging (Favre averaging) of any quantity $\xi$, denoted by $\hat\xi$ is given as $\hat\xi=\overline{\rho\xi}/\bar\rho$, where,
\begin{align}
&\hat H=\hat E+\frac{\bar p}{\bar\rho},\\
&\bar\sigma_{ij}=\mu_t\left(\frac{\partial\hat u_i}{\partial x_j}+\frac{\partial\hat u_j}{\partial x_i}-\frac{2}{3}\frac{\partial\hat u_k}{\partial x_k}\delta_{ij}\right),\\
&\tau_{ij}=-\overline{\rho u_i'' u_j''},\\
&k=\frac{\widehat{u_i''^2}+\widehat{v_i''^2}+\widehat{w_i''^2}}{2},\\
&\bar p = (\gamma-1)\bar\rho\left[\hat E - \frac{\hat u^2+\hat v^2+\hat w^2}{2} - k\right].
\end{align}

To provide closure to the above equations, we use the model proposed by~\cite{Spalart:1992}. In this closure, the Boussinesq hypothesis relates the Reynolds stress and the effect of turbulence as an eddy viscosity $\mu_t$.
Employing the Boussinesq approach, and Reynolds Analogy a transport equation for a working variable $\tilde\nu$ is solved to estimate the eddy viscosity field at every iteration.
\begin{align}
\label{eq:eq10}
&\frac{\partial \tilde{\nu}}{\partial t}+u_j\frac{\partial \tilde{\nu}}{\partial x_j} =  C_{b1}\left[1-f_{t2}\right]\tilde{S}\tilde{\nu} + \\
\nonumber &~\frac{1}{\sigma}\left\lbrace\nabla\cdot\left[\left(\nu+\tilde{\nu}\right)\nabla\tilde{\nu}\right]+C_{b2}\left|\nabla\tilde{\nu}\right|^2\right\rbrace
-\left[C_{w1}f_w-\frac{C_{b1}}{\kappa^2}f_{t2}\right]\left(\frac{\tilde{\nu}}{d}\right)^2.
\end{align}
The turbulent eddy viscosity is computed as $\mu_t=\bar\rho\tilde\nu f_{v1}$, where,
\begin{align}
&\nonumber f_{v1}=\frac{\chi^3}{\chi^3+C_{v1}^3},~\chi=\frac{\tilde{\nu}}{\nu},~\nu=\frac{\mu}{\bar\rho},\\
&\nonumber f_{t2}=C_{t3} \exp \left(-C_{t4}\chi^2\right),\\
&\nonumber \tilde{S}=S+\frac{\tilde{\nu}}{\kappa^2 d^2}f_{v2},\\
&\nonumber S=\sqrt{2\Omega_{ij}\Omega{ij}},~f_{v2}=1-\frac{\chi}{1+\chi f_{v1}},\\
&\nonumber f_w=g\left[\frac{1+C_{w3}^6}{g^6+C_{w3}^6}\right]^{1/6}, \\ 
&\nonumber g=r+C_{w2}(r^6-r),~
r=\frac{\tilde{\nu}}{\tilde{S}\kappa^2d^2}, \\
&\nonumber C_{w1}=\frac{C_{b1}}{\kappa^2}+\frac{1+C_{b2}}{\sigma},~	C_{b1}=0.1355,~\sigma=2/3,~C_{b2}=0.622,\\
&\nonumber \kappa=0.41,~C_{w2}=0.3,~
C_{w3}=2.0,~C_{v1}=7.1,~C_{t3}=1.2,~C_{t4}=0.5.
\end{align}
The first term on the right hand side of this Eq.~\ref{eq:eq10} is the production term for $\tilde\nu$ while the second term represents dissipation.


\begin{thebibliography}{48}
\providecommand{\natexlab}[1]{#1}
\providecommand{\url}[1]{{#1}}
\providecommand{\urlprefix}{URL }
\expandafter\ifx\csname urlstyle\endcsname\relax
  \providecommand{\doi}[1]{DOI~\discretionary{}{}{}#1}\else
  \providecommand{\doi}{DOI~\discretionary{}{}{}\begingroup
  \urlstyle{rm}\Url}\fi
\providecommand{\eprint}[2][]{\url{#2}}

\bibitem[{Aksoy and Haralick(2000)}]{Aksoy:2000}
Aksoy S, Haralick RM (2000) Feature normalization and likelihood-based
  similarity measures for image retrieval. Pattern Recognition Letters
  22:563--582

\bibitem[{Amidror(2002)}]{Amidror:2002}
Amidror I (2002) Scattered data interpolation methods for electronic imaging
  systems: a survey. J Electronic Imaging 11(2):157--176

\bibitem[{{Aranake} et~al(2012){Aranake}, {Lakshminarayan}, and
  {Duraisamy}}]{Aranake:2012}
{Aranake} A, {Lakshminarayan} V, {Duraisamy} K (2012) Assessment of transition
  model and cfd methodology for wind turbine flows. In: 42nd AIAA Fluid
  Dynamics Conference and Exhibit, American Institute of Aeronautics and
  Astronautics, p 2720

\bibitem[{Bengio(2009)}]{Bengio:2009}
Bengio Y (2009) Learning deep architectures for ai. Foundations and Trends in
  Machine Learning 2(1):1--127

\bibitem[{Carr et~al(2001)Carr, Beatson, Cherrie, Mitchell, Fright, McCallum,
  and Evans}]{Carr:2001}
Carr JC, Beatson RK, Cherrie JB, Mitchell TJ, Fright WR, McCallum BC, Evans TR
  (2001) Reconstruction and representation of 3d objects with radial basis
  functions. In: Proceedings of the 28th Annual Conference on Computer Graphics
  and Interactive Techniques, ACM, New York, NY, USA, SIGGRAPH '01, pp 67--76

\bibitem[{{Chollampatt} and {Tou Ng}(2018)}]{Chollampatt:2018}
{Chollampatt} S, {Tou Ng} H (2018) {A Multilayer Convolutional Encoder-Decoder
  Neural Network for Grammatical Error Correction}. arXiv e-prints
  arXiv:1801.08831, \eprint{1801.08831}

\bibitem[{Duraisamy(2005)}]{Duraisamy:2005}
Duraisamy K (2005) Studies in tip vortex formation, evolution and contro, dept
  of aerospace engineering, univ of maryland. PhD thesis, University of
  Maryland

\bibitem[{Duraisamy et~al(2019)Duraisamy, Iaccarino, and Xiao}]{Duraisamy:2019}
Duraisamy K, Iaccarino G, Xiao H (2019) Turbulence modeling in the age of data.
  Annual Review of Fluid Mechanics 51(1):357--377

\bibitem[{{Fernando} et~al(2015){Fernando}, {Karaoglu}, and
  {Saha}}]{Fernando:2015}
{Fernando} B, {Karaoglu} S, {Saha} SK (2015) {Object Class Detection and
  Classification using Multi Scale Gradient and Corner Point based Shape
  Descriptors}. arXiv e-prints arXiv:1505.00432, \eprint{1505.00432}

\bibitem[{Foley et~al(1995)Foley, van Dam, Feiner, and Hughes}]{Foley:1995}
Foley JD, van Dam A, Feiner SK, Hughes JF (1995) Computer Graphics: Principles
  and Practice in C, second edition edn. Addison-Wesley

\bibitem[{Fuhrmann and Goesele(2014)}]{Fuhrmann:2014}
Fuhrmann S, Goesele M (2014) Floating scale surface reconstruction. ACM Trans
  Graph 33(4):46:1--46:11

\bibitem[{Guo et~al(2016)Guo, Li, and Iorio}]{Guo:2016}
Guo X, Li W, Iorio F (2016) Convolutional neural networks for steady flow
  approximation. In: Proceedings of the 22Nd ACM SIGKDD International
  Conference on Knowledge Discovery and Data Mining, ACM, KDD '16, pp 481--490

\bibitem[{{Hennigh}(2017)}]{hennigh:2017b}
{Hennigh} O (2017) {Lat-Net: Compressing Lattice Boltzmann Flow Simulations
  using Deep Neural Networks}. arXiv e-prints arXiv:1705.09036,
  \eprint{1705.09036}

\bibitem[{Hoppe et~al(1992)Hoppe, DeRose, Duchamp, McDonald, and
  Stuetzle}]{Hoppe:1992}
Hoppe H, DeRose T, Duchamp T, McDonald J, Stuetzle W (1992) Surface
  reconstruction from unorganized points. SIGGRAPH Comput Graph 26(2):71--78

\bibitem[{J\"{u}rgen(2015)}]{Schmidhuber:2015}
J\"{u}rgen S (2015) Deep learning in neural networks: An overview. Neural
  Networks 61:85--117

\bibitem[{Kazhdan et~al(2006)Kazhdan, Bolitho, and Hoppe}]{Kazhdan:2006}
Kazhdan M, Bolitho M, Hoppe H (2006) {Poisson Surface Reconstruction}. In:
  Sheffer A, Polthier K (eds) Symposium on Geometry Processing, The
  Eurographics Association, pp 61--70

\bibitem[{Koren(1993)}]{Koren:1993}
Koren B (1993) A robust upwind discretization method for advection, diffusion
  and source terms, Vieweg, pp 117--138

\bibitem[{Lakshminarayan and Baeder(2010)}]{Lakshminarayan:2010}
Lakshminarayan VK, Baeder JD (2010) Computational investigation of microscale
  coaxial-rotor aerodynamics in hover. Journal of Aircraft 47(3):940--955

\bibitem[{Lecun et~al(1998)Lecun, Bottou, Bengio, and Haffner}]{Lecun:1998}
Lecun Y, Bottou L, Bengio Y, Haffner P (1998) Gradient-based learning applied
  to document recognition. Proceedings of the IEEE 86(11):2278--2324

\bibitem[{{Lee} and {You}(2017)}]{Lee:2017}
{Lee} S, {You} D (2017) {Prediction of laminar vortex shedding over a cylinder
  using deep learning}. arXiv e-prints arXiv:1712.07854, \eprint{1712.07854}

\bibitem[{{Lee} and {You}(2018)}]{Lee:2018}
{Lee} S, {You} D (2018) {Data-driven prediction of unsteady flow fields over a
  circular cylinder using deep learning}. arXiv e-prints arXiv:1804.06076,
  \eprint{1804.06076}

\bibitem[{van Leer(1979)}]{VanLeer:1979}
van Leer B (1979) Towards the ultimate conservative difference scheme. v. a
  second-order sequel to godunov's method. Journal of Computational Physics
  32(1):101--136

\bibitem[{Ling and Jacobs(2007)}]{Ling:2007}
Ling H, Jacobs DW (2007) Shape classification using the inner-distance. IEEE
  Transactions on Pattern Analysis and Machine Intelligence 29(2):286--299

\bibitem[{{Mathieu} et~al(2015){Mathieu}, {Couprie}, and
  {LeCun}}]{Mathieu:2015}
{Mathieu} M, {Couprie} C, {LeCun} Y (2015) {Deep multi-scale video prediction
  beyond mean square error}. arXiv e-prints arXiv:1511.05440,
  \eprint{1511.05440}

\bibitem[{Medida and Baeder(2011)}]{Medida:2011}
Medida S, Baeder J (2011) Application of the Correlation-based Gamma-Re Theta t
  Transition Model to the Spalart-Allmaras Turbulence Model, American Institute
  of Aeronautics and Astronautics

\bibitem[{{Miyanawala} and {Jaiman}(2017)}]{Miyanawala:2017}
{Miyanawala} TP, {Jaiman} RK (2017) {An Efficient Deep Learning Technique for
  the Navier-Stokes Equations: Application to Unsteady Wake Flow Dynamics}.
  arXiv e-prints arXiv:1710.09099, \eprint{1710.09099}

\bibitem[{Pandya et~al(2003)Pandya, Venkateswaran, and Pulliam}]{Pandya:2003}
Pandya S, Venkateswaran S, Pulliam T (2003) Implementation of Preconditioned
  Dual-Time Procedures in OVERFLOW, American Institute of Aeronautics and
  Astronautics

\bibitem[{{Prantl} et~al(2017){Prantl}, {Bonev}, and {Thuerey}}]{Prantl:2017}
{Prantl} L, {Bonev} B, {Thuerey} N (2017) {Generating Liquid Simulations with
  Deformation-aware Neural Networks}. arXiv e-prints arXiv:1704.07854,
  \eprint{1704.07854}

\bibitem[{Pulliam and Chaussee(1981)}]{Pulliam:1981}
Pulliam T, Chaussee D (1981) A diagonal form of an implicit
  approximate-factorization algorithm. Journal of Computational Physics
  39(2):347--363

\bibitem[{Raissi and Karniadakis(2018)}]{Raissi:2018b}
Raissi M, Karniadakis GE (2018) Hidden physics models: Machine learning of
  nonlinear partial differential equations. Journal of Computational Physics
  357:125--141

\bibitem[{{Raissi} et~al(2018){Raissi}, {Yazdani}, and
  {Karniadakis}}]{Raissi:2018}
{Raissi} M, {Yazdani} A, {Karniadakis} GE (2018) {Hidden Fluid Mechanics: A
  Navier-Stokes Informed Deep Learning Framework for Assimilating Flow
  Visualization Data}. arXiv e-printse-prints p arXiv:1808.04327

\bibitem[{Raissi et~al(2019)Raissi, Perdikaris, and Karniadakis}]{Raissi:2019}
Raissi M, Perdikaris P, Karniadakis G (2019) Physics-informed neural networks:
  A deep learning framework for solving forward and inverse problems involving
  nonlinear partial differential equations. Journal of Computational Physics
  378:686--707

\bibitem[{{Ramachandran} et~al(2017){Ramachandran}, {Zoph}, and
  {Le}}]{Ramachandran:2017}
{Ramachandran} P, {Zoph} B, {Le} QV (2017) {Searching for Activation
  Functions}. arXiv e-prints arXiv:1710.05941, \eprint{1710.05941}

\bibitem[{Roe(1986)}]{Roe:1986}
Roe PL (1986) Characteristic-based schemes for the euler equations. Annual
  Review of Fluid Mechanics 18(1):337--365

\bibitem[{Sethian(1996)}]{Sethian:1996}
Sethian JA (1996) A fast marching level set method for monotonically advancing
  fronts. Proc\ Natl\ Acad\ Sci\ USA 93:1591--1595

\bibitem[{{Shijie} et~al(2017){Shijie}, {Ping}, {Peiyi}, and
  {Siping}}]{Shijie:2017}
{Shijie} J, {Ping} W, {Peiyi} J, {Siping} H (2017) Research on data
  augmentation for image classification based on convolution neural networks.
  In: 2017 Chinese Automation Congress (CAC), pp 4165--4170

\bibitem[{Somers(1997{\natexlab{a}})}]{Somers:1997a}
Somers D (1997{\natexlab{a}}) Design and experimental results for the s805
  airfoil. Tech. Rep. NREL/SR-440-6917

\bibitem[{Somers(1997{\natexlab{b}})}]{Somers:1997b}
Somers DM (1997{\natexlab{b}}) Design and experimental results for the s809
  airfoil. Tech. Rep. NREL/SR-440-6918

\bibitem[{Somers(2004)}]{Somers:2004}
Somers DM (2004) S814 and s815 airfoils: October 1991--july 1992. Tech. rep.

\bibitem[{Spalart and Allmaras(1992)}]{Spalart:1992}
Spalart P, Allmaras S (1992) A one-equation turbulence model for aerodynamic
  flows, American Institute of Aeronautics and Astronautics

\bibitem[{Taylor et~al(2010)Taylor, Fergus, LeCun, and Bregler}]{Taylor:2010}
Taylor GW, Fergus R, LeCun Y, Bregler C (2010) Convolutional learning of
  spatio-temporal features. In: Computer Vision -- ECCV 2010, pp 140--153

\bibitem[{{Tompson} et~al(2016){Tompson}, {Schlachter}, {Sprechmann}, and
  {Perlin}}]{Tompson:2016}
{Tompson} J, {Schlachter} K, {Sprechmann} P, {Perlin} K (2016) {Accelerating
  Eulerian Fluid Simulation With Convolutional Networks}. arXiv e-prints
  arXiv:1607.03597, \eprint{1607.03597}

\bibitem[{Turkel(1999)}]{Turkel:1999}
Turkel E (1999) Preconditioning techniques in computational fluid dynamics.
  Annual Review of Fluid Mechanics 31(1):385--416

\bibitem[{{Xu} et~al(2015){Xu}, {Kim}, {Huang}, and {Kalogerakis}}]{Xu:2015}
{Xu} K, {Kim} VG, {Huang} Q, {Kalogerakis} E (2015) {Data-Driven Shape Analysis
  and Processing}. arXiv e-prints arXiv:1502.06686, \eprint{1502.06686}

\bibitem[{Zhang and Lu(2004)}]{Zhang:2004a}
Zhang D, Lu G (2004) Review of shape representation and description techniques.
  Pattern Recognition 37(1):1--19, \urlprefix\url{doi:
  10.1016/j.patcog.2003.07.008}

\bibitem[{{Zhang} et~al(2017){Zhang}, {Sung}, and {Mavris}}]{Zhang:2017}
{Zhang} Y, {Sung} WJ, {Mavris} D (2017) {Application of Convolutional Neural
  Network to Predict Airfoil Lift Coefficient}. arXiv e-prints
  arXiv:1712.10082, \eprint{1712.10082}

\bibitem[{Zhao(2005)}]{Zhao:2005}
Zhao H (2005) A fast sweeping method for {E}ikonal equations. Math\ Comput
  74(250):603--627

\bibitem[{Zuo et~al(2015)Zuo, Shuai, Wang, Liu, Wang, Wang, and
  Chen}]{Zuo:2015}
Zuo Z, Shuai B, Wang G, Liu X, Wang X, Wang B, Chen Y (2015) Convolutional
  recurrent neural networks: Learning spatial dependencies for image
  representation. In: 2015 IEEE Conference on Computer Vision and Pattern
  Recognition Workshops (CVPRW), pp 18--26

\end{thebibliography}
\bibliographystyle{spbasic}

\end{document}